\documentclass[twocolumn]{aastex63}
\usepackage{color}
\usepackage[titletoc]{appendix}
\usepackage[fleqn]{amsmath}
\usepackage{amssymb}
\usepackage{mathtools}
\usepackage{upgreek}
\usepackage{float}
\usepackage{comment}
\usepackage{enumitem}
\usepackage{natbib}
\usepackage{graphicx}
\usepackage{bm}
\usepackage{totcount}
\usepackage{multirow}

\newtotcounter{citnum} 
\def\oldbibitem{} \let\oldbibitem=\bibitem
\def\bibitem{\stepcounter{citnum}\oldbibitem}

\shortauthors{Millholland et al.}
\shorttitle{The Kepler Continuum}

\begin{document} 

\defcitealias{2016ApJS..225....9H}{H16}
\defcitealias{2019MNRAS.490.4575H}{H19}
\defcitealias{2020AJ....160..276H}{H20}

\title{Evidence for a Non-Dichotomous Solution to the Kepler Dichotomy: \\ Mutual Inclinations of Kepler Planetary Systems from Transit Duration Variations} 

\author[0000-0003-3130-2282]{Sarah C. Millholland}
\altaffiliation{NASA Sagan Fellow}
\affiliation{Department of Astrophysical Sciences, Princeton University, 4 Ivy Lane, Princeton, NJ 08544, USA}
\email{sarah.millholland@princeton.edu}

\author[0000-0002-5223-7945]{Matthias Y. He}
\affiliation{Department of Astronomy \& Astrophysics, 525 Davey Laboratory, The Pennsylvania State University, University Park, PA 16802, USA}
\affiliation{Center for Exoplanets \& Habitable Worlds, 525 Davey Laboratory, The Pennsylvania State University, University Park, PA 16802, USA}
\affiliation{Center for Astrostatistics, 525 Davey Laboratory, The Pennsylvania State University, University Park, PA 16802, USA}
\affiliation{Institute for Computational \& Data Sciences, 525 Davey Laboratory, The Pennsylvania State University, University Park, PA 16802, USA}

\author[0000-0001-6545-639X]{Eric B. Ford}
\affiliation{Department of Astronomy \& Astrophysics, 525 Davey Laboratory, The Pennsylvania State University, University Park, PA 16802, USA}
\affiliation{Center for Exoplanets \& Habitable Worlds, 525 Davey Laboratory, The Pennsylvania State University, University Park, PA 16802, USA}
\affiliation{Center for Astrostatistics, 525 Davey Laboratory, The Pennsylvania State University, University Park, PA 16802, USA}
\affiliation{Institute for Computational \& Data Sciences, 525 Davey Laboratory, The Pennsylvania State University, University Park, PA 16802, USA}
\affiliation{Institute for Advanced Study, Einstein Drive, Princeton, NJ 08540, USA}

\author[0000-0003-1080-9770]{Darin Ragozzine}
\affiliation{Department of Physics \& Astronomy, N283 ESC, Brigham Young University, Provo, UT 84602, USA}

\author[0000-0003-3750-0183]{Daniel Fabrycky}
\affiliation{Department of Astronomy and Astrophysics, University of Chicago, 5640 S Ellis Ave, Chicago, IL 60637, USA}

\author[0000-0002-4265-047X]{Joshua N.\ Winn}
\affiliation{Department of Astrophysical Sciences, Princeton University, 4 Ivy Lane, Princeton, NJ 08544, USA}

\begin{abstract}
Early analyses of exoplanet statistics from the Kepler Mission revealed that a model population of multiple-planet systems with low mutual inclinations (${\sim1^{\circ}-2^{\circ}}$) adequately describes the multiple-transiting systems but underpredicts the number of single-transiting systems. This so-called ``Kepler dichotomy’’ signals the existence of a sub-population of multi-planet systems possessing larger mutual inclinations. However, the details of these inclinations remain uncertain. In this work, we derive constraints on the intrinsic mutual inclination distribution by statistically exploiting Transit Duration Variations (TDVs) of the Kepler planet population. When planetary orbits are mutually inclined, planet-planet interactions cause orbital precession, which can lead to detectable long-term changes in transit durations. These TDV signals are inclination-sensitive and have been detected for roughly two dozen Kepler planets. We compare the properties of the Kepler observed TDV detections to TDV detections of simulated planetary systems constructed from two population models with differing assumptions about the mutual inclination distribution. We find strong evidence for a continuous distribution of relatively low mutual inclinations that is well-characterized by a power law relationship between the median mutual inclination ($\tilde{\mu}_{i,n}$) and the intrinsic multiplicity ($n$): $\tilde{\mu}_{i,n} = \tilde{\mu}_{i,5}(n/5)^{\alpha}$, where $\tilde{\mu}_{i,5} = 1.10^{+0.15}_{-0.11}$ and $\alpha = -1.73^{+0.09}_{-0.08}$. These results suggest that late-stage planet assembly and possibly stellar oblateness are the dominant physical origins for the excitation of Kepler planet mutual inclinations.
\\
\end{abstract}

\section{Introduction}
\label{sec: Introduction}

One of the fundamental features of our Solar System is the near-coplanarity of the planetary orbits.  The orbital inclinations of the Solar System planets relative to the invariable plane
are a few degrees or less, with the exception of Mercury, which is inclined by $6^{\circ}$. This coplanarity was one of the key observables leading to the nebular hypothesis \citep{Kant1755, Laplace1796} and the modern-day paradigm of planet formation in thin gaseous disks \citep[e.g.][]{2011ARA&A..49..195A}. It is unclear whether most planetary systems retain their primordially small inclinations (like the Solar System) or develop larger misalignments. Accordingly, constraining the mutual inclination distribution of exoplanetary systems is a large-scale objective with fundamental implications for planet formation theory \citep[e.g.][]{2013ApJ...775...53H} and population statistics \citep[e.g.][]{2012AJ....143...94T}. 

Although the transit and Doppler methods do not generally reveal the
mutual inclinations of multiple-planet systems
without further assumptions,
there have still been several advancements in our knowledge of the mutual inclinations of the population of short-period ($P \lesssim 1$ yr), tightly-packed systems discovered by the Kepler mission \citep{2010Sci...327..977B, 2013ApJS..204...24B, 2011ApJS..197....8L, 2014ApJ...790..146F}. (See \citealt{2021arXiv210302127Z} for a recent review.) Roughly half of Sun-like stars host these short-period planets \citep{2015ARA&A..53..409W, 2018ApJ...860..101Z, 2019MNRAS.490.4575H, 2020AJ....159..164Y}, most of which are likely in multi-planet (but not necessarily multi-transiting) systems \citep{2018AJ....156...24M, 2020AJ....160..276H}. Such systems will be our primary focus in this paper. 

Kepler systems of multiple-transiting planets (``Kepler multis'') are nearly coplanar on average \citep[e.g.][]{2011ApJS..197....8L}.\footnote{Ultra-short-period planets are a notable exception \citep{2018ApJ...864L..38D}.} This conclusion has
been reached by analyzing, for instance, the distribution of ratios of transit chord lengths between pairs of planets within the same system
\citep{2010ApJ...725.1226S, 2012ApJ...761...92F, 2014ApJ...790..146F}. According to these studies, at least half of Kepler systems have mutual inclinations that are consistent with a Rayleigh distribution with scale parameter $\sigma_i \sim 1^{\circ} - 2^{\circ}$.

Within the Kepler sample, there is also evidence for a subset of systems with larger mutual inclinations, based on analyses of the observed transiting multiplicity distribution. Inferences from this distribution are difficult, though, because it depends on both the intrinsic multiplicity distribution and the mutual inclination distribution \citep[e.g.][]{2011ApJS..197....8L, 2012AJ....143...94T}. To break the degeneracy,
additional assumptions
or observations are required, such as inputs from radial velocity (RV) surveys or the incidence of transit-timing variations (TTVs).

Several studies used statistical planet population models to fit the observed transiting multiplicity distribution of Kepler systems \citep[e.g.][]{2011ApJS..197....8L, 2012AJ....143...94T, 2012ApJ...761...92F, 2016ApJ...816...66B}. These studies identified an apparent excess of single transiting systems (``singles''), compared to
the number of singles one would expect if all planetary systems have the same $\sim1^{\circ}-2^{\circ}$ mutual inclination dispersion (and intrinsic multiplicity distribution) of the transit multis. This apparent discrepancy between the observed and expected number of singles is known as the ``Kepler dichotomy'', and it is sometimes interpreted as evidence for two (or more) planet populations with distinct architectures \citep[e.g.][]{2011ApJS..197....8L, 2012ApJ...758...39J, 2013ApJ...775...53H, 2016ApJ...816...66B}. Biases due to detection order in multiple-transiting systems likely contribute to the Kepler dichotomy \citep{2019MNRAS.483.4479Z} but cannot entirely explain it \citep{2019MNRAS.490.4575H}.

One proposed resolution to the Kepler dichotomy is to invoke a substantial fraction of intrinsically single-planet systems \citep{2012ApJ...761...92F, 2019MNRAS.489.3162S}. While such a model can fit the observed multiplicity distribution, \cite{2019MNRAS.490.4575H} showed that it is an unlikely
explanation because it requires too many stars to have only one planet. Specifically, after assigning nearly coplanar multi-planet systems to target stars with an occurrence derived from the transiting multi-planet systems, the remaining proportion of stars is not large enough to account for the observed number of transit
singles. Moreover, the observed incidence of TTVs of Kepler planets does not provide evidence for a large population of intrinsic singles \citep{2011ApJS..197....2F, 2016ApJS..225....9H}.

\begin{figure*}
\epsscale{1.1}
\plotone{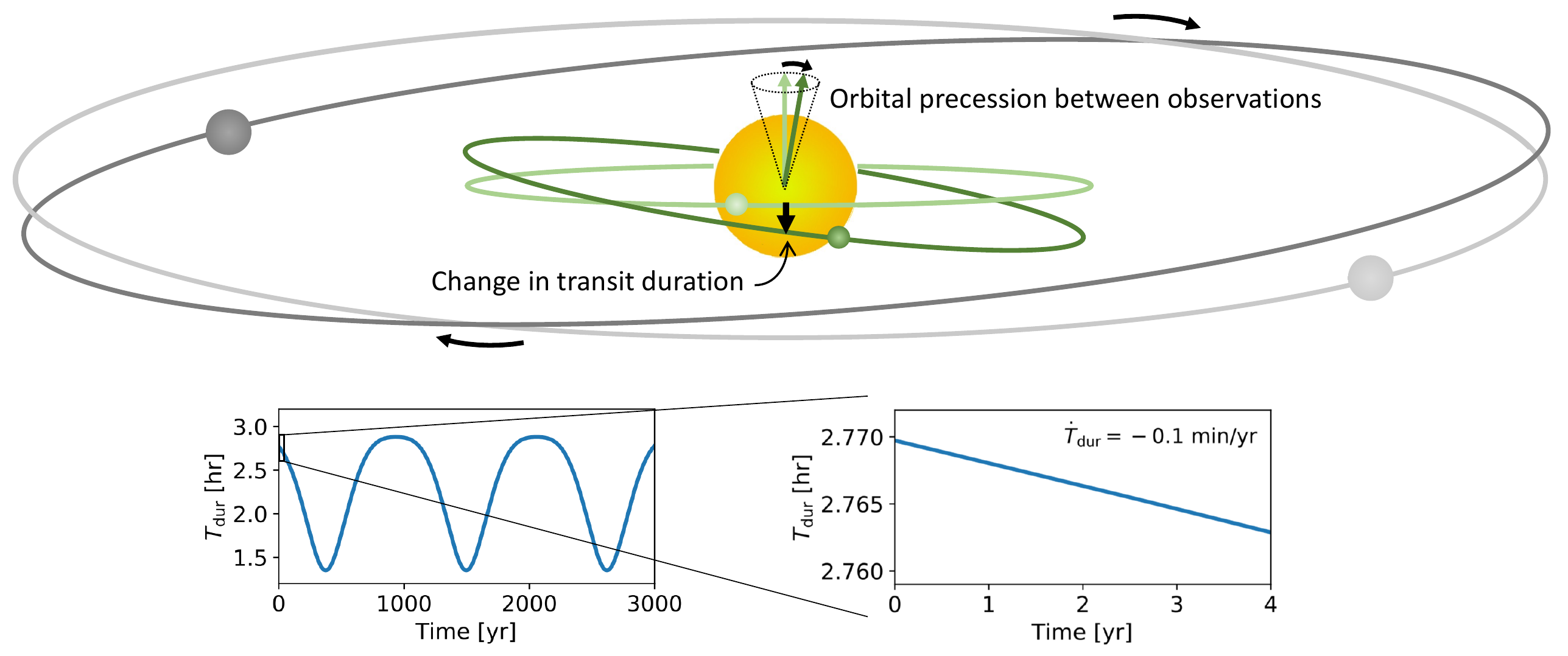}
\caption{Schematic illustration of the physical set-up and concept of this study. A transiting planet experiences orbital precession induced by secular perturbations from a companion planet, which is non-transiting in this illustration. The orbital precession leads to changes in the transiting planet's inclination with respect to the line-of-sight (but not to the invariable plane) and the transit duration, $T_{\mathrm{dur}}$, here defined as the duration during which the center of the planet is projected in front of the stellar disk. Over long timescales ($10^{2-3}$ years), these transit duration variations (TDVs) manifest as an oscillation at the orbital precession period, as indicated with the example time evolution plot on the bottom left. However, over short timescales, such as the $\sim 4$ year baseline of the Kepler prime mission, the TDVs manifest as a linear trend with slope $\dot{T}_{\mathrm{dur}}$. This is indicated with the zoom-in on the bottom right. The TDV slope depends on the mutual orbital inclinations. Examining the TDVs of an ensemble of planets thus allows us to place constraints on the distribution of mutual inclinations.} 
\label{fig: geometric diagram}
\end{figure*}

The leading explanation of the Kepler dichotomy is that there is a distribution of mutual inclinations that includes a substantial fraction of systems with low mutual inclinations and a separate fraction of systems with significantly larger mutual inclinations.   For example, there might be two (or more) sub-populations, one with $1^{\circ}-2^{\circ}$ mutual inclinations and one with $>30^{\circ}$ mutual inclinations.  An alternative is a smooth distribution, such as a model in which the inclinations are drawn from a Rayleigh distribution with a width parameter that is itself a  variable drawn from a smooth distribution \citep[e.g.][]{2011ApJS..197....8L} or that depends on the number 
of planets in the system
\citep[e.g.][]{2016ApJ...832...34M, 2018ApJ...860..101Z}. However, both the statistical properties and physical origins of these variations are uncertain.

Recently, \cite{2019MNRAS.490.4575H, 2020AJ....160..276H} built a forward modeling framework capable of fitting a statistical description of the underlying planet population to the survey data by ``observing'' simulated planetary systems with the Kepler detection pipeline. \cite{2019MNRAS.490.4575H}, hereafter \citetalias{2019MNRAS.490.4575H}, showed that the observed multiplicity distribution (together with several other aspects of the Kepler population, including the distributions of orbital periods, period ratios, transit depths, depth ratios, and transit durations) can be described by two populations consisting of a low and high mutual inclination component, with $\sigma_{i,\mathrm{low}}\approx1^{\circ}-2^{\circ}$, $\sigma_{i,\mathrm{high}}\approx30^{\circ}-65^{\circ}$, and  $\sim40
\%$ of systems having high mutual inclinations. \citetalias{2019MNRAS.490.4575H}'s findings are similar to \cite{2018AJ....156...24M}, who used an analogous forward modeling framework and also found the data to be compatible with a dichotomous two-population model.

A more recent model by \cite{2020AJ....160..276H}, hereafter \citetalias{2020AJ....160..276H}, showed that the observations can also be reproduced when the mutual inclinations (and eccentricities) are distributed according to a stability limit dictated by the system's total angular momentum deficit \citep{2017A&A...605A..72L}. An interesting prediction of this model is an inverse correlation between the mutual inclination dispersion and the intrinsic multiplicity that is well-described by a power law, $\sigma_{i,n}\propto n^{\alpha}$ (with $n$ the multiplicity and $\alpha<0$). This feature is qualitatively similar to the model of \cite{2018ApJ...860..101Z}, and both  lead to the same key conclusion that the mutual inclination distribution does not necessarily have to be dichotomous, but rather it can be characterized by a broad and multiplicity-dependent distribution.

The \citetalias{2019MNRAS.490.4575H} and \citetalias{2020AJ....160..276H} models provide comparably good fits to the observations, which again reflects the fundamental degeneracy between the intrinsic multiplicity and mutual inclination distributions. Breaking this degeneracy requires an extra source of information, such as data from RVs \citep{2012AJ....143...94T, 2012A&A...541A.139F} or TTVs \citep{2018ApJ...860..101Z}. In this work, we consider transit duration variations (TDVs) as a hitherto unexploited source of extra information that is highly sensitive to the mutual inclination distribution. Secular (long-term average) planet-planet interactions lead to apsidal and nodal precession of the orbits on a timescale of $10^{2-3}$ years for typical Kepler planets \citep{2019NatAs...3..424M}. As first pointed out by \cite{2002ApJ...564.1019M}, this orbital precession leads to variations in the transit duration of transiting planets that manifest as a slow drift on observable timescales (see Figure \ref{fig: geometric diagram}). The drift timescale is sensitive to mutual inclinations because the signal goes as $\dot{T}_{\mathrm{dur}}\propto \dot{\Omega}P\sin{i}$ \citep{2002ApJ...564.1019M}, where $i$ is the transiting planet's inclination with respect to the invariable plane and $\Omega$ is the longitude of the ascending node. Moreover, TDVs of planets in single-transiting systems encode information about inclined, non-transiting companion planets. 

TDV signals have been detected for $\sim30$ Kepler planets \citep{2016ApJS..225....9H, 2019AJ....157..171K, 2021MNRAS.tmp.1312S}. Notable examples include Kepler-108 \citep{2017AJ....153...45M}, Kepler-693 \citep{2017AJ....154...64M}, and Kepler-9 \citep{2018A&A...618A..41F, 2019MNRAS.484.3233B}. The TDVs have led to mutual inclination constraints in these systems. In other examples \citep[e.g.][]{2012Natur.487..449S}, an absence of TDVs has been used as evidence for low mutual inclinations. Recently, \cite{2020AJ....159..207B} investigated observed planets that are the best candidates for exhibiting detectable TDV signals in near-future observations. In addition, \cite{2020AJ....159..223D} described how to use TDVs to infer systems' three-dimensional architectures. While this paper was in review, \cite{2021MNRAS.tmp.1312S} published a new comprehensive analysis of TDVs of Kepler planets, building off of \cite{2016ApJS..225....9H}. Overall, their results are complementary to this work; where relevant, we will discuss specific comparisons.

Here we use TDVs to constrain the mutual inclination distribution of Kepler systems. We accomplish this by comparing the TDV statistics of the observed planet population to expectations from simulated planet populations constructed by the forward models of \citetalias{2019MNRAS.490.4575H} and \citetalias{2020AJ....160..276H}. Both models reproduce many aspects of the Kepler survey statistics, but they have significantly different intrinsic mutual inclination distributions. In Section \ref{sec: analytic calculation}, we describe the relevant equations for orbital precession-induced TDVs, including an analytic calculation of $\dot{T}_{\mathrm{dur}}$. In Section \ref{sec: methods}, we describe our methods for calculating the TDVs of the simulated planets and comparing their properties to the observed Kepler planets with TDVs. We summarize our results in Section \ref{sec: results}. Namely, we show that the TDV statistics of the observed planets strongly support the non-dichotomous model of \citetalias{2020AJ....160..276H} and disfavor the dichotomous model of \citetalias{2019MNRAS.490.4575H}. In Section \ref{sec: discussion}, we review the proposed theories for generating mutual inclinations among Kepler planets and discuss which are most consistent with our results. 

\section{Analytic Calculation of Transit Duration Variations}
\label{sec: analytic calculation}

We begin with an analytic calculation of a planet's transit duration variation (TDV), which is related to the time derivative of its transit duration, $T_{\mathrm{dur}}$. Our derivation bears resemblance to that of \cite{2002ApJ...564.1019M}, but we allow for arbitrary eccentricities rather than assume circular orbits. We will consider the orbital evolution to be secular, with the semi-major axis, $a$, approximately constant. In this case, the time evolution of $T_{\mathrm{dur}}$ is driven by orbital precession of the longitude of the ascending node, $\Omega$, and the argument of periapse, $\omega$, as well as changes in the orbital eccentricity, $e$. Our goal is to relate $\dot{T}_{\mathrm{dur}}$ to $\dot{\Omega}$, $\dot{\omega}$, and $\dot{e}$. 

Within the following derivation, there are three distinct planes to keep in mind: (1) the orbital plane, which is perpendicular to the planet's orbital angular momentum vector; (2) the invariable plane, which is perpendicular to the system's total angular momentum vector; and (3) the sky plane, which is perpendicular to the line-of-sight. Specifically, we define the invariable plane to be $xy$, with the $x$-axis the intersection of the invariable plane and the sky plane. The planet's orbital plane has an inclination, $i$, with respect to the invariable plane, and the line of nodes of the orbit forms an angle $\Omega$ with the $x$-axis. Note that here we will assume $i$ is constant. This is strictly only valid for a two-planet system, but it is a good approximation when considering TDVs over a baseline as short as the $\sim4$ year Kepler mission, which is much shorter than the secular timescales over which $i$ varies. For simplicity, we will also assume $R_p \ll R_{\star}$, a good approximation for the sub-Neptune-sized planets we are focusing on here. It is straightforward to generalize the treatment without this assumption \citep{2002ApJ...564.1019M}.

For $a\gg R_\star$, a planet's transit duration is given by
\begin{equation}
T_{\mathrm{dur}} = \frac{2 R_{\star} \sqrt{1-b^2}}{v_{\mathrm{mid}}}, 
\label{eq: Tdur}
\end{equation}
where $R_{\star}$ is the stellar radius, $b$ is the dimensionless impact parameter, and $v_{\mathrm{mid}}$ is the sky-projected orbital velocity at mid-transit, given by
\begin{equation}
v_{\mathrm{mid}} = \frac{n a(1+e\sin\omega_{\mathrm{sky}})}{\sqrt{1-e^2}}. 
\label{eq: vmid}
\end{equation}
Here, $n = 2\pi/P$ is the mean-motion, $e$ is the orbital eccentricity, and $\omega_{\mathrm{sky}}$ is the argument of periapse measured with respect to the sky plane. The argument $\omega_{\mathrm{sky}}$ is related to the corresponding invariable plane angle $\omega$ via the expression \citep{2020AJ....160..195J}
\begin{equation}
\tan(\omega_{\mathrm{sky}} - \omega) = \frac{\sin{\Omega}\cos{\beta}}{\cos{\Omega}\cos{i}\cos{\beta} + \sin{i}\sin{\beta}}.
\label{eq: omega_sky and omega relationship}
\end{equation}
Here, $\beta$ is the fixed angle between the invariable plane and the line of sight. Put another way, $\beta = \pi/2 - i_{\mathrm{inv}}$, where $i_{\mathrm{inv}}$ is the angle between the invariable plane and the sky plane.

Taking the time derivative of equation \ref{eq: Tdur} yields
\begin{equation}
\dot{T}_{\mathrm{dur}} = - T_{\mathrm{dur}}\bigg[\dot{b}\frac{b}{(1-b^2)} + \frac{\dot{v}_{\mathrm{mid}}}{v_{\mathrm{mid}}}\bigg].
\label{eq: dTdur/dt}
\end{equation}
The time derivatives $\dot{b}$ and $\dot{v}_{\mathrm{mid}}$ are driven by orbital precession of the longitude of the ascending node and the argument of periapse, as well as changes in the orbital eccentricity. These are parameterized by $\dot{\omega}$, $\dot{\Omega}$, and $\dot{e}$, respectively.  As for the first term in equation \ref{eq: dTdur/dt}, we can compute $\dot{b}$ by using the definition of the dimensionless impact parameter,
\begin{equation}
b = \frac{r_{\mathrm{mid}}\sin{\alpha}}{R_{\star}},
\label{eq: b}
\end{equation}
where $r_{\mathrm{mid}}$ is the star-planet separation at mid-transit,
\begin{equation}
r_{\mathrm{mid}} = \frac{a(1-e^2)}{1+e\sin{\omega_{\mathrm{sky}}}}.
\label{r_mid}
\end{equation}
The quantity $\alpha$ is the angle between the orbital plane of the planet and the line of sight. (Alternatively, $\alpha = \pi/2 - i_{\mathrm{sky}}$, where $i_{\mathrm{sky}}$ is the orbital inclination with respect to the sky plane.) This angle is related to the other angles in the problem via \citep{2020AJ....160..195J}
\begin{equation}
\label{eq: sin(alpha)}
\sin\alpha = -\sin{i}\cos{\Omega}\cos{\beta} + \cos{i}\sin{\beta}.
\end{equation}
Using these relationships, we obtain
\begin{equation}
\label{eq: db/dt}
\dot{b} = \frac{1}{R_{\star}}\left[\dot{r}_{\mathrm{mid}}\sin{\alpha}+r_{\mathrm{mid}}\dot{\Omega}\sin{i}\cos{\beta}\sin{\Omega}\right],
\end{equation}
where $\dot{r}_{\mathrm{mid}}$ can be calculated from equation \ref{r_mid},
\begin{equation}
\dot{r}_{\mathrm{mid}} = \frac{a[-2e\dot{e}(1+e\sin\omega_{\mathrm{sky}}) - (1-e^2)\frac{\mathrm{d}}{\mathrm{d}t}(e\sin\omega_{\mathrm{sky}})]}{(1+e\sin\omega_{\mathrm{sky}})^2}.
\end{equation}

As for the second term in the expression for $\dot{T}_{\mathrm{dur}}$ (equation \ref{eq: dTdur/dt}), $\dot{v}_{\mathrm{mid}}$ can be calculated by differentiating equation \ref{eq: vmid}
\begin{equation}
\dot{v}_{\mathrm{mid}} = na\frac{\dot{e}(e + \sin{\omega_{\mathrm{sky}}}) +  \dot{\omega}_{\mathrm{sky}}e\cos{\omega_{\mathrm{sky}}} (1-e^2)}{(1-e^2)^{\frac{3}{2}}}.
\end{equation}
We can relate this expression to $\omega$ and $\dot{\omega}$ by using equation \ref{eq: omega_sky and omega relationship}, from which $\dot{\omega}$ can be calculated straightforwardly, recalling that both $\beta$ and $i$ are held fixed. 

\subsection{Lagrange's Planetary Equations}
Equipped with the analytic formula for $\dot{T}_{\mathrm{dur}}$ in terms of $\dot{\Omega}$, $\dot{\omega}$, and $\dot{e}$, we now move on to the analytic calculations of these three latter quantities. 
Lagrange's planetary equations yield \citep{1999ssd..book.....M}
\begin{equation}
\label{eq: Lagrange's planetary equations}
\begin{split}
\dot{e} &= -\frac{\sqrt{1-e^2}}{n a^2 e}\frac{\partial\mathcal{R}}{\partial\varpi} \\
\dot{\Omega} &= \frac{1}{n a^2 \sqrt{1-e^2}\sin{i}}\frac{\partial\mathcal{R}}{\partial i} \\
\dot{\varpi} &= \dot{\Omega} + \dot{\omega} = \frac{\sqrt{1-e^2}}{n a^2 e}\frac{\partial\mathcal{R}}{\partial e} + \frac{\tan{\frac{1}{2}i}}{n a^2 \sqrt{1-e^2}}\frac{\partial\mathcal{R}}{\partial i}.
\end{split}
\end{equation}
Here, $\mathcal{R}$ is the disturbing function, or the non-Keplerian perturbing gravitational potential experienced by the planets due to their mutual interactions. We adopt a secular expansion of $\mathcal{R}$ to fourth-order in $e$ and ${s\equiv\sin(i/2)}$ using the Appendix tables from \cite{1999ssd..book.....M}. The resulting disturbing function expansion is provided in Appendix \ref{sec: Appendix disturbing function expansion}.

\subsection{Comparison of analytic TDV to \textit{N}-body}
\label{sec: comparison to N-body}

In Appendix \ref{sec: Appendix comparison of analytic TDV to N-body}, we assess the accuracy of the analytic calculation of $\dot{T}_{\mathrm{dur}}$ by showing how it compares to a direct $N$-body computation. We demonstrate that the analytic calculation performs well when inclinations are low, $i \lesssim 10^{\circ}$. In this regime, the analytic $\dot{T}_{\mathrm{dur}}$ is within $50\%$ of the $N$-body $\dot{T}_{\mathrm{dur}}$ in roughly three quarters of cases. However, when inclinations are larger than $\sim10^{\circ}$, there can be orders-of-magnitude discrepancies between the analytic calculation and the $N$-body result. This occurs because the disturbing function expansion breaks down for large $e$ and $i$, leading to perturbations in the equations for $\dot{e}$, $\dot{\Omega}$, and $\dot{\varpi}$. As we will describe in Section \ref{sec: TDV calculation}, this accuracy problem ultimately limits the applicability of the analytic $\dot{T}_{\mathrm{dur}}$ calculation. 

\section{Methods}
\label{sec: methods}

Our goal is to constrain the underlying distribution of Kepler planet mutual inclinations by comparing the TDVs of the observed Kepler planets to those of various simulated planet populations with different mutual inclination distributions. In this section, we describe our methods. We define our observed and simulated planet populations in Sections \ref{sec: observations} and \ref{sec: simulations}, respectively. We then explain the procedure for identifying detectable TDVs of the simulated planets in Section \ref{sec: name this subsection}.

\subsection{Observations: TDVs of Kepler Planets}
\label{sec: observations}

\setlength{\extrarowheight}{2pt}
\setlength\tabcolsep{15pt}
\begin{table*}[t]
\centering
\caption{\textbf{KOIs with Significant TDV Slopes.} List of KOIs from the \citetalias{2016ApJS..225....9H} TDV catalog with $|\dot{T}_{\mathrm{dur}}|> 3 \  \sigma_{\dot{T}_{\mathrm{dur}}}$ that pass our additional cuts on planet and stellar properties. $N_{\mathrm{obs}}$ is the observed transit multiplicity of the system according to the Kepler DR25 catalog \citep{2018ApJS..235...38T}. Orbital periods are drawn from DR25; planetary radii are from \cite{2020AJ....160..108B}; and the $\dot{T}_{\mathrm{dur}}$ measurements are from the \citetalias{2016ApJS..225....9H} TDV catalog.}
\begin{tabular}{c c c c c c c}
\hline
\hline
KOI & Kepler name & $N_{\mathrm{obs}}$ & $P$ [days] & $R_p$ [$R_{\oplus}$] & $\dot{T}_{\mathrm{dur}}$ [min/yr] \\
\hline
103.01 & -- & 1 & 14.911 & $2.55^{+0.06}_{-0.05}$ & $5.0\pm1.1$\\
137.02 & Kepler-18 d & 3 & 14.859 & $5.16^{+0.09}_{-0.11}$ & $1.73\pm0.41$ \\
139.01 & Kepler-111 c & 2 & 224.779 & $6.91^{+0.16}_{-0.17}$ & $8.1\pm2.0$  \\
142.01 & Kepler-88 b & 1 & 10.916 & $3.84^{+0.08}_{-0.31}$ & $1.98\pm0.58$ \\
209.02 & Kepler-117 b & 2 & 18.796 & $8.36^{+0.33}_{-0.52}$ & $2.98\pm0.82$  \\
377.01 & Kepler-9 b & 3 & 19.271 & $7.9^{+0.16}_{-0.16}$ & $1.64\pm0.3$  \\
377.02 & Kepler-9 c & 3 & 38.908 & $8.14^{+0.18}_{-0.18}$ & $-4.58\pm0.56$  \\ 
460.01 & Kepler-559 b & 2 & 17.588 & $3.41^{+0.11}_{-0.09}$ & $6.0\pm1.5$  \\ 
806.01 & Kepler-30 d & 3 & 143.206 & $8.79^{+0.49}_{-0.31}$ & $6.3\pm1.9$ \\
841.02 & Kepler-27 c & 5 & 31.33 & $6.24^{+0.24}_{-0.28}$ & $3.6\pm1.1$ \\
872.01 & Kepler-46 b & 2 & 33.601 & $7.45^{+0.2}_{-0.26}$ & $5.78\pm0.83$  \\ 
1320.01 & Kepler-816 b & 1 & 10.507 & $9.44^{+0.46}_{-0.36}$ & $-1.88\pm0.44$  \\
1423.01 & Kepler-841 b & 1 & 124.42 & $5.02^{+0.25}_{-0.19}$ & $3.7\pm1.1$ \\
1856.01 & -- & 1 & 46.299 & $2.24^{+0.08}_{-0.05}$ & $-11.8\pm3.1$  \\ 
2698.01 & Kepler-1316 b & 1 & 87.972 & $3.4^{+0.11}_{-0.08}$ & $18.1\pm4.5$  \\ 
2770.01 & -- & 1 & 205.386 & $2.26^{+0.13}_{-0.08}$ & $19.4\pm6.4$ \\
\hline
\end{tabular}
\label{tab: KOIs with TDVs}
\end{table*}

We first identify a set of Kepler planets with significant TDVs. These planets will later be compared to the simulated planets. As part of their comprehensive catalog of TTV measurements for 2599 Kepler Objects of Interest (KOIs) across the full 17 quarters of the Kepler mission, \cite{2016ApJS..225....9H} (hereafter \citetalias{2016ApJS..225....9H}) also measured transit depths and durations. The duration and depth measurements were limited to cases with $T_{\mathrm{dur}} > 1.5$ hr and $\mathrm{SNR} > 10$, where SNR is the signal-to-noise per transit.
As a result, a total of 779 KOIs have duration and depth measurements. These were then analyzed for any TDVs or TPVs (transit depth variations) by identifying potentially significant periodicities or long term trends. The long term trends, quantified by the TDV slope $\dot{T}_{\mathrm{dur}}$, are the most relevant for our analysis, as these are the expected signal of duration drifts induced by orbital precession.

\begin{figure}[t]
\epsscale{1.2}
\plotone{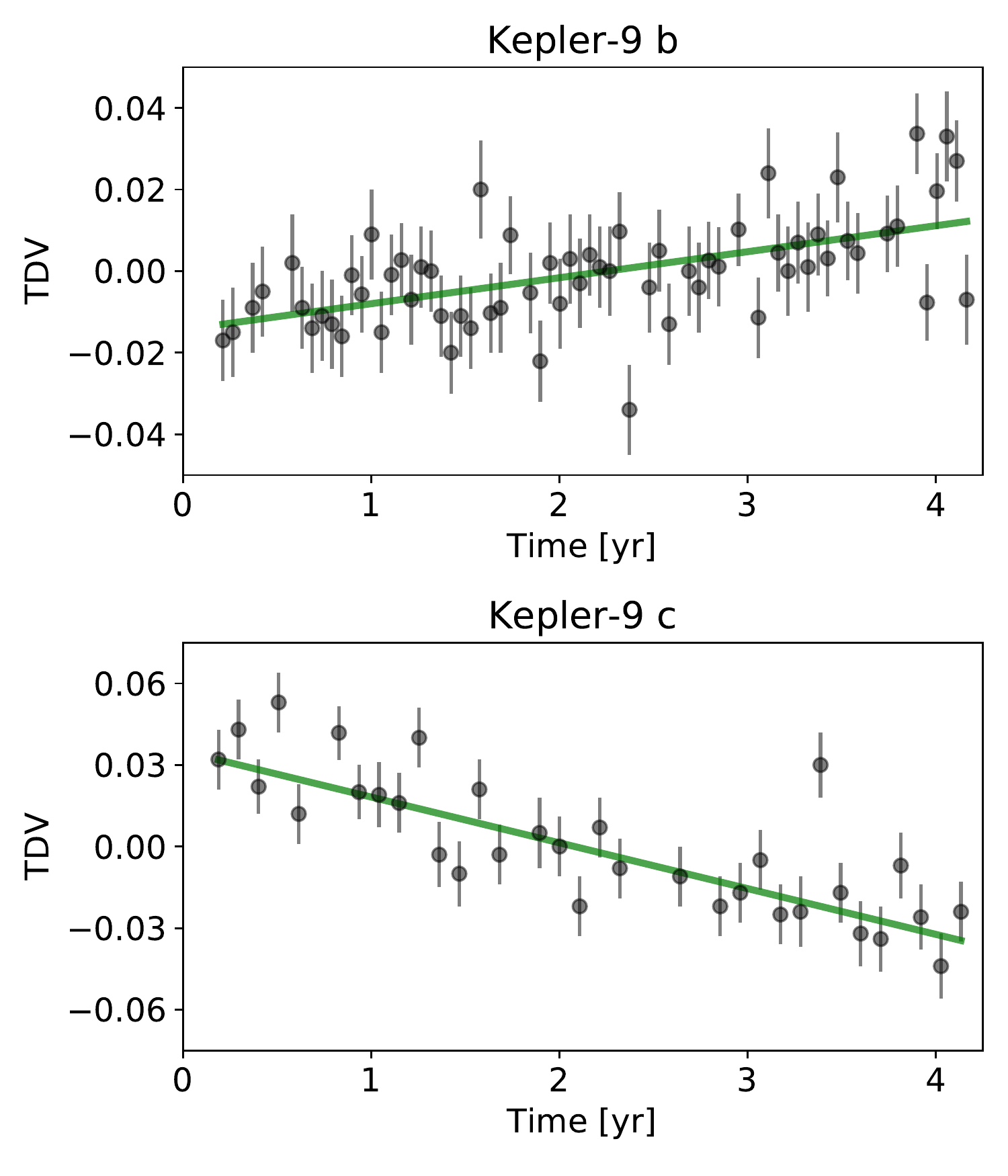}
\caption{Examples of the TDV drift signal, shown here for Kepler-9 b and c (KOI 377.01 and 377.02). The data for the TDVs, given by $\mathrm{TDV} = (T_{\mathrm{dur}} - \bar{T}_{\mathrm{dur}})/\bar{T}_{\mathrm{dur}}$, is taken from \citetalias{2016ApJS..225....9H}. The green curves represents the best-fit lines, with slopes equal to those from the \citetalias{2016ApJS..225....9H} TDV catalog.} 
\label{fig: Kepler-9 TDVs}
\end{figure}

\citetalias{2016ApJS..225....9H} summarized their results for TDVs in the in the \texttt{TDV\_statistics.txt} 
file at \href{ftp://wise-ftp.tau.ac.il/pub/tauttv/TTV/ver_112}{ftp://wise-ftp.tau.ac.il/pub/tauttv/TTV/ver\_112}. We will refer to this as the ``\citetalias{2016ApJS..225....9H} TDV catalog''. In particular, the quantities that are relevant to our analysis are the slope of the TDV measurements and the estimated error of the slope. We apply several cuts to the \citetalias{2016ApJS..225....9H} TDV catalog to establish a list of observed planets with detected TDVs. 
\begin{enumerate}
    \item \textit{Significant TDV slope:} We consider a detectable transit duration drift to be one with ${|\dot{T}_{\mathrm{dur}}|> 3 \  \sigma_{\dot{T}_{\mathrm{dur}}}}$. A total of 31 KOIs pass this criterion.
    \footnote{\cite{2021MNRAS.tmp.1312S} updated the calculation of $\dot{T}_{\mathrm{dur}}$ from the \citetalias{2016ApJS..225....9H} TDV data and used a different definition of a significant TDV slope. This yielded a slightly different set of KOIs with significant TDV slopes but an identical number (when including those which \cite{2021MNRAS.tmp.1312S} labeled ``intermediate significance'').}
    \item \textit{Cuts on planet properties:} The simulated planet population that we will introduce in Section \ref{sec: simulations} is restricted to planets with $3 \ \mathrm{days} < P < 300 \ \mathrm{days}$ and $0.5 \ R_{\oplus} < R_p < 10 \ R_{\oplus}$. Applying these cuts leaves 21 remaining KOIs. Most of the removed planets are cut for having $R_p > 10 \ R_{\oplus}$. Three have $P < 3$ days.
    \item \textit{Cuts on stellar properties:} The simulated planet population is built using a curated sample of FGK dwarf stars, developed using a series of quality cuts on the Kepler DR25 target list in conjunction with target information from Gaia DR2 \citepalias{2019MNRAS.490.4575H, 2020AJ....160..276H}. We require the host stars of the observed planets to be in this cleaned stellar input catalog. After this requirement, we are left with 16 remaining KOIs.
\end{enumerate} 
The final list of 16 KOIs with significant TDV slopes are shown in Table \ref{tab: KOIs with TDVs}. The 16 KOIs are part of 15 systems. Of these, seven are single-transiting systems; four are double-transiting systems; three are triple-transiting systems; and one is a quintuple-transiting system. Examples of TDV drift signals observed for Kepler-9 b and c are shown in Figure \ref{fig: Kepler-9 TDVs}. 

\subsection{Simulations: SysSim Planet Populations}
\label{sec: simulations}

We consider populations of simulated planets from the SysSim (short for ``Planetary Systems Simulator'') forward modeling framework. SysSim is an empirical model that generates simulated planetary systems according to flexibly parameterized statistical descriptions. It was first introduced by \cite{2018AJ....155..205H} and has undergone continuous development by \cite{2019AJ....158..109H}, \citetalias{2019MNRAS.490.4575H}, \citetalias{2020AJ....160..276H}, and \cite{2021AJ....161...16H}. The model has been calibrated using a range of summary statistics of the observed Kepler planet population, including (but not limited to) the observed distributions of multiplicities, orbital periods and period ratios, transit depths and depth ratios, and transit durations. SysSim is implemented in the Julia programming language \citep{2014arXiv1411.1607B} within the ExoplanetsSysSim.jl package.  Both the core SysSim code  (\href{https://github.com/ExoJulia/ExoplanetsSysSim.jl}{https://github.com/ExoJulia/ExoplanetsSysSim.jl}) and the specific forward models explored in this study (\href{https://github.com/ExoJulia/SysSimExClusters}{https://github.com/ExoJulia/SysSimExClusters}) are publicly available. More information can be found in the key papers mentioned above. We provide a brief description of the model below.

The process of the SysSim forward modeling framework is to first generate an underlying population of planetary systems (``physical catalog'') according to a statistical model of the intrinsic distribution of systems. The physical catalog is a simulated realization designed to resemble the entire population of Kepler planetary systems, including planets that are not observed. The next step is to generate an observed population of planetary systems (``observed catalog'') from the physical catalog by simulating the full Kepler detection pipeline and determining which planets would be detected by the pipeline and labeled as planet candidates during the automated vetting process. This simulated observed catalog is then compared with the true Kepler planet population using a set of summary statistics and a distance function. The preceding steps are repeated iteratively in order to optimize the distance function and identify the best-fit parameters of the statistical model. Finally, Approximate Bayesian Computing is applied to approximate the posterior distributions of model parameters.

In this work, we do not implement extensions of SysSim or fit new statistical models. Rather, we examine sets of simulated catalogs from two previously optimized models from \citetalias{2019MNRAS.490.4575H} and \citetalias{2020AJ....160..276H}. These models effectively describe many aspects of the Kepler population statistics, and they fit the data nearly equally well. By construction, the two models are similar in most ways, but they differ primarily in the distributions of eccentricities, inclinations, and number of planets per star.

\subsubsection{Two-Rayleigh model}
\label{sec: Two-Rayleigh model}

\citetalias{2019MNRAS.490.4575H} considered independent distributions of eccentricities and inclinations. The eccentricities were drawn from a Rayleigh distribution, $e \sim \mathrm{Rayleigh}(\sigma_e)$. The mutual inclinations were drawn from one of two Rayleigh distributions, representing a low and high inclination population, with the fraction of systems belonging to the high inclination population being $f_{\sigma_{i,\mathrm{high}}}$.\footnote{The model also assigns mutual inclinations drawn from the low scale ($\sigma_{i,\rm low}$) for all planets near a first-order mean motion resonance (MMR) with another planet (defined as having a period ratio within 5\% greater than $(i+1):i$ for any $i = 1,2,3,4$), regardless of whether the system belongs to the high or low inclination population. Around $30\%$ of the simulated planets are near a first-order MMR, and approximately half ($f_{\rm mmr} \simeq 0.49$) of all simulated systems with at least one planet contain such a planet pair. Thus, the fraction of planetary systems where near-MMR planets are reassigned low mutual inclinations is $f_{\rm mmr} \times f_{\sigma_{i,\rm high}} = 0.21_{-0.05}^{+0.07}$.} That is, 
\begin{equation}
i\sim 
\begin{cases}
\mathrm{Rayleigh}(\sigma_{i,\mathrm{high}}), \ \ u<f_{\sigma_{i,\mathrm{high}}} \\ 
\mathrm{Rayleigh}(\sigma_{i,\mathrm{low}}), \ \ u \geq f_{\sigma_{i,\mathrm{high}}}, \\ 
\end{cases}
\end{equation}
where $u \sim \mathrm{Unif}(0,1)$.
Because of the dichotomous nature of the inclination distribution, we will call this the ``two-Rayleigh model''. 

After fitting to the Kepler data, \citetalias{2019MNRAS.490.4575H} identified best-fit parameters equal to $\sigma_{i,\mathrm{low}} = 1.40^{+0.54}_{-0.39}$ deg,  $\sigma_{i,\mathrm{high}} = 48^{+17}_{-17}$ deg, and $f_{\sigma_{i,\mathrm{high}}} = 0.42^{+0.08}_{-0.07}$. The high inclination component makes up nearly half of the total population of systems (and nearly a third of all planets), but the mutual inclination scale $\sigma_{i,\mathrm{high}}$ is poorly constrained, although clearly greater than $\sim 10^{\circ}$. We note that, in this work, we use the two-Rayleigh model of \cite{2021AJ....161...16H}, which is similar to \citetalias{2019MNRAS.490.4575H} but with an additional dependence of the fraction of stars with planets on the host star color. This feature is also present in the model introduced in the next section, allowing for a better comparison.

\subsubsection{Maximum AMD model}
\label{sec: Maximum AMD model}

The second model uses a joint distribution in which the eccentricity and inclination distributions are dependent and correlated with the number of planets in a system. 
This contrasts with the first model, where eccentricities and inclinations are independent of one another and of the number of planets in a system. 
The second approach is based on the argument that a system's long-term orbital stability is governed by its angular momentum deficit (AMD). The AMD is defined as the difference between the total orbital angular momentum of the system and what it would be if all orbits were circular and coplanar \citep[e.g.][]{2017A&A...605A..72L}. For a system of $N$ planets, the AMD can be written as 
\begin{equation}
\label{eq: AMD}
\mathrm{AMD} = \sum_{k=1}^{N}\Lambda_k(1-\sqrt{1-e_k^2}\cos{i_k}),
\end{equation}
where $\Lambda_k = M_{p,k}\sqrt{G M_{\star} a_k}$. 
This quantity is effectively a measure of the dynamical excitation of the system, and it is a conserved quantity when the orbits are evolving secularly. The AMD has been shown to be a reasonable predictor of long-term stability, at least when mean-motion resonances are absent and when distant and dynamically-detached planets are not included in the AMD calculation \citep{2017A&A...605A..72L}.

The assumption of \citetalias{2020AJ....160..276H} is that all systems have the critical (i.e.\ maximum) AMD for stability, calculated using the analytic criteria from \cite{2017A&A...605A..72L} and \cite{2017A&A...607A..35P}. The physical justification for this assumption is that collisional events during planet formation decrease the total AMD, such that systems may generally evolve from outside the stability limit to just inside after a sequence of collisions. While this ``maximum AMD model'' assumption breaks down at a detailed level (i.e.\ not all systems are exactly at the critical AMD), it is a useful physical framework for assigning system orbital properties and has been shown to reproduce many aspects of the Kepler data \citepalias{2020AJ....160..276H}. In this work, we are primarily interested in the model's mutual inclination and eccentricity distributions. 

Given a set of planet masses and orbital periods of a system, \citetalias{2020AJ....160..276H} assigned orbital properties by (1) calculating the system's critical AMD, (2) distributing the AMD among the planets per unit mass, and (3) further randomly dividing each planet's AMD to eccentricity and inclination components. This approach results in eccentricities and inclinations that are correlated with one another at the population level. Another key feature that emerges from the model is an inverse trend of inclination and eccentricity dispersion with intrinsic multiplicity. In particular, \citetalias{2020AJ....160..276H} found that the median mutual inclination, $\tilde{\mu}_{i,n}$, of systems with $n = 2, 3, ..., 10$ planets is well modeled by a power-law relationship of the form 
\begin{equation}
\label{eq: mutual inclination power law}
\tilde{\mu}_{i,n} = \tilde{\mu}_{i,5}\left(\frac{n}{5}\right)^{\alpha},
\end{equation}
with $\tilde{\mu}_{i,5} = 1.10^{+0.15}_{-0.11}$ and $\alpha = -1.73^{+0.09}_{-0.08}$. This power-law relationship is qualitatively similar but shallower than that inferred by \cite{2018ApJ...860..101Z}. Equation \ref{eq: mutual inclination power law} illustrates that the typical mutual inclinations of systems within the maximum AMD model are less than a few degrees, in sharp contrast with the two-Rayleigh model.

\subsubsection{Model comparison}
\label{sec: model comparison}

\begin{figure}
\epsscale{1.}
\plotone{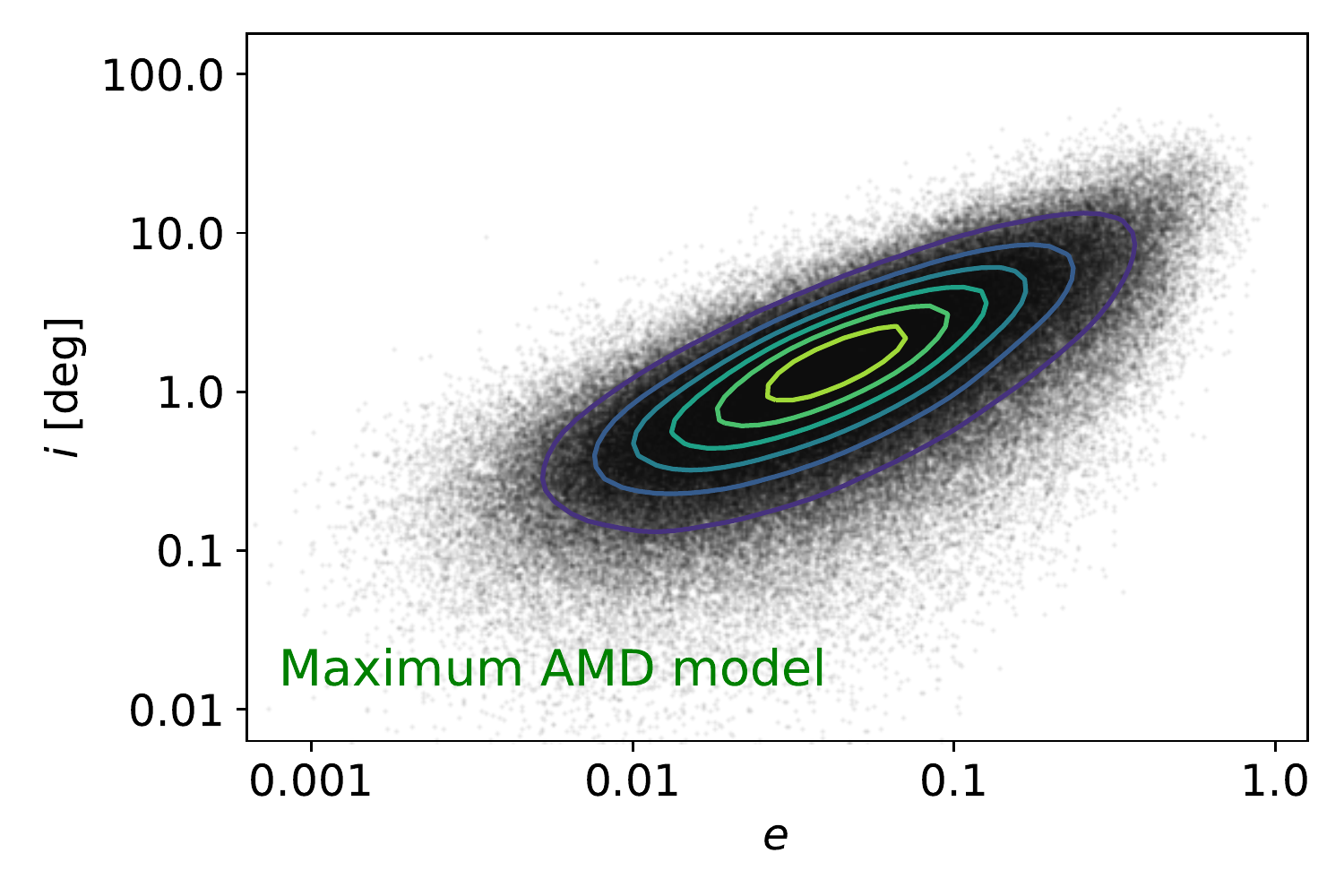}
\plotone{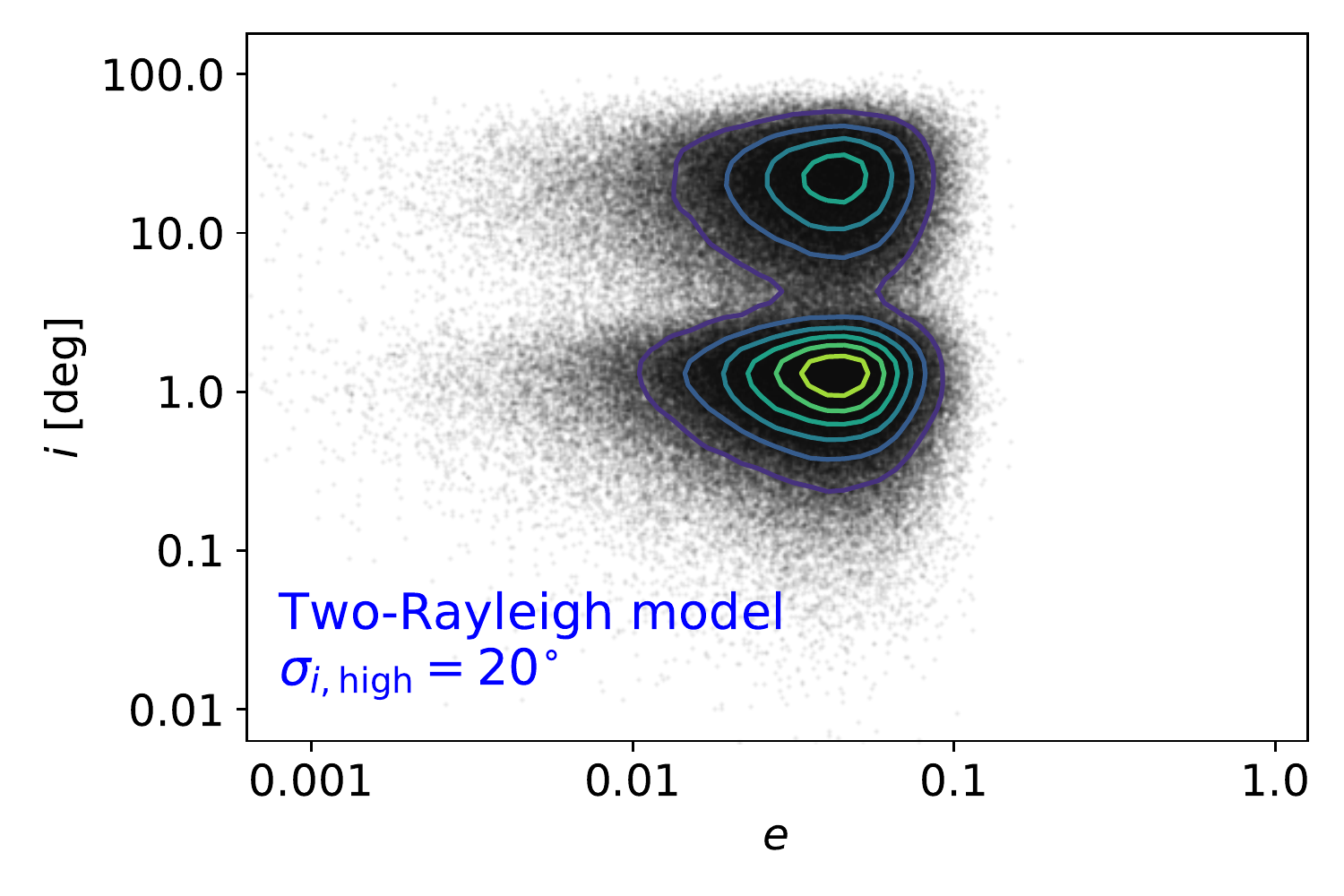}
\plotone{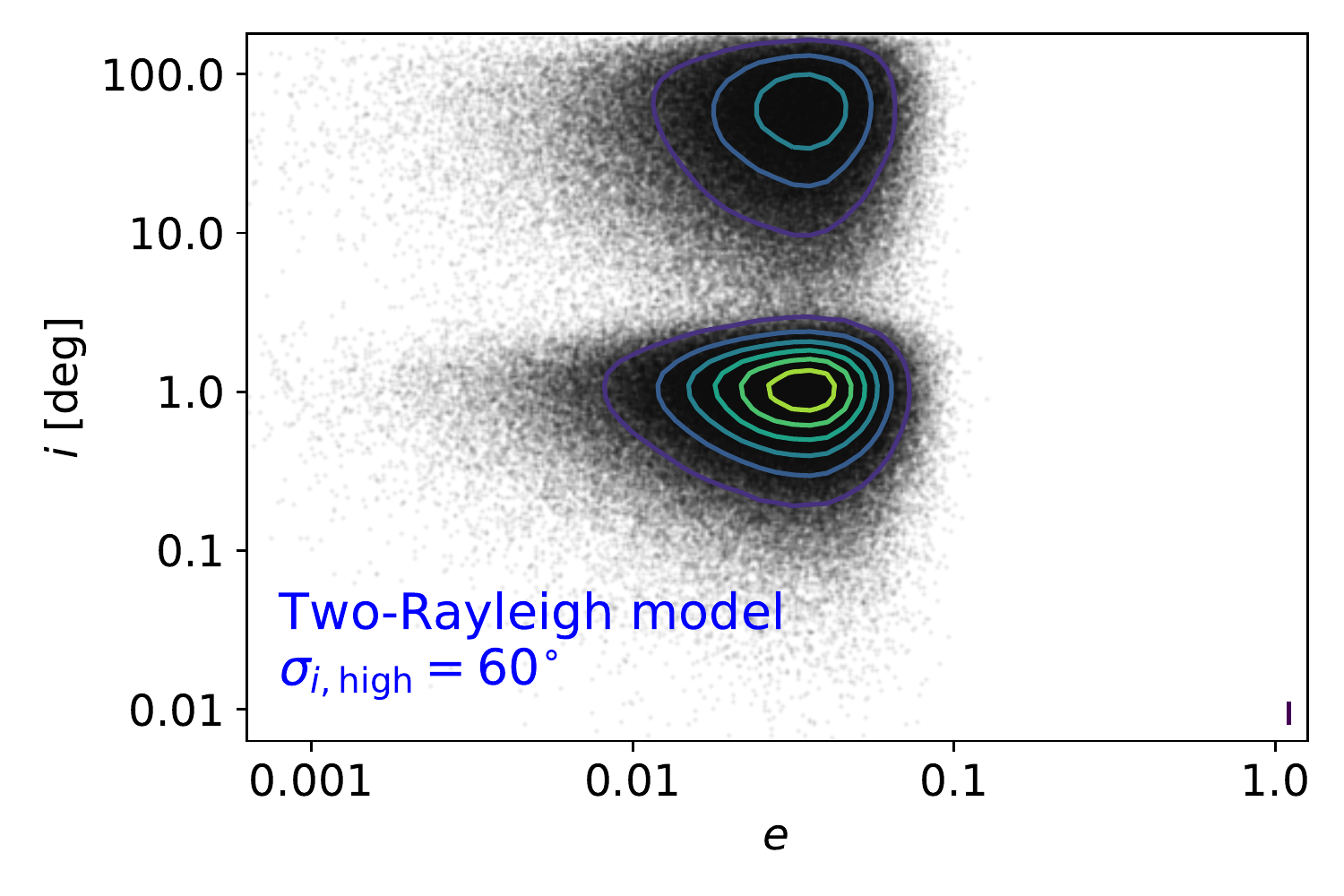}
\caption{Scatter plots of orbital inclinations referenced to the invariable plane ($i$) versus eccentricity ($e$) for intrinsic multi-planet systems in SysSim physical catalogs. The top panel considers a single physical catalog from the maximum AMD model. The middle and bottom panels consider physical catalogs from the two-Rayleigh model using $\sigma_{i,\mathrm{high}} = 20^{\circ}$ and $\sigma_{i,\mathrm{high}} = 60^{\circ}$, respectively. To illustrate the scatter plot density, we display a set of contours calculated using a Gaussian kernel density estimation. } 
\label{fig: i vs e scatterplots}
\end{figure}

The two-Rayleigh model and maximum AMD model both fit the Kepler data roughly equally well (see \citetalias{2020AJ....160..276H} Section 3.1),
but they have very different underlying inclination and eccentricity distributions. To illustrate these differences, Figure \ref{fig: i vs e scatterplots} displays scatter plots of inclinations and eccentricities of intrinsic multi-planet systems in SysSim physical catalogs. We observe that eccentricities and inclinations are strongly correlated when distributed according to the maximum AMD model. Moreover, the maximum AMD model contains considerably fewer systems with very high inclination configurations than the two-Rayleigh model. For instance, $97\%$ of the mutual inclinations are below $10^{\circ}$ in the maximum AMD example plotted in Figure \ref{fig: i vs e scatterplots}, whereas $68\%$ and $63\%$ of the inclinations are below $10^{\circ}$ in the $\sigma_{i,{\mathrm{high}}} = 20^{\circ}$ and $\sigma_{i,{\mathrm{high}}} = 60^{\circ}$ two-Rayleigh model examples, respectively. 

As a caveat, we note that the simulated systems have not been evaluated for long-term orbital stability in either model. The systems in the maximum AMD model obey the AMD stability criterion by construction, but that doesn't guarantee that they are all long-term stable \citep{2021arXiv210104117C}. They are, however, more likely to be stable than the systems in the two-Rayleigh model, which frequently have very high inclinations (sometimes even $> 90^{\circ}$). This difference means that a direct comparison of the two models is slightly misleading. A stability analysis would require significant computation but would be interesting to address in future SysSim models.

\subsection{TDV calculations of SysSim Planets}
\label{sec: name this subsection}

For each the two-Rayleigh model and maximum AMD model, we aim to compare the TDVs of the simulated planets with those of the observed Kepler planets. We thus need to simulate the TDV detection process for the SysSim planets in a similar way as the true observations, which are obtained from the \citetalias{2016ApJS..225....9H} TDV catalog (Section \ref{sec: observations}). Clearly, we can only measure TDVs for planets in the observed catalog (i.e.\ that have been ``detected'' in transit), but we also need to consider whether they have parameters suitable for individual transit duration measurements and whether their TDV signal has a detectable slope. We will discuss this process in three steps: TDV measurability (Section \ref{sec: TDV measurability}), TDV calculation (Section \ref{sec: TDV calculation}), and TDV slope detectability (Section \ref{sec: TDV slope detectability}). We will then summarize the process end-to-end in Section \ref{sec: summary of the full process}.

\subsubsection{TDV measurability}
\label{sec: TDV measurability}

We must first determine which planets in a given observed catalog have transits with sufficiently high signal-to-noise (SNR) such that individual transit durations would be measurable and the planet would enter into the \citetalias{2016ApJS..225....9H} TDV catalog. We summarize this as a planet having ``TDV measurability''. \citetalias{2016ApJS..225....9H} derived individual transit durations only when $T_{\mathrm{dur}} > 1.5$ hr and $\mathrm{SNR} > 10$; we thus apply these same thresholds to the SysSim planets. 

\begin{figure}
\epsscale{1.1}
\plotone{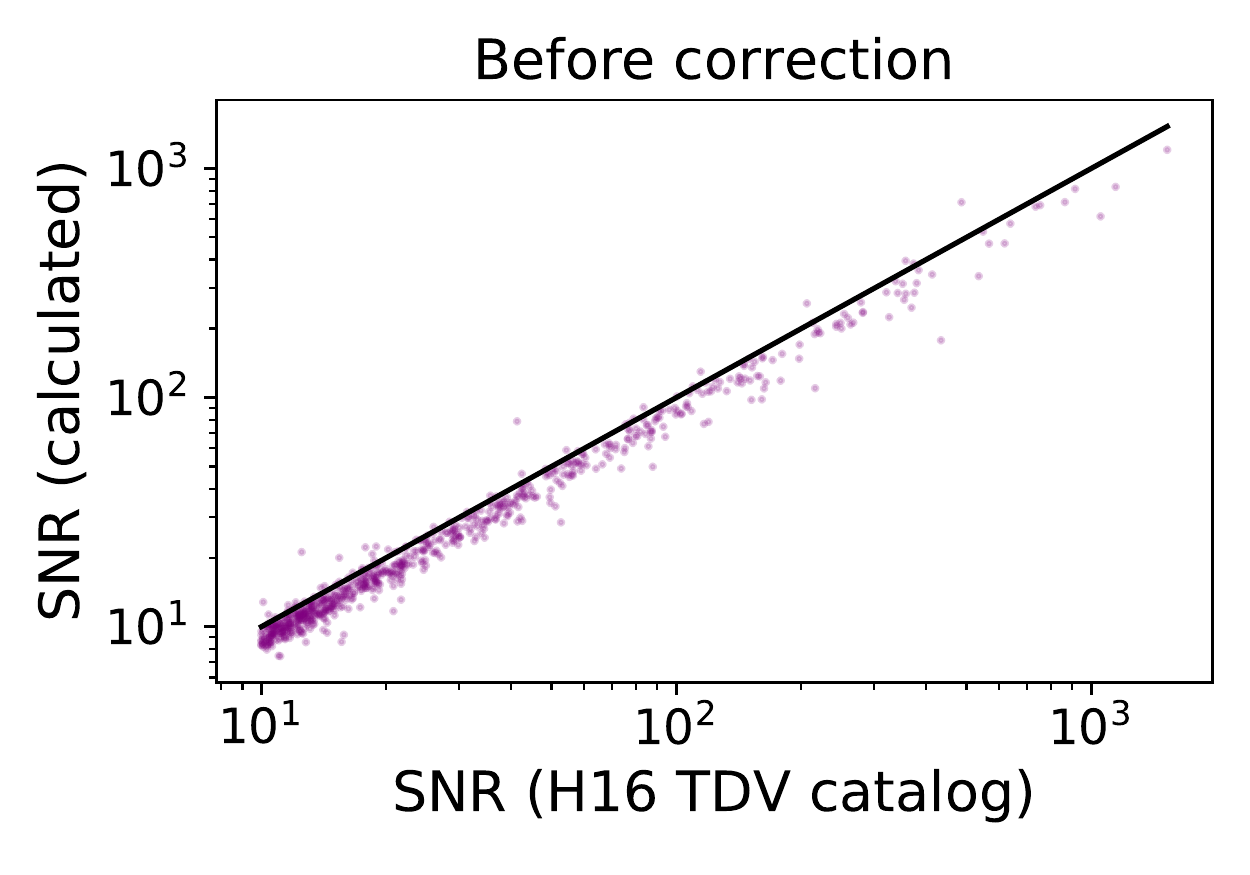}
\plotone{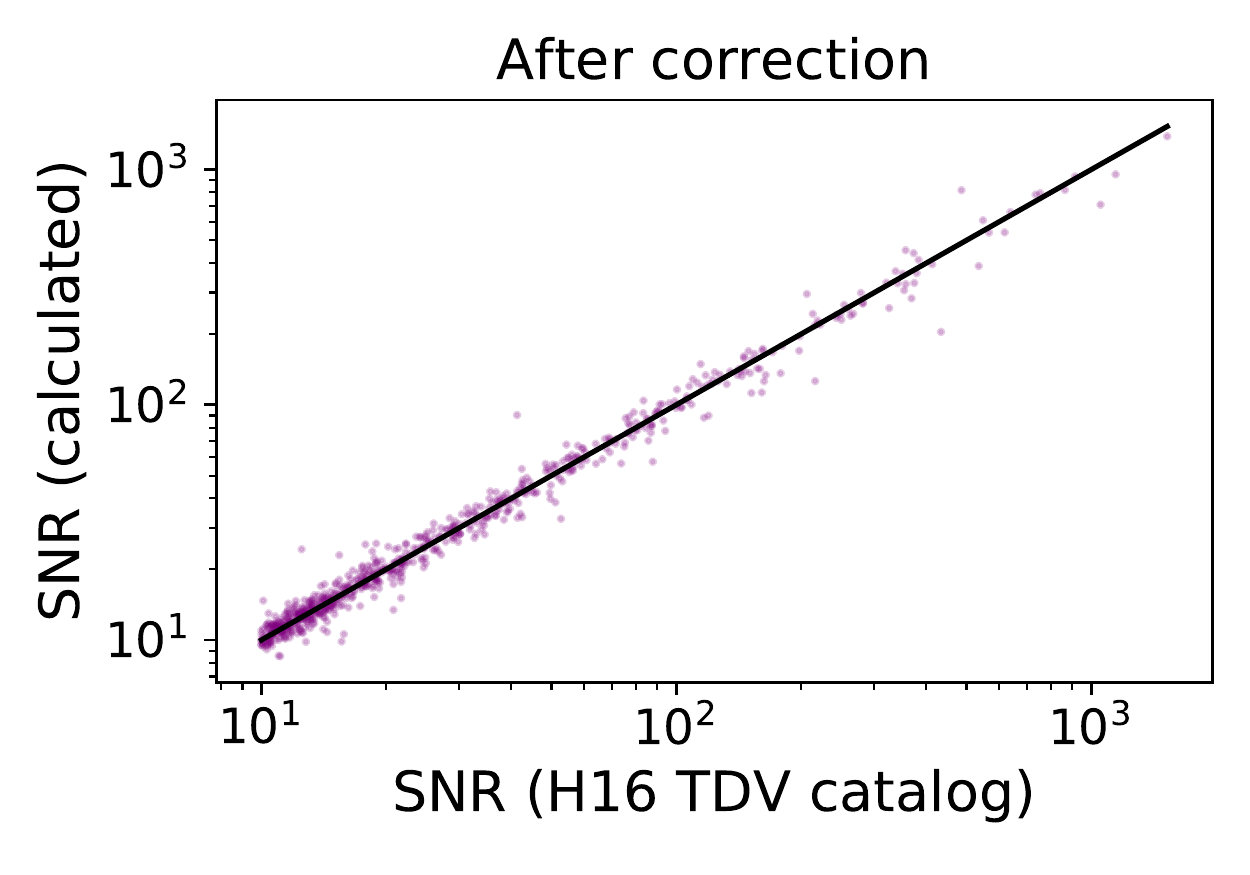}
\caption{Calibration of the SNR calculation. We show the calculated SNR (using equation \ref{eq: SNR}) versus the \citetalias{2016ApJS..225....9H} SNR for the planets in the \citetalias{2016ApJS..225....9H} TDV catalog. The top panel is before the scaling correction ($f_{\mathrm{CDPP}} = 1$), and the bottom panel is after the correction ($f_{\mathrm{CDPP}} = 0.8707$). The black line is one-to-one.} 
\label{fig: SNR scatterplots}
\end{figure}

We calculate the individual transit SNR of planets in the SysSim observed catalogs as 
\begin{equation}
\label{eq: SNR}
\mathrm{SNR} = \frac{\delta}{\mathrm{CDPP_{eff}}}.
\end{equation}
Here $\delta$ is the transit depth and $\mathrm{CDPP_{eff}}$ is the effective combined differential photometric precision, given by 
\begin{equation}
\label{eq: CDPP_eff}
\mathrm{CDPP_{eff}} = f_{\mathrm{CDPP}} \times \mathrm{CDPP}_{1.5 \ \mathrm{hr}}\sqrt{\frac{1.5 \ \mathrm{hr}}{T_{\mathrm{dur}}}}.
\end{equation}
$\mathrm{CDPP}_{1.5 \ \mathrm{hr}}$ is the mission average of the 1.5 hr duration combined differential photometric precision (CDPP) for the target star, taken from the Kepler Q1-Q17 DR 25 stellar catalog \citep{2018ApJS..235...38T}. Equation \ref{eq: CDPP_eff} also contains a scaling factor, $f_{\mathrm{CDPP}} < 1$, that accounts for a systematic offset in the calculated SNR compared to that of \citetalias{2016ApJS..225....9H}. The systematic offset arises because \citetalias{2016ApJS..225....9H} derived the transit SNR using measurement uncertainties, rather than the CDPP. The CDPP is larger because it includes both photon noise and variability due to the star and the instrument. We identify the optimal scaling factor by calculating the SNR for the planets in the \citetalias{2016ApJS..225....9H} TDV catalog and minimizing the difference with respect to \citetalias{2016ApJS..225....9H}'s reported SNR values. The resulting value is $f_{\mathrm{CDPP}} = 0.8707$. Figure \ref{fig: SNR scatterplots} shows the calculated SNR versus the \citetalias{2016ApJS..225....9H} SNR, before and after the scaling factor correction.

\newpage
\subsubsection{TDV calculation}
\label{sec: TDV calculation}

After a planet is deemed to have sufficiently high transit SNR such that the individual transit durations would be measurable, we obtain an estimate of its TDV slope, $\dot{T}_{\mathrm{dur}}$. All planets in the system (transiting or not) are accounted for in this calculation. For systems with two or more planets, we calculate $\dot{T}_{\mathrm{dur}}$ in two ways. The first is the analytic calculation outlined in Section~\ref{sec: analytic calculation}. The second is a numerical approach aided by the \texttt{REBOUND} $N$-body gravitational dynamics software \citep{2012A&A...537A.128R}. We use the \texttt{REBOUND} Wisdom-Holman integrator \citep{2015MNRAS.452..376R} to calculate a short orbital evolution over a 4 year baseline (approximately the length of the Kepler prime mission). The integration uses a timestep equal to $0.1 P_1$. We do not account for general relativity (or any other additional forces), since the perturbations on $\dot{T}_{\mathrm{dur}}$ from general relativistic apsidal precession are negligible relative to the influences of nodal precession. Following the integration, we calculate $T_{\mathrm{dur}}$ as a function of time using equation~\ref{eq: Tdur} (along with the other relevant relationships for $b$ and $v_{\mathrm{mid}}$). Finally, we perform a least-squares linear fit to $T_{\mathrm{dur}}$ vs. time to calculate $\dot{T}_{\mathrm{dur}}$.

As discussed in Section \ref{sec: comparison to N-body} and Appendix \ref{sec: Appendix comparison of analytic TDV to N-body}, the analytic calculation of $\dot{T}_{\mathrm{dur}}$ is adequate (relative to $N$-body) when inclinations are low (i.e.\ $i\lesssim10^{\circ}$, as in the maximum AMD model) but can have orders-of-magnitude discrepancies at high inclinations (i.e.\ in the two-Rayleigh model). In order to avoid systematic errors in our study of the two models, we opt to use the $N$-body $\dot{T}_{\mathrm{dur}}$ calculation for all analyses going forward. The analytic derivation, however, is still useful for physical insight and for quick calculations in low inclination systems.

\subsubsection{TDV slope detectability}
\label{sec: TDV slope detectability}

The next step is to determine whether the calculated TDV slope, $\dot{T}_{\mathrm{dur}}$, would be detectable according to the $|\dot{T}_{\mathrm{dur}}| > 3 \ \sigma_{\dot{T}_{\mathrm{dur}}}$ threshold identified above for the observed systems. We thus require an estimate of the TDV slope uncertainty, $\sigma_{\dot{T}_{\mathrm{dur}}}$, to use in conjunction with the calculated $\dot{T}_{\mathrm{dur}}$. Using generalized least squares, $\sigma_{\dot{T}_{\mathrm{dur}}}$ can be related to the individual transit duration uncertainty, $(\sigma_{T_{\mathrm{dur}}})_{\mathrm{ind}}$, via the expression
\begin{equation}
\label{eq: sigma_dotT v1}
\sigma_{\dot{T}_{\mathrm{dur}}}^2 \approx \frac{(\sigma_{T_{\mathrm{dur}}}^2)_{\mathrm{ind}}}{P^2 f_0 \sum_{j=-M}^{M}j^2}.
\end{equation}
Here $f_0$ is the duty cycle (the fraction of data cadences with valid data), and the transits are assumed to run from $-M$ to $M$ in the absence of data gaps.
(In principle, one could estimate the increased uncertainty due to some transits not being observed by not including the corresponding $j^2$ terms in the denominator of equation \ref{eq: sigma_dotT v1}.)
We approximate $M \approx t_{\mathrm{obs}}/(2P)$, where $t_{\mathrm{obs}}$ is the total observation time span. 
Next, we can simplify equation \ref{eq: sigma_dotT v1} by expressing $(\sigma_{T_{\mathrm{dur}}})_{\mathrm{ind}}$ in terms of the composite transit duration uncertainty, $\sigma_{T_{\mathrm{dur}}}$, and the number of transits, $N_{\mathrm{tr}}$, yielding
$(\sigma_{T_{\mathrm{dur}}}^2)_{\mathrm{ind}} = \sigma_{T_{\mathrm{dur}}}^2N_{\mathrm{tr}} = \sigma_{T_{\mathrm{dur}}}^2 t_{\mathrm{obs}}f_0/P$. Substituting this into equation \ref{eq: sigma_dotT v1} and adding a scaling factor, $f_{\sigma_{\dot{T}_{\mathrm{dur}}}}$, yields
\begin{equation}
\label{eq: sigma_dotT v2}
\sigma_{\dot{T}_{\mathrm{dur}}}^2 = \frac{\sigma_{T_{\mathrm{dur}}}^2 t_{\mathrm{obs}}f_{\sigma_{\dot{T}_{\mathrm{dur}}}}^2}{P^3 \sum_{j=-M}^{M}j^2}.
\end{equation}
Similar to the calculation of $\mathrm{CDPP_{eff}}$ in equation \ref{eq: CDPP_eff}, the scaling factor is required to calibrate $\sigma_{\dot{T}_{\mathrm{dur}}}$ to the \citetalias{2016ApJS..225....9H} TDV catalog. Using equation \ref{eq: sigma_dotT v2} without a scaling factor leads to a systematic overestimation of the calculated $\sigma_{\dot{T}_{\mathrm{dur}}}$ compared to the \citetalias{2016ApJS..225....9H} values because of differences in the $\sigma_{T_{\mathrm{dur}}}$ calculation.  It appears that \citetalias{2016ApJS..225....9H}'s use of measurement uncertainties rather than the CDPP led them to underestimate $\sigma_{T_{\mathrm{dur}}}$ and that this effect was greater than the impact of Kepler's duty cycle being less than 100\%.  We solve for the optimal value of $f_{\sigma_{\dot{T}_{\mathrm{dur}}}}$ by calculating $\sigma_{\dot{T}_{\mathrm{dur}}}$ for the planets in the \citetalias{2016ApJS..225....9H} TDV catalog and minimizing the difference with respect to \citetalias{2016ApJS..225....9H}'s reported values. The resulting value is ${f_{\sigma_{\dot{T}_{\mathrm{dur}}}} = 0.7378}$.

We use equation \ref{eq: sigma_dotT v2} to estimate $\sigma_{\dot{T}_{\mathrm{dur}}}$ for the SysSim planets with calculated TDVs. Finally, we use the $|\dot{T}_{\mathrm{dur}}| > 3 \ \sigma_{\dot{T}_{\mathrm{dur}}}$ criterion to determine whether a planet qualifies as having a detectable TDV signal.

\subsection{Summary of the full process}
\label{sec: summary of the full process}

We synthesize the preceding series of steps and calculate TDVs for simulated planet populations in both the two-Rayleigh model and maximum AMD model. We consider a set of 100 physical/observed catalog pairs for each model with the parameters of the statistical models sampled according to their posterior distributions (see \citetalias{2020AJ....160..276H}, \citealt{2021AJ....161...16H}). The physical catalogs contain orbital elements (inclinations, longitudes of ascending node, etc.) referenced to the sky plane. We transform these orbital elements to be referenced to the invariable plane using equations \ref{eq: omega_sky and omega relationship} and \ref{eq: sin(alpha)}. 

The observed catalogs are first used to specify which planets to perform the TDV calculation for based on whether their TDVs are measurable (Section \ref{sec: TDV measurability}). The physical catalogs determine the details of a given planet's TDV calculation (Section \ref{sec: TDV calculation}). Finally, transit properties from the observed catalog then help quantify whether the resulting TDV signal is detectable (Section \ref{sec: TDV slope detectability}). Each physical/observed catalog pair yields a subset of the planet population with detectable TDV slopes.

\section{Results}
\label{sec: results}

\subsection{Number of planets with detected TDVs}
\label{sec: number of planets with detected TDVs}

The most direct way of comparatively evaluating the two-Rayleigh model and maximum AMD model is to examine each model's total number of simulated planets with detected TDV signals. This quantity can then be compared to the number of observed planets with detected TDV signals, thereby determining which model is a better fit in terms of TDV statistics. In this section, we consider this simple tabulation; in the next section, we look deeper into the properties of the simulated and observed planets with detected TDVs.

\begin{figure}
\epsscale{1.2}
\plotone{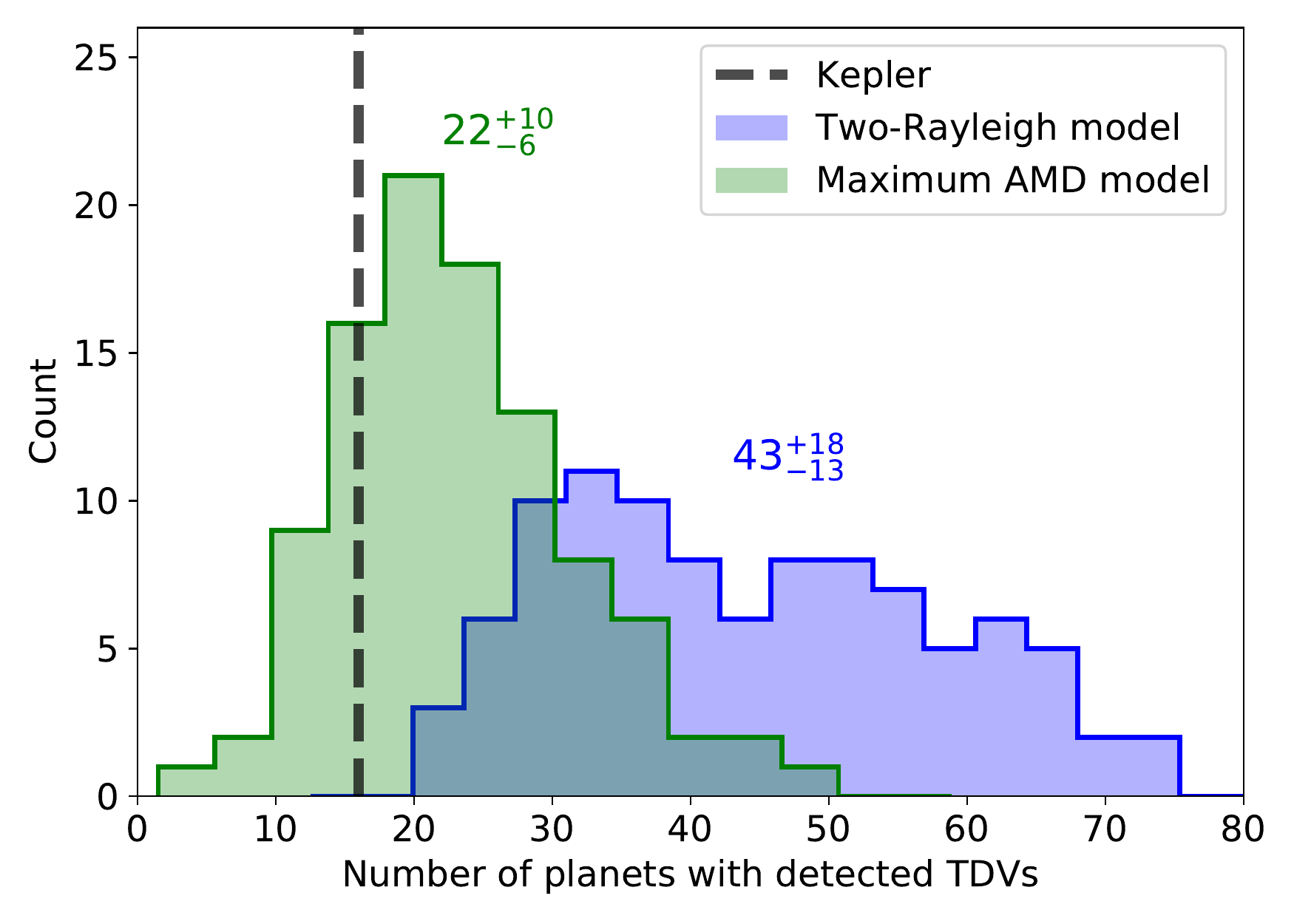}
\caption{Distributions of the number of planets with detected TDVs in the two-Rayleigh model (blue histogram) and the maximum AMD model (green histogram). The histograms correspond to the 100 sets of physical and observed catalog pairs for each model (see Section \ref{sec: summary of the full process}). The medians of the distributions and confidence intervals representing the 16th and 84th percentiles are listed above the corresponding histograms. The number of planets with detected TDVs in the Kepler observations is shown with the dashed vertical line. The maximum AMD model agrees very well with the observed number of planets with detected TDVs, while the two-Rayleigh model produces too many detected TDVs.}
\label{fig: histograms with number of detected TDVs}
\end{figure}

Figure \ref{fig: histograms with number of detected TDVs} presents the tabulation of TDV detections. We show histograms of the number of planets with detected TDVs for 100 physical and observed catalog pairs of each model. (That is, each physical/observed catalog pair corresponds to a single number of planets with detected TDVs.) The medians of the distributions and confidence intervals representing the 16th and 84th percentiles are shown in the figure; these values are $43^{+18}_{-13}$ planets with detected TDVs for the two-Rayleigh model and $22^{+10}_{-6}$ for the maximum AMD model. The observed number of planets with detected TDVs according to Kepler observations (16; see Section \ref{sec: observations}) is also shown. 
The maximum AMD model yields a quantity of planets with detected TDVs that is in agreement with the observations. In contrast, the two-Rayleigh model yields too many planets with detected TDVs to be compatible with the observations. Thus, the TDV statistics support the maximum AMD model (and its associated mutual inclination distribution) over the two-Rayleigh model. We will revisit this conclusion in the Discussion (Section \ref{sec: discussion}).

\begin{figure*}
\epsscale{1.1}
\plotone{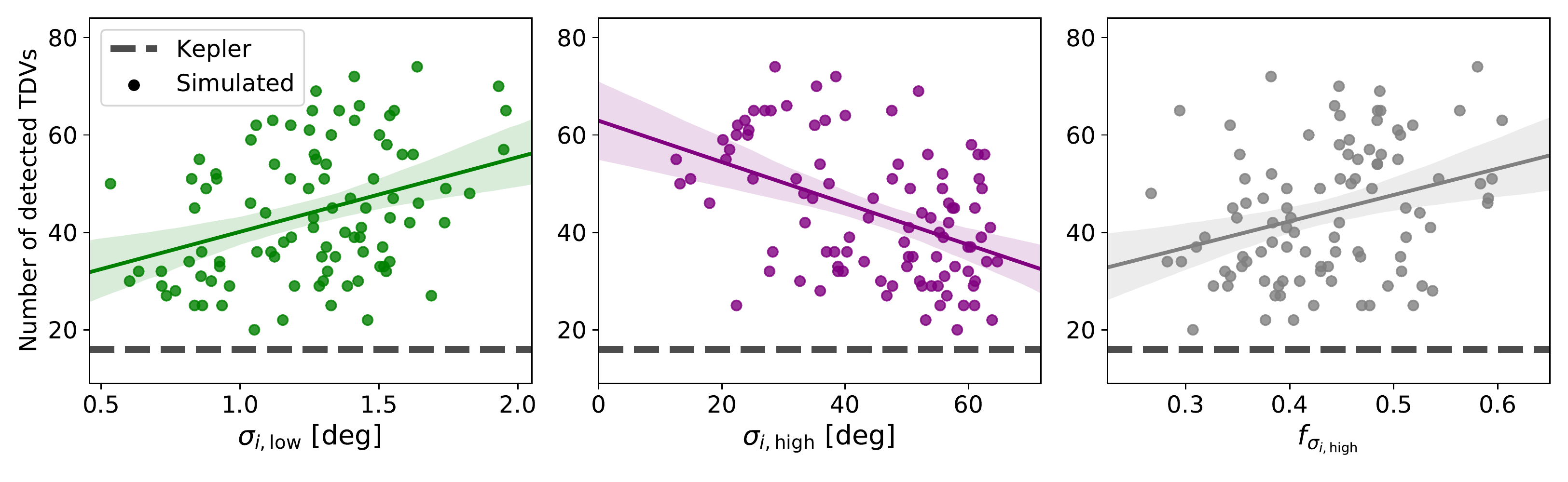}
\caption{Scatter plots of the correlations between the number of planets with detected TDVs in the two-Rayleigh model and the parameters of the model. The $x$-axes of the left and middle panels are the Rayleigh distribution scale parameters of the low and high inclination components, $\sigma_{i,\mathrm{low}}$ and $\sigma_{i,\mathrm{high}}$. The $x$-axis of the right panel is the fraction of systems belonging to the high inclination population, $f_{\sigma_{i,\mathrm{high}}}$. To guide the eye, we include linear regression lines for each scatter plot. The horizontal dashed line corresponds to the number of planets with detected TDVs in the Kepler observations.} 
\label{fig: number of detected TDVs vs sigma_i scatterplots}
\end{figure*}

Although the majority of outcomes for the two-Rayleigh model lead to more simulated systems with detectable TDVs than are observed, the low end of the distribution ($\sim 20$ planets with detected TDVs) is close to the observed number. This suggests that it may be difficult to rule out the two-Rayleigh model entirely. It is helpful to better understand these cases within the context of the broader distribution. To do this, we can study the relationships between the number of planets with detected TDVs and the parameters of the low and high inclination population components. These include the fraction of systems belonging to the high inclination population, $f_{\sigma_{i,\mathrm{high}}}$, and the Rayleigh distribution scale parameters of the low and high components, $\sigma_{i,\mathrm{low}}$ and $\sigma_{i,\mathrm{high}}$.

\begin{figure}
\epsscale{1.1}
\plotone{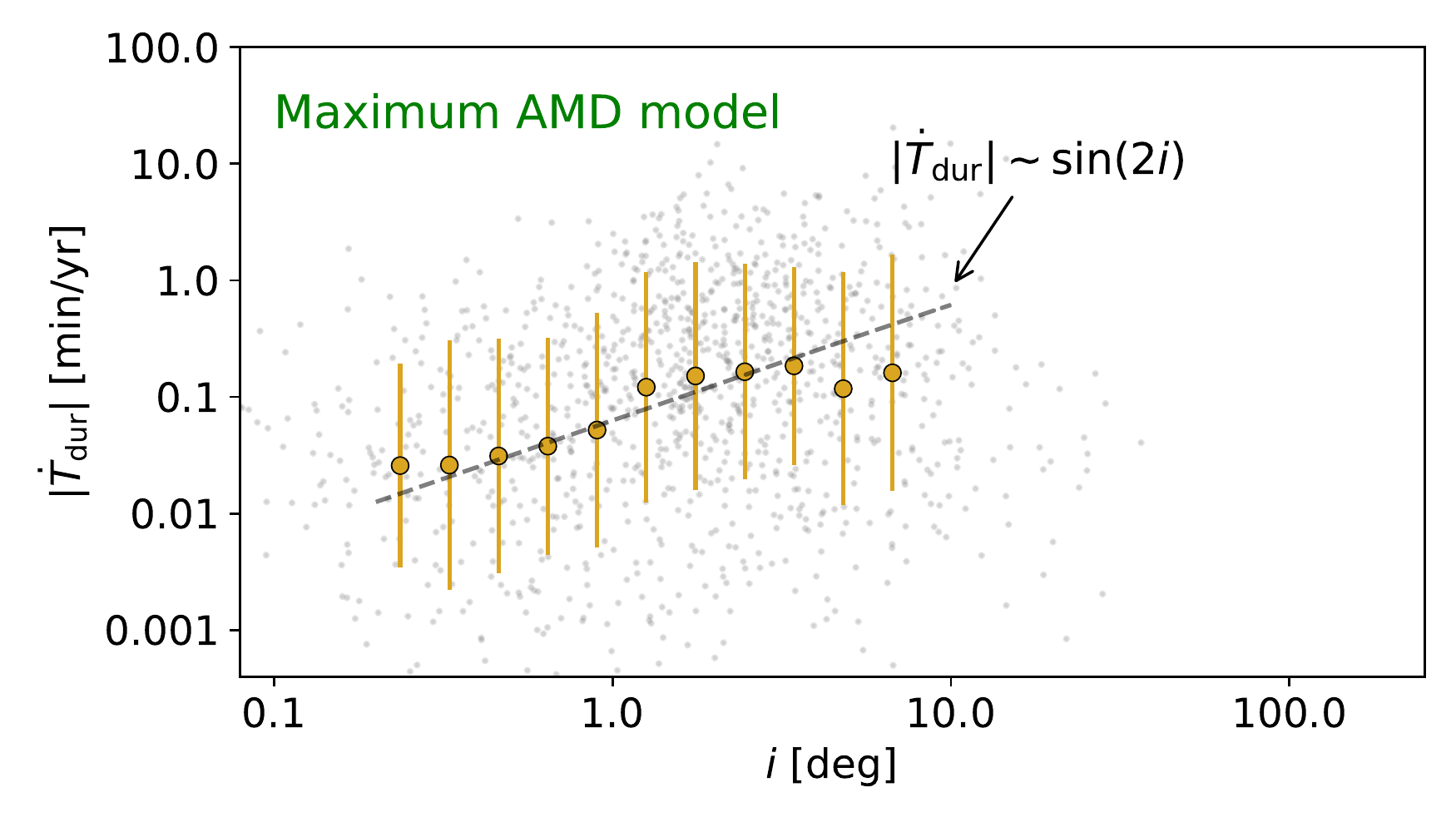}
\plotone{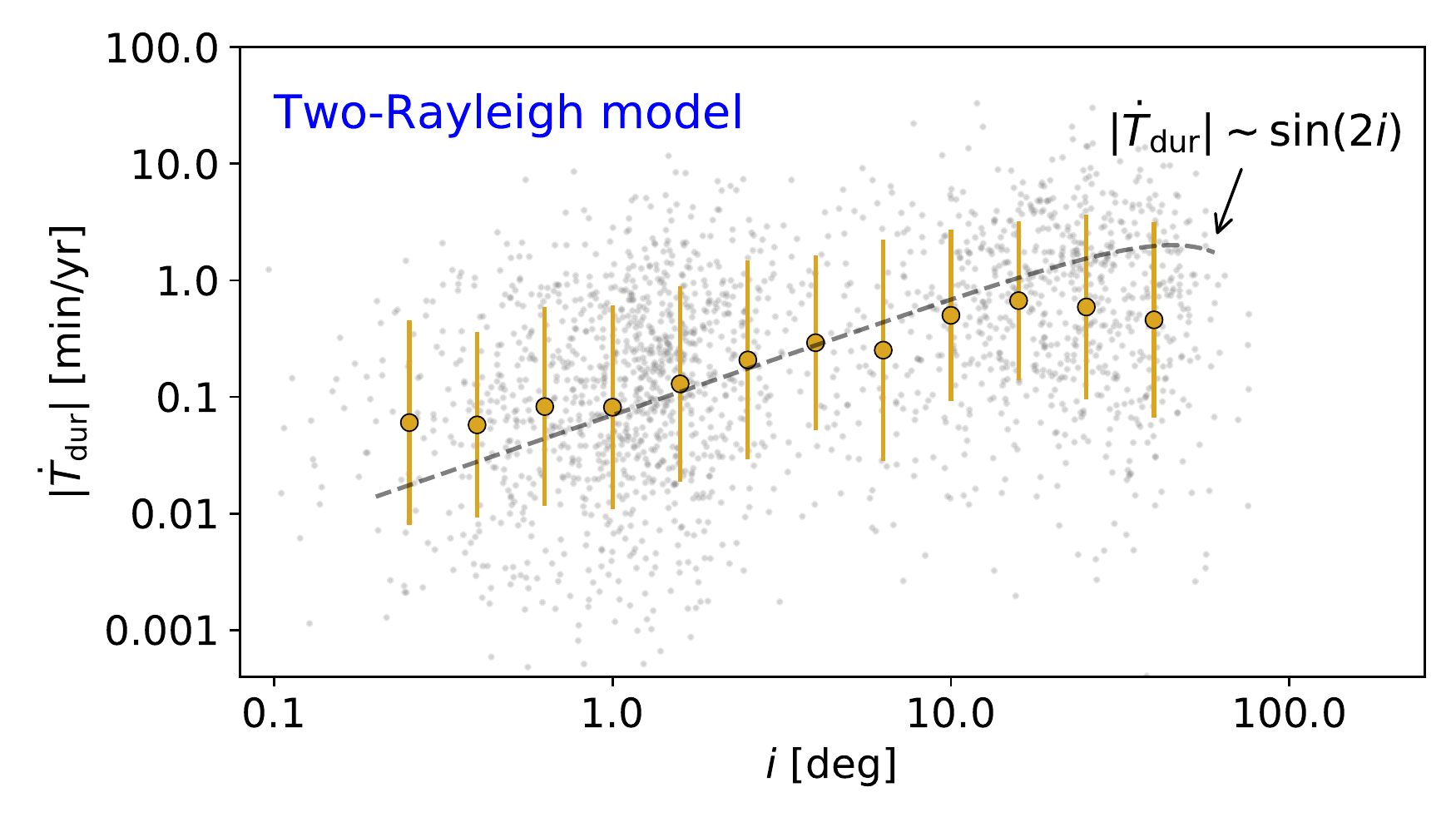}
\caption{Absolute value of the TDV slope, $|\dot{T}_{\mathrm{dur}}|$, as a function of the inclination of the planet's orbit relative to the invariable plane. The gray points indicate all TDV calculations (without regard to TDV detection) within a single observed catalog for the maximum AMD model (top panel) and two-Rayleigh model (bottom panel). The yellow points represent the mean and standard deviation of the gray points within log-uniform bins of $i$. The dashed black lines represent $|\dot{T}_{\mathrm{dur}}| \sim \sin(2i)$ with a normalization such that the curve approximately passes through the data. While $|\dot{T}_{\mathrm{dur}}|\sim\sin(2i)$ is a good rough approximation, there is significant scatter due to the other factors within the $\dot{T}_{\mathrm{dur}}$ calculation (planet masses, semi-major axes, etc.).  }
\label{fig: TDV slope vs inc}
\end{figure}

Figure \ref{fig: number of detected TDVs vs sigma_i scatterplots} shows scatter plots of these relationships. We plot the number of planets with detected TDVs versus $\sigma_{i,\mathrm{low}}$, $\sigma_{i,\mathrm{high}}$, and $f_{\sigma_{i,\mathrm{high}}}$. The number of planets with detected TDVs is positively (albeit weakly) correlated with $\sigma_{i,\mathrm{low}}$ and $f_{\sigma_{i,\mathrm{high}}}$, while it is negatively correlated with $\sigma_{i,\mathrm{high}}$. These relationships can be understood by noting that $|\dot{\Omega}| \sim \cos{i}$, such that $|\dot{T}_{\mathrm{dur}}|\sim\sin{i}\cos{i} \sim \sin{2i}$ (equations \ref{eq: dTdur/dt}, \ref{eq: db/dt}, and \ref{eq: Lagrange's planetary equations}; {see also Figure \ref{fig: TDV slope vs inc}}). When $i$ is small, $|\dot{T}_{\mathrm{dur}}| \sim i$, creating the positive correlation between the number of planets with detected TDVs and $\sigma_{i,\mathrm{low}}$ (left panel of Figure \ref{fig: number of detected TDVs vs sigma_i scatterplots}). Since $|\dot{T}_{\mathrm{dur}}|\sim\sin{2i}$ peaks at $45^{\circ}$, the high inclination component of the two-Rayleigh model has larger average TDV signals, which results in the positive trend between the number of TDV detections and $f_{\sigma_{i,\mathrm{high}}}$ (right panel of Figure \ref{fig: number of detected TDVs vs sigma_i scatterplots}). However, we might also expect that the number of TDV detections would peak at $\sigma_{i,\mathrm{high}} \sim 45^{\circ}$ rather than show a negative correlation (middle panel of Figure \ref{fig: number of detected TDVs vs sigma_i scatterplots}). The negative correlation arises because $f_{\sigma_{i,\mathrm{high}}}$ is inversely correlated with $\sigma_{i,\mathrm{high}}$, such that a low $\sigma_{i,\mathrm{high}}$ is associated with many more systems in the high inclination population, and these systems show more detectable TDVs on average. 

The middle panel of Figure \ref{fig: number of detected TDVs vs sigma_i scatterplots} shows that in most cases where the number of planets with detected TDVs is on the low end of the distribution ($\lesssim 20$), the scale parameter of the high inclination population is near its maximum, $\sigma_{i,\mathrm{high}} \gtrsim 50^{\circ}$. However, systems with extreme mutual inclinations of the level $\sigma_{i,\mathrm{high}} \gtrsim 50^{\circ}$ are disfavored from a physical standpoint, since the orbits are sometimes retrograde and are more likely to be unstable due to secular planet-planet interactions, as noted by \citetalias{2020AJ....160..276H}. The trend shown in the middle panel of Figure \ref{fig: number of detected TDVs vs sigma_i scatterplots} thus further disfavors the two-Rayleigh model; although the trend itself is weak, it is clear that obtaining a plausible number of TDV detections generally requires less plausible values of $\sigma_{i,\mathrm{high}}$ for stability.

\subsection{Properties of planets with detected TDVs}
\label{sec: properties of planets with detected TDVs}

\begin{figure}
\epsscale{1.1}
\plotone{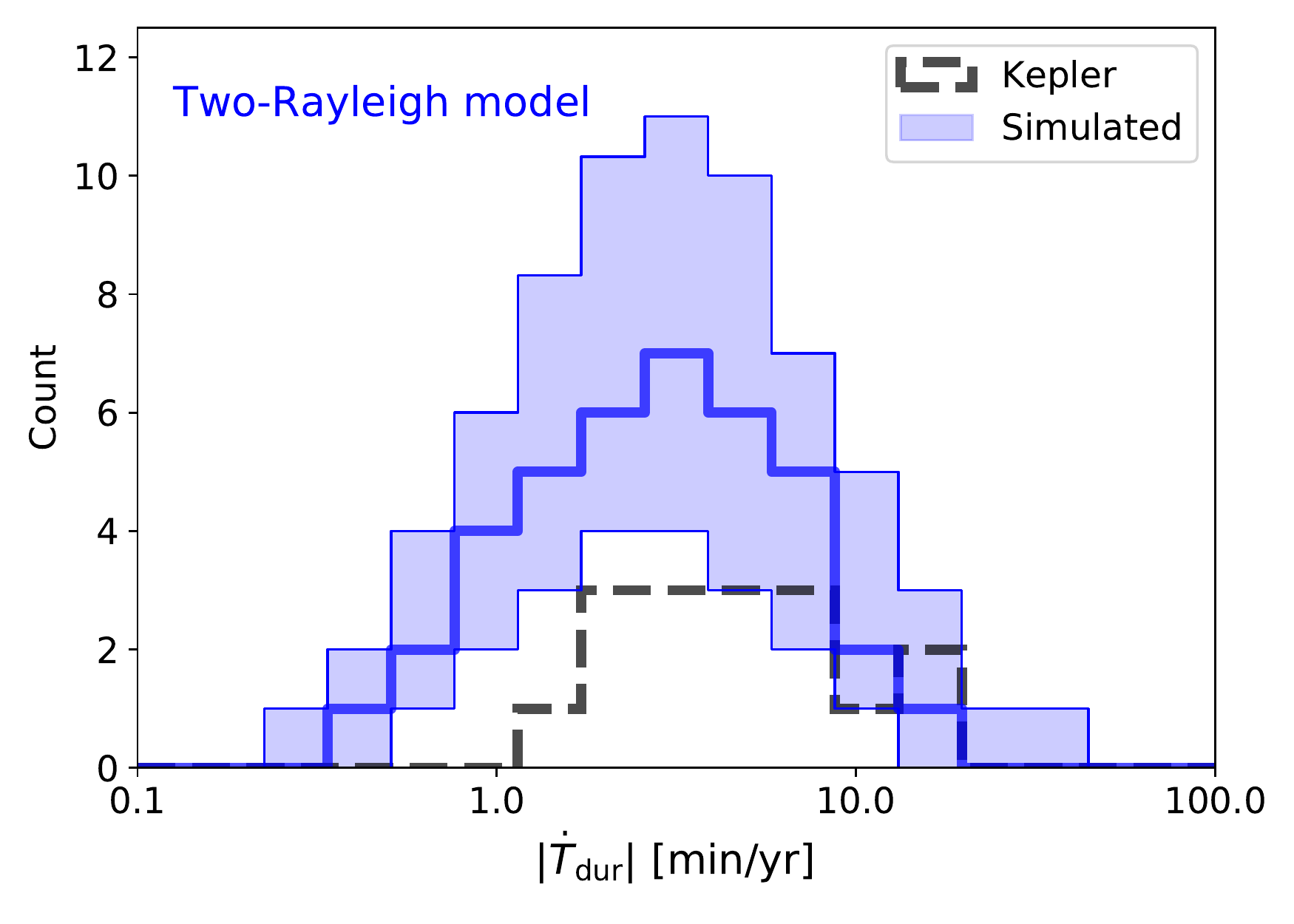}
\plotone{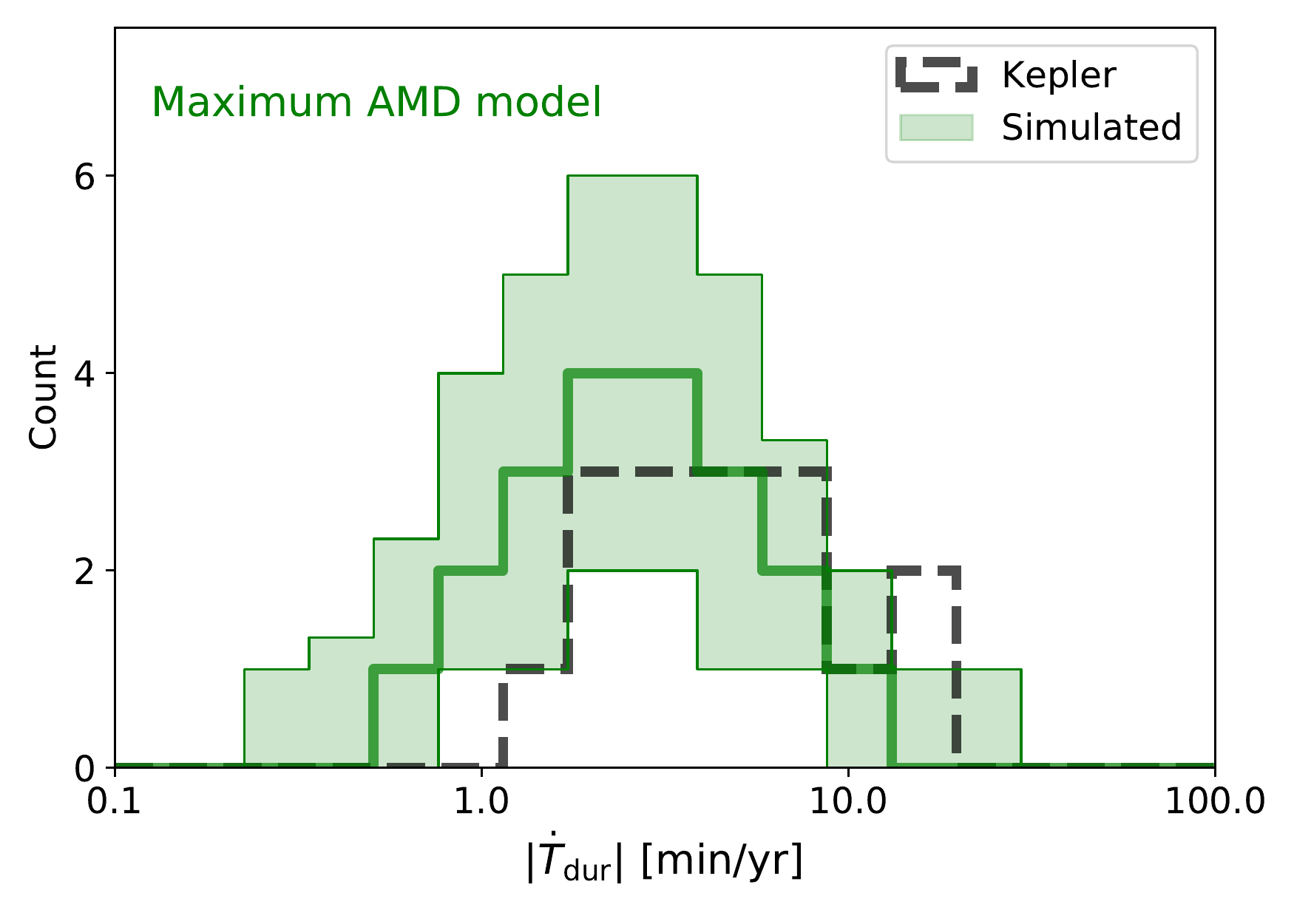}
\caption{Distributions of the absolute value of the TDV slope, $|\dot{T}_{\mathrm{dur}}|$, for planets with detected TDVs. The top and bottom panels show the results for the two-Rayleigh model and the maximum AMD model, respectively. The solid thick line represents the median of the individual histograms for each of the 100 sets of physical and observed catalog pairs. The shaded regions represent the intervals between the 16th and 84th percentiles. The histogram of $|\dot{T}_{\mathrm{dur}}|$ values for the observations is shown with the dashed thick gray line. } 
\label{fig: TDV slope histograms}
\end{figure}

\begin{figure}
\epsscale{1.25}
\plotone{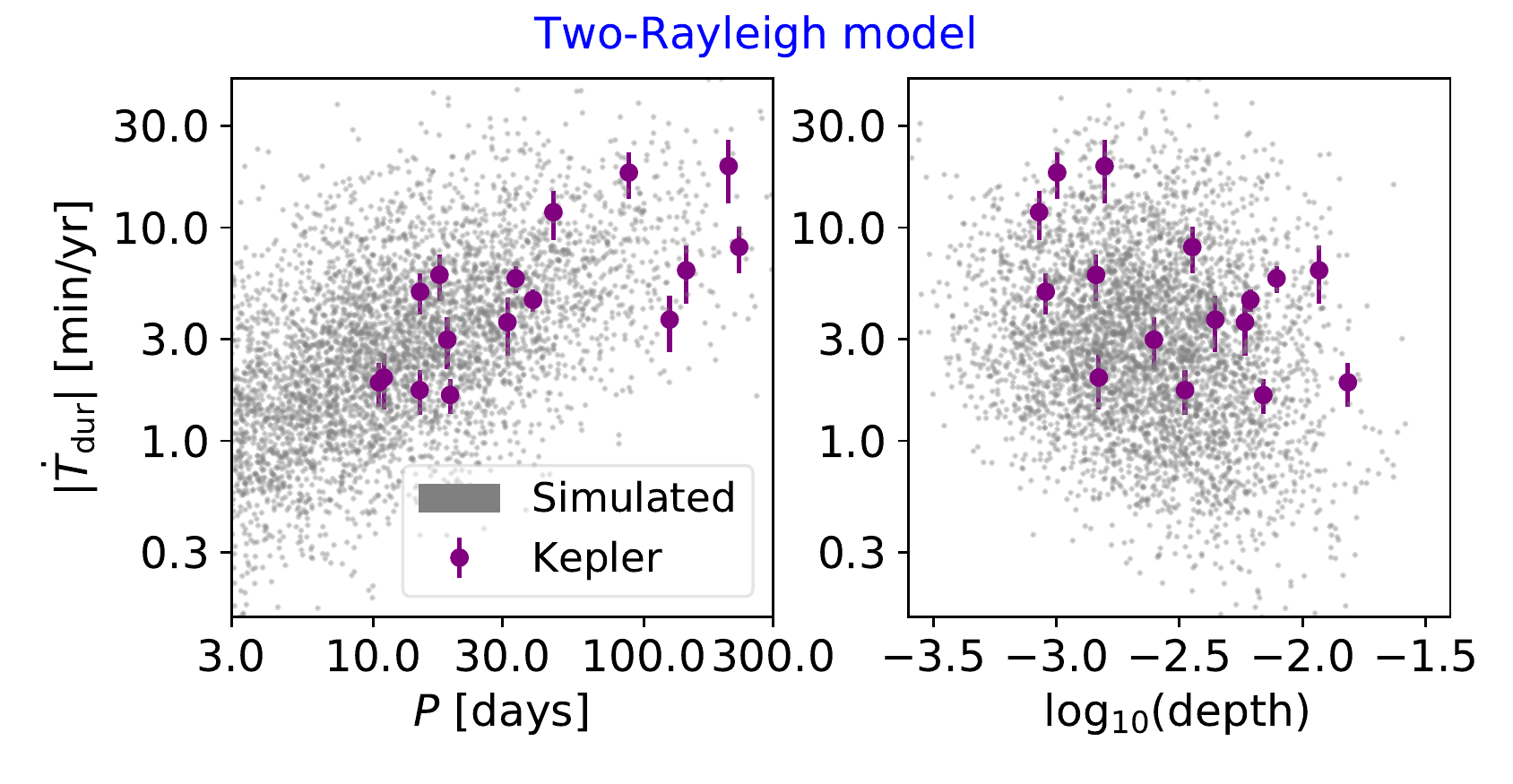}
\plotone{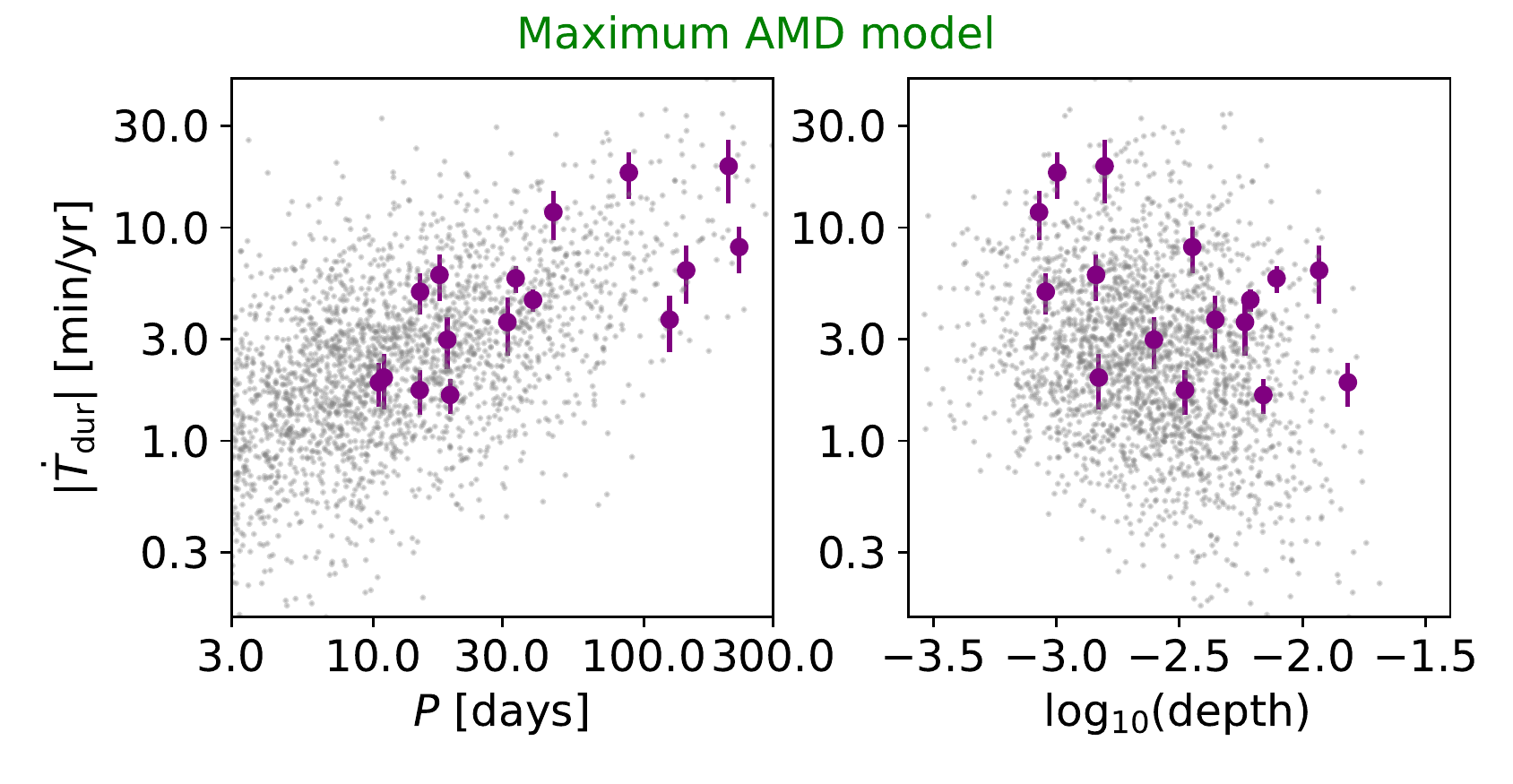}
\caption{Absolute value of the TDV slope, $|\dot{T}_{\mathrm{dur}}|$, as a function of $P$ (left column) and $\log_{10}(\mathrm{depth})$ (right column) of the planets with detected TDVs. The small gray points correspond to the simulated planets with detected TDVs across all 100 sets of physical and observed catalog pairs from the two-Rayleigh model (top row) and maximum AMD model (bottom row). The purple points indicate the observed Kepler planets with detected TDVs.  } 
\label{fig: TDV slope vs P and depth}
\end{figure}

In addition to the simple tabulation of the number of planets with detected TDVs, it is also valuable to compare the specific properties of the simulated planets to the corresponding observed planets with detected TDVs. Relevant properties include the magnitude of the TDV slopes, as well the planets' orbital periods and transit depths. Figures \ref{fig: TDV slope histograms} and \ref{fig: TDV slope vs P and depth} (as well as Table \ref{tab: average properties comparison}, discussed later) show these properties. 

Figure \ref{fig: TDV slope histograms} presents histograms of the absolute value of the TDV slope, $|\dot{T}_{\mathrm{dur}}|$, for planets with detected TDVs. The range and typical values of $|\dot{T}_{\mathrm{dur}}|$ for both the two-Rayleigh model and the maximum AMD model agree well with the observed planets. In all three distributions (observations and two models), the majority of values of $|\dot{T}_{\mathrm{dur}}|$ fall in the range of $1 - 10$ min/yr. The simulations have a slightly greater proportion of detections with $|\dot{T}_{\mathrm{dur}}| \lesssim 1$ min/yr.
The two-Rayleigh model is inconsistent in terms of the observed count, but this is just a restatement of the finding from Figure \ref{fig: histograms with number of detected TDVs} that the two-Rayleigh model produces too many detected TDVs.

Figure \ref{fig: TDV slope vs P and depth} shows additional dimensions of this comparison between simulated and observed planets. We plot $|\dot{T}_{\mathrm{dur}}|$ versus $P$ and $\log_{10}(\mathrm{depth})$ for planets with detected TDVs in the maximum AMD model, the two-Rayleigh model, and the Kepler observations. The left column illustrates that $|\dot{T}_{\mathrm{dur}}|$ is positively correlated with $P$ for both the simulated and observed planets. 

It is illuminating to discuss the physical origins of this positive correlation. In the limit of circular orbits, the expression for $\dot{T}_{\mathrm{dur}}$ (equation \ref{eq: dTdur/dt}) becomes
\begin{equation}
\label{eq: circular dTdur/dt}
\dot{T}_{\mathrm{dur}} \approx -P\left(\frac{b}{\pi\sqrt{1-b^2}}\right)\dot{\Omega}\sin{i}\cos{\beta}\sin{\Omega}.
\end{equation}
If the typical period ratio between adjacent planets, $P_{i+1}/P_i$, is independent of $P$, then $|\dot{\Omega}|\propto P^{-1}$, and the dependence of $|\dot{T}_{\mathrm{dur}}|$ on $P$ vanishes.\footnote{The $|\dot{\Omega}|\propto P^{-1}$ dependence can be seen by combining equations \ref{eq: Lagrange's planetary equations} and \ref{eq: R and R'} or by examining the Laplace-Lagrange secular solution of a two-planet system \citep{1999ssd..book.....M}.} Indeed, $|\dot{T}_{\mathrm{dur}}|$ is completely uncorrelated with $P$ in the full distribution (that includes the TDV slopes that are too small to be detected). However, $\sigma_{\dot{T}_{\mathrm{dur}}}$ is positively correlated with $P$ (due to the decreasing number of transits with increasing $P$), and since TDV detection requires $|\dot{T}_{\mathrm{dur}}| > 3 \sigma_{\dot{T}_{\mathrm{dur}}}$ (Section \ref{sec: observations}), this conspires to yield a positive correlation between $|\dot{T}_{\mathrm{dur}}|$ and $P$. That is, for large $P$, only the steepest values of $|\dot{T}_{\mathrm{dur}}|$ are detectable. 

\cite{2021MNRAS.tmp.1312S} also identified a positive slope between $|\dot{T}_{\mathrm{dur}}|$ and $P$ within the Kepler detected TDVs, and they additionally pointed to a paucity of large $|\dot{T}_{\mathrm{dur}}|$ detections at small $P$, the latter of which might not be explained by the $\sigma_{\dot{T}_{\mathrm{dur}}}$ bias. The simulated planets at small $P$ in Figure \ref{fig: TDV slope vs P and depth} show a slight deficiency of large $|\dot{T}_{\mathrm{dur}}|$ detections ($\sim10$ min/yr) compared to more moderate slopes ($\sim1-3$ min/yr), which is due to the fact that moderate values are more common in the underlying distribution. However, it is not clear that the deficiency in the simulations matches that of the data. Altogether, the simulations reproduce the correlation, which is the dominant feature of the data, but they do not clearly reproduce the paucity of large $|\dot{T}_{\mathrm{dur}}|$ at small $P$.

The $|\dot{T}_{\mathrm{dur}}|$ versus $P$ distribution also shows that the simulated planets are more concentrated at smaller orbital periods in the range ${P=3-10}$ days than the observed planets, which are mostly at ${P>10}$ days. In particular, the observations have more planets with detected TDVs with ${P=100-300}$ days than represented in the simulations. This may be small-number statistics of the data, or it could be because SysSim is less well-calibrated in the ${P=100-300}$ day range due to fewer Kepler planet detections there. In addition, the observed systems may contain perturbing companion planets with ${P>300}$ days, but the simulated systems do not. We will return to this in Section \ref{sec: model assumptions}.

The right column of Figure \ref{fig: TDV slope vs P and depth} shows that $|\dot{T}_{\mathrm{dur}}|$ versus $\log_{10}(\mathrm{depth})$ exhibits a weak negative correlation, which is due to the fact that a larger transit SNR can allow for the detection of a smaller TDV slope. The typical transit depths of the simulated planets are in general agreement with the observed planets, although the observed planets appear to have a larger spread in $\log_{10}(\mathrm{depth})$ than the simulated planets. Altogether, these results indicate that the agreement between the simulated and observed planets with detected TDVs is satisfactory.

\setlength{\extrarowheight}{4pt}
\setlength\tabcolsep{10pt}
\begin{table*}
\centering
\caption{\textbf{Average Properties of Planets with Detected TDVs.} Means of the distributions of several planetary and system properties of planets with detected TDVs. The left column corresponds to the observed KOIs listed in Table \ref{tab: KOIs with TDVs}. We report the mean and standard error of the mean. The middle and right columns correspond to the SysSim simulated planets in the maximum AMD model and two-Rayleigh model, respectively. We calculate the mean values associated with each of the 100 catalogs and report the medians and 16th and 84th percentiles of the distributions of means. Here $N_{\mathrm{obs}}$ is the observed transit multiplicity of the system, and $N_{\mathrm{phys}}$ is the intrinsic multiplicity of the system. The quantity $\sigma_i$ is the standard deviation of the system's orbital inclinations with respect to the invariable plane, including non-detected planets. $N_{\mathrm{phys}}$ and $\sigma_i$ are unknown for the observed systems. The final row is the ratio of the number of TDV detections that are in single-transiting systems compared to multiple-transiting systems. } 
\begin{tabular}{c | c c c }
\hline
\hline
& \multicolumn{3}{c}{\underline{Average values}} \\
& Observations & Maximum AMD model & Two-Rayleigh model \\
\hline
Planet properties &  &  \\
$R_p$ [$R_{\oplus}$] & $5.7 \pm 0.6$ & $4.7 ^{+ 0.4 }_{- 0.4 }$ & $4.8 ^{+ 0.3 }_{- 0.4 }$ \\
$M_p$ [$M_{\oplus}$] & -- & $13.1 ^{+ 10.3 }_{- 3.4 }$ & $18.1 ^{+ 7.1 }_{- 6.1 }$ \\
depth [ppm] & $4327 \pm 929$ & $2975 ^{+ 672 }_{- 531 }$ & $3096 ^{+ 420 }_{- 437 }$ \\
$P$ [days]  & $65.2 \pm 17.8$ & $21.8 ^{+ 7.7 }_{- 5.9 }$ & $22.5 ^{+ 3.7 }_{- 4.6 }$ \\
$T_{\mathrm{dur}}$ [hr] & $6.1 \pm 0.8$ & $3.0 ^{+ 0.3 }_{- 0.2 }$ & $3.2 ^{+ 0.2 }_{- 0.3 }$ \\
$|\dot{T}_{\mathrm{dur}}|$ [min/yr] & $6.4 \pm 1.4$ & $3.6 ^{+ 0.8 }_{- 0.9 }$ & $4.2 ^{+ 0.9 }_{- 0.8 }$ \\ \\
\hline
System properties &  &  \\
$N_{\mathrm{obs}}$ & $2.0 \pm 0.3$ & $1.8 ^{+ 0.2 }_{- 0.2 }$ & $1.6 ^{+ 0.3 }_{- 0.2 }$ \\
$N_{\mathrm{phys}}$ & -- & $3.8 ^{+ 0.5 }_{- 0.4 }$ & $5.6 ^{+ 0.4 }_{- 0.4 }$ \\
$\sigma_{i}$ [deg] & -- & $1.3 ^{+ 0.3 }_{- 0.2 }$ & $16.7 ^{+ 6.5 }_{- 6.2 }$ \\ \\
\hline
Catalog properties & & \\ 
single-to-multi ratio &  0.78 & $0.91 ^{+ 0.5 }_{- 0.3 }$ & $1.77 ^{+ 0.8 }_{- 0.6 }$ \\
\end{tabular}
\label{tab: average properties comparison}
\end{table*}

Table \ref{tab: average properties comparison} shows a quantitative comparison of the average properties of the planets with detected TDVs (and their systems) between the SysSim simulated planets and the observed KOIs. The table summarizes several of the key takeaways from Figures \ref{fig: TDV slope histograms} and \ref{fig: TDV slope vs P and depth}, including the general agreement of $|\dot{T}_{\mathrm{dur}}|$ between simulations and observations and the larger average $P$ for the observed planets. In addition, we note that the observed transit multiplicity of systems with detected TDVs shows good agreement, with $N_{\mathrm{obs}} = 1.8 ^{+ 0.2 }_{- 0.2 }$ for the simulated systems in the maximum AMD model versus $2.0 \pm 0.3$ for the observed systems. The average intrinsic multiplicity of the maximum AMD model simulated systems hosting planets with detected TDVs is $N_{\mathrm{phys}} = 3.8 ^{+ 0.5 }_{- 0.4 }$ (but clearly unknown for the observed systems). This is marginally smaller than the average of the full population of simulated systems with observed planets, which is $N_{\mathrm{phys}} \sim 4.5$.\footnote{This quantity is larger than the average number of planets per planetary system found by \citetalias{2020AJ....160..276H}, $3.12^{+0.36}_{-0.28}$, because it is conditional upon the system having at least one \textit{observed} planet.} The maximum AMD model simulated systems hosting planets with detected TDVs have an average standard deviation of the inclinations (measured with respect to the invariable plane) equal to $\sigma_i = 1.3 ^{+ 0.3 }_{- 0.2 }$ deg, marginally larger than the average of the full population, $\sigma_i \sim 1$ deg.    

\newpage
\subsection{Sensitivity of results to model assumptions}
\label{sec: model assumptions}

\begin{figure}
\epsscale{1.1}
\plotone{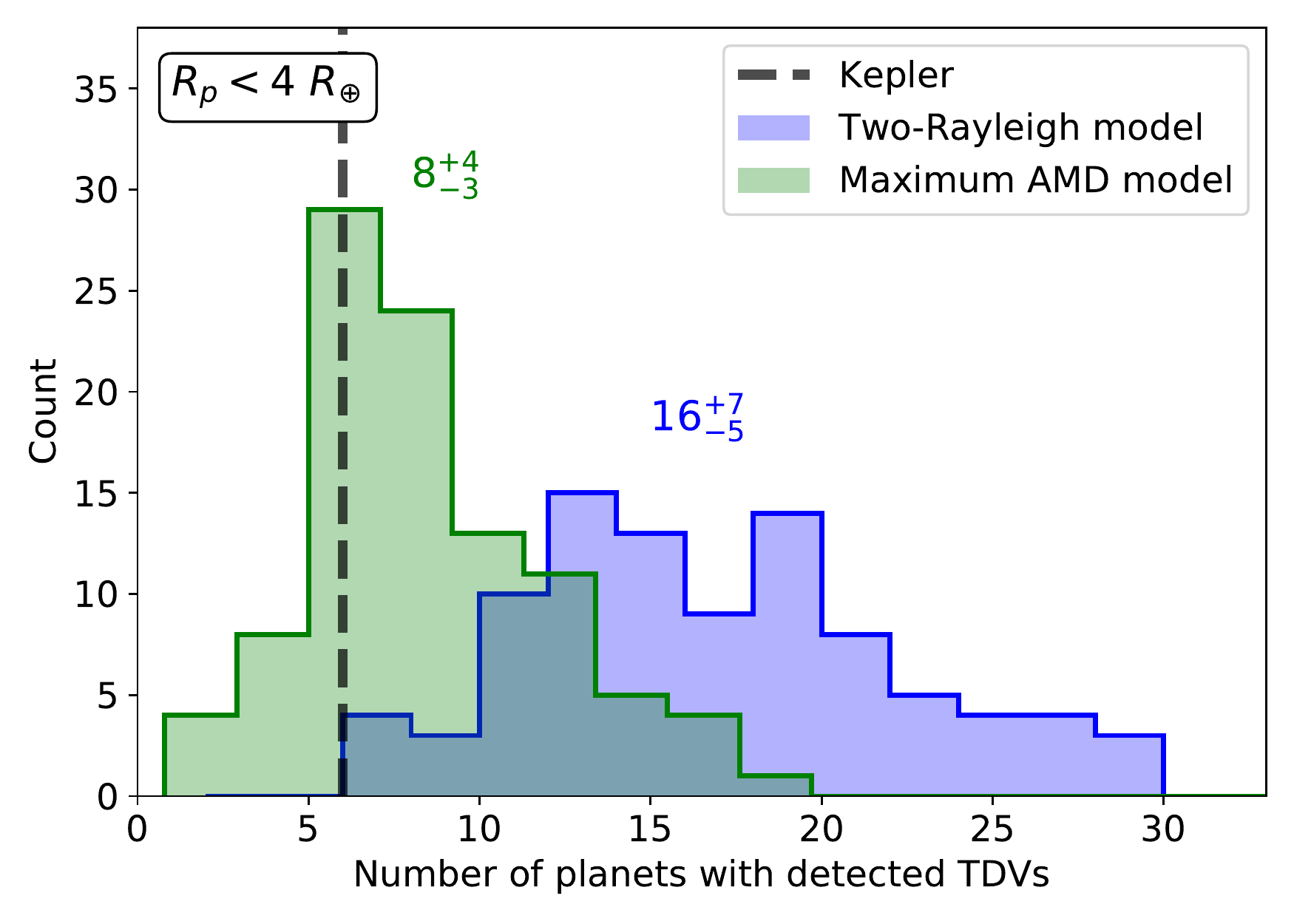}
\plotone{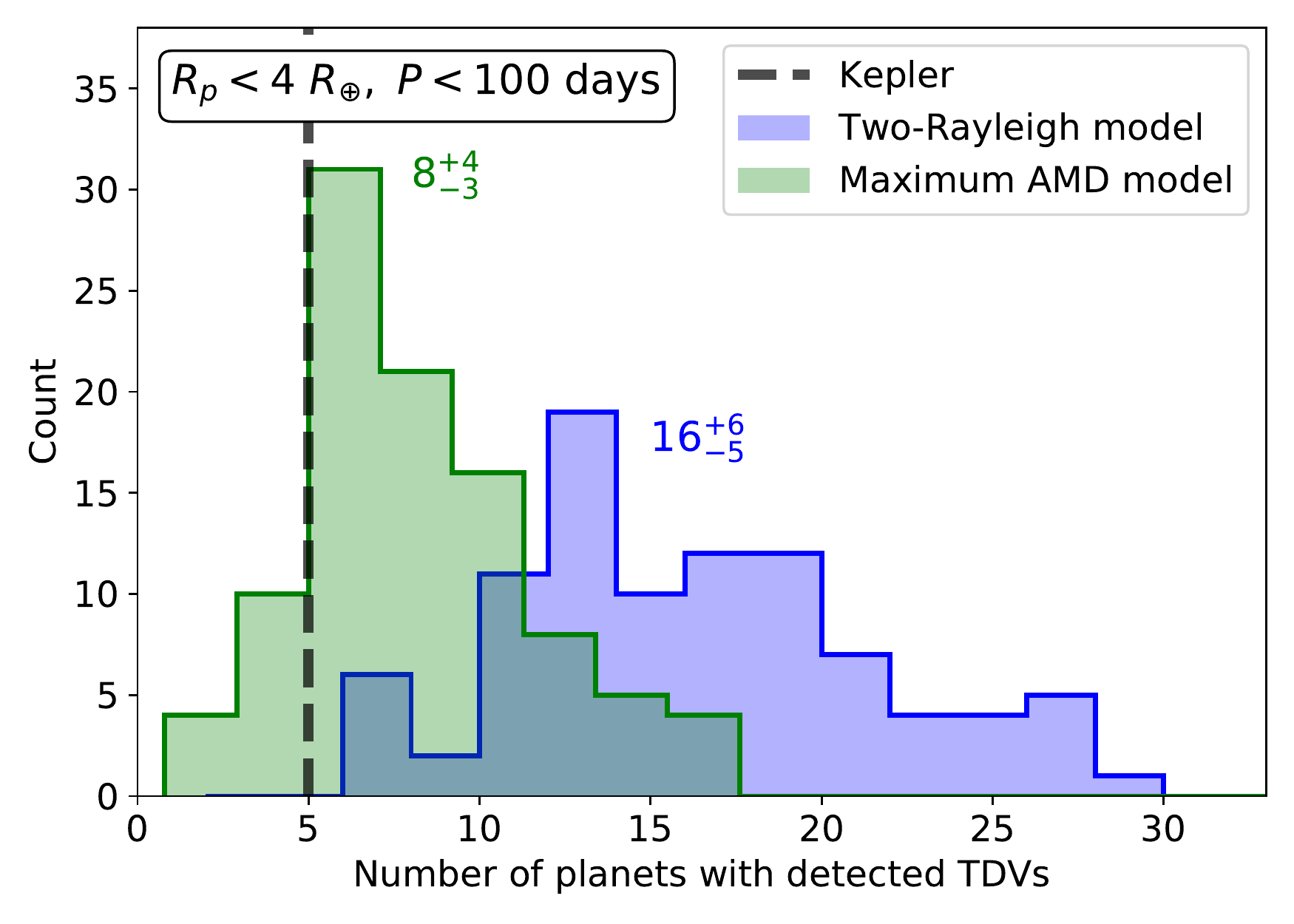}
\caption{Same as Figure \ref{fig: histograms with number of detected TDVs}, except with the planet radius upper limit equal to $4 \ R_{\oplus}$ rather than $10 \ R_{\oplus}$ (both panels) and with the orbital period upper limit equal to 100 days rather than 300 days (bottom panel).}
\label{fig: histograms with number of detected TDVs for 4 Rearth cut}
\end{figure}

In this section, we investigate the robustness of our results with respect to variations in the models, specifically the planet radius and period ranges, and \citetalias{2020AJ....160..276H}’s assumption that systems are at the critical AMD. To begin, we note that SysSim considers planet radii in the range $0.5 \ R_{\oplus} < R_p < 10 \ R_{\oplus}$, but it is best constrained for $R_p < 4 \ R_{\oplus}$ due to the large number of detections of sub-Neptune-sized planets. We can repeat our analysis by changing the $R_p$ upper limit from $10 \ R_{\oplus}$ to $4 \ R_{\oplus}$. This reduces the number of observed Kepler TDV detections from 16 (Section \ref{sec: observations}) to 6.  Figure \ref{fig: histograms with number of detected TDVs for 4 Rearth cut} (top panel) displays the distributions of the number of simulated planets with detected TDVs, as in Figure \ref{fig: histograms with number of detected TDVs}. We observe a similar or perhaps even better agreement between the maximum AMD model and the observed number of planets with detected TDVs. Meanwhile, the two-Rayleigh model again produces too many detected TDVs. Our results are therefore robust with respect to the choice of the upper limit on $R_p$.

In addition to the finite planet radius range, SysSim also uses a finite range in orbital periods, ${3 \ \mathrm{days} < P < 300 \ \mathrm{days}}$. This leads to an important limitation of our analysis, in that our population of simulated TDV detections is missing cases that would have arisen from planetary perturbers with $P < 3$ days or $P > 300$ days. This may be responsible for the underabundance of simulated TDV detections with $P \sim 100-300$ days relative to the observations (Figure \ref{fig: TDV slope vs P and depth}). Accounting for this underprediction of planets with detected TDVs in the simulated population would tend to further disfavor the two-Rayleigh model, although it may also lead to a tension with the maximum AMD model if the difference is significant. The best way to address this would be to generalize the \citetalias{2020AJ....160..276H} model to include planets with orbital periods beyond the current cutoffs and then use SysSim to generate new simulated catalogs for comparison.  Generalizing the \citetalias{2020AJ....160..276H} model and performing parameter estimation on the new model's  parameters is beyond the scope of this study. 
However, we can repeat the analysis on a restricted sample of TDV detections with $P < 100$ days (bottom panel of Figure \ref{fig: histograms with number of detected TDVs for 4 Rearth cut}), which is less affected by the $300$ day cutoff. We find that the overall conclusions are unchanged. 

Another simulation feature to consider is the assumption of the maximum AMD model that all multi-planet systems are at the AMD-stability limit, with ${\mathrm{AMD} = \mathrm{AMD_{crit}}}$. \citetalias{2020AJ....160..276H} tested variations of their model in which ${\mathrm{AMD} = f_{\mathrm{crit}}\times\mathrm{AMD_{crit}}}$, where $f_{\mathrm{crit}}$ is some factor in the range $[0,2]$. They found that values between 0.4 and 2 are all acceptable and that the fit did not significantly improve with the extra $f_{\mathrm{crit}}$ parameter. However, it is possible that this factor could affect the TDV distribution more strongly than the other Kepler observables. We find this to be unlikely. As shown by Figure 7 in \citetalias{2020AJ....160..276H}, when allowing for $f_{\mathrm{crit}} \ne 1$, the mutual inclination distribution maintains the same shape and shifts by less than a factor of two. Specifically, the median mutual inclination shifts to $\sim 0.82^\circ$ ($1.64^\circ$) for $f_{\rm crit} = 0.5$ (2). Meanwhile, $|\dot{T}_{\mathrm{dur}}|$ varies by up to three orders-of-magnitude among the planets with detected TDVs (with $|\dot{T}_{\mathrm{dur}}|$ ranging from $\sim 0.1 - 100$ min/yr; see Figure \ref{fig: TDV slope vs P and depth}). Including the non-detected TDVs, there is an even larger range, with $|\dot{T}_{\mathrm{dur}}|$ as low as $\sim10^{-4}$ min/yr. Given the dynamic range in $|\dot{T}_{\mathrm{dur}}|$, the small changes in inclinations cannot significantly change the distribution of TDVs.

\section{Discussion}
\label{sec: discussion}

\subsection{Small mutual inclinations for the Kepler planets}
\label{sec: small mutual inclinations for the Kepler planets}

The TDV statistics of the Kepler population agree very well with the simulated planet population constructed by the maximum AMD model, whereas the two-Rayleigh model produces too many detected TDVs (Figures \ref{fig: histograms with number of detected TDVs} and \ref{fig: histograms with number of detected TDVs for 4 Rearth cut}). Given that the primary difference between these two models is the mutual inclination distribution, which TDVs are sensitive to (e.g. Figures \ref{fig: number of detected TDVs vs sigma_i scatterplots} and \ref{fig: TDV slope vs inc}), we conclude that the TDV statistics support the mutual inclination distribution of the maximum AMD model. We emphasize that the TDV statistics are not capable of assessing \citetalias{2020AJ....160..276H}'s assumption that systems are at the critical AMD. However, they appear consistent with the implications of that assumption.

As discussed in Section \ref{sec: Maximum AMD model} (and originally in Section 3.3 of \citetalias{2020AJ....160..276H}), the maximum AMD model yields median mutual inclinations, $\tilde{\mu}_{i,n}$, of systems with $n= 2, 3, ..., 10$ planets that are well modeled by a power-law of the form
$\tilde{\mu}_{i,n} = 1.10^{\circ}(n/5)^{-1.73}$. This equation evaluates to $5.4^{\circ}$, $2.7^{\circ}$, and $1.6^{\circ}$ for two-, three- and four-planet systems, respectively. These inclinations are quite modest relative to the high mutual inclinations that have previously been invoked to explain the Kepler dichotomy (e.g. \citealt{2018AJ....156...24M}, \citetalias{2019MNRAS.490.4575H}). Of course, systems with higher mutual inclinations do exist (e.g. Kepler-108, \citealt{2017AJ....153...45M}), but according to our results they do not make up a large fraction of the population. 

A final point to emphasize about \citetalias{2020AJ....160..276H}'s mutual inclination distribution is that it is non-dichotomous; it does not bifurcate the population into ``dynamically cool'' and ``dynamically hot'' systems. The fact that the TDV statistics support this model is therefore evidence for a continuum of architectures rather than a dichotomy.
As a caveat, we note that there are many possible combinations in the true multiplicity/inclination distribution that are not encapsulated in the two models we tested. Reality is likely more complicated than ``there is'' or ``there is not'' a dichotomy. However, we can say that (1) the continuous maximum AMD model is consistent with the data, and (2) replacing any of the model's low inclination systems with high inclination configurations would tend to increase the number of TDV detections, thus leading to eventual inconsistency with the small number of observed detections. The two models tested here serve as a basis of comparison for future, more complex models. For now, we focus on the implications of the favored maximum AMD model.

The mutual inclinations of the maximum AMD model are consistent with previous works that constrained the distribution by supplementing the Kepler transit statistics with additional observations. The most similar result is that of \cite{2018ApJ...860..101Z} (see also \citealt{2020AJ....159..164Y}), who used Kepler TTV statistics as their additional constraint. \cite{2018ApJ...860..101Z}'s model assumed a power-law of the mutual inclination dispersion with the intrinsic multiplicity, $\sigma_{i,n} = \sigma_{i,5}(n/5)^{\alpha}$, and identified best-fit parameters equal to $\alpha = -3.5$ and $\sigma_{i,5} = 0.8^{\circ}$. This is steeper than that found by \citetalias{2020AJ....160..276H} but a similar result indicating relatively small inclinations that depend on the intrinsic multiplicity. 

Rather than TTVs, \cite{2012AJ....143...94T} and \cite{2012A&A...541A.139F} used RV survey data as their extra observational constraint. By combining statistical information from Kepler and RV surveys, they found evidence that suggested that the mean mutual inclinations are $\lesssim5^{\circ}$. 
Detailed analyses of RVs of individual multiple planet systems with strong gravitational interactions also provide evidence for small mutual inclinations \citep{2002ApJ...579..455L, 2014MNRAS.441..442N, 2016MNRAS.455.2484N, 2018AJ....155..106M}.  

Finally, the inference of small mutual inclinations of the Kepler planets is also consistent with population-level constraints on Kepler stellar obliquities. Studies using photometric variability observations \citep{2015ApJ...801....3M} and $v\sin i$ measurements \citep{2017AJ....154..270W, 2021AJ....161...68L} showed that stellar spin-orbit misalignments are small for Kepler planets orbiting cool stars ($T_{\mathrm{eff}} \lesssim 6250$ K). All of these results are broadly consistent with \citetalias{2020AJ....160..276H}'s maximum AMD model.  Thus, we are seeing agreement between Kepler transit statistics, TTVs, RVs, stellar obliquities, and TDVs, indicating a collection of robust evidence that the vast majority of inner planetary systems around Sun-like stars have small mutual inclinations.

\subsection{Physical origins of the mutual inclinations}
\label{sec: physical origins}

The predominantly small (but non-zero) inclinations, $i\lesssim5^{\circ}-10^{\circ}$, must have been excited through one or more physical processes. While several dynamical mechanisms have been proposed, it is still unclear which dominate. Here we will review the relevant theories and discuss which are favored by the observational evidence for small inclinations.

\subsubsection{Excitation during late-stage planet assembly}
\label{sec: primordial excitation}
To begin, we note that the requisite mutual inclinations can probably be acquired primordially during the planet formation epoch. Many studies have used $N$-body simulations to study the formation of close-in, compact systems in their final assembly stage, during which planetary embryos undergo mutual scatterings and collisional growth. Some of these models have considered \textit{in situ} formation in a gas-poor or gas-empty environment, typically starting with a dense planetesimal disk with various disk masses and radial profiles \citep[e.g.][]{2013ApJ...775...53H, 2016ApJ...822...54D, 2016ApJ...832...34M, 2017AJ....154...27M, 2020MNRAS.491.5595P}. In contrast to \textit{in situ} formation models, migration models consider formation of protoplanets beyond $\sim 1$ AU. Disk-planet interactions lead to inward migration and the formation of long chains of short-period protoplanets in mean-motion resonances. Once the gas disk disperses, these compact resonant chains can become dynamically unstable and collide, promoting further planetary growth  \citep[e.g.][]{2007ApJ...654.1110T, 2014A&A...569A..56C, 2014MNRAS.445..479C,  2017MNRAS.470.1750I, 2018ApJ...866..104C,  2019arXiv190208772I, 2019MNRAS.486.3874C}.

In both the \textit{in situ} and migration scenarios, gravitational scattering of planetary embryos leads to dynamical excitation of the mutual inclinations and eccentricities of the final planetary system, potentially allowing for the production of systems that agree with the Kepler multiplicity distribution.\footnote{This dynamical excitation must occur among planetary embryos during the formation epoch. Dynamical instabilities among fully-formed Kepler multi-planet systems are unlikely to produce sufficient excitation \citep[e.g.][]{2012ApJ...758...39J}.} These models are generally capable of producing inclinations in the range of $\sim1^{\circ}-10^{\circ}$ \citep[e.g.][]{2013ApJ...775...53H, 2016ApJ...832...34M, 2019arXiv190208772I, 2019MNRAS.486.3874C, 2020MNRAS.491.5595P}, in agreement with the mutual inclinations of the maximum AMD model from \citetalias{2020AJ....160..276H}. For instance, the migration simulations of \cite{2019arXiv190208772I} find an excess of single-transiting systems that matches the Kepler multiplicity distribution and arises primarily from systems of $2-3$ planets with inclinations between $\sim4^{\circ}-10^{\circ}$. In contrast to \cite{2019arXiv190208772I}, some other variations of late-stage planet formation simulations do not reproduce the large number of single-transiting systems, at least not with a uniform underlying model \citep[e.g.][]{2013ApJ...775...53H, 2020MNRAS.491.5595P}. However, this appears equally if not more influenced by the intrinsic multiplicities of the simulated systems being too high (average of $\sim5$ planets per system rather than $\sim3$ as found by \citetalias{2020AJ....160..276H}) than the inclinations being too low, and the multiplicities are strongly influenced by the initial conditions of the simulations. Finally, in addition to the proper range of inclinations, these models also naturally produce an inverse correlation between inclination (and eccentricity) dispersion and intrinsic multiplicity, as shown in the analysis of \cite{2018ApJ...866..104C}'s simulations  by \citetalias{2020AJ....160..276H}.

\subsubsection{External perturbations: stellar oblateness and distant giant planets}
\label{sec: external perturbations}
While dynamical instabilities during the late-stage planet assembly process appear capable of delivering the required level of mutual inclination excitation, this is not guaranteed, and the primordial formation conditions are poorly-constrained. This has led some to postulate the importance of external gravitational perturbations from the star or other planets in the system as sources of dynamical excitation. These external perturbations additionally offer solutions to systems with very high ($\gtrsim20^{\circ}$) mutual inclinations \citep[e.g.][]{2017AJ....153...45M} and/or misaligned stellar obliquities \citep[e.g.][]{2021MNRAS.502.2893K}, which are observed in a handful of systems and cannot be attributed to primordial collisional excitation among the inner Kepler planets. 

One source of external perturbation is the quadrupolar gravitational potential of a tilted host star, which is particularly strong when the star is young and rapidly rotating \citep{2016ApJ...830....5S}. Shortly after disk dispersal, a tilted star can drive differential nodal precession of the inner planetary orbits that leads to mutual inclinations comparable to the magnitude of the stellar obliquity. The excitation also depends on the coupling of the Kepler planets compared to the stellar forcing. To match the $\lesssim10^{\circ}$ mutual inclinations, spin-orbit misalignments of this magnitude must be widespread during the disk-hosting stage. There are numerous pathways for generating star-disk tilts \citep[e.g.][]{2012Natur.491..418B, 2014MNRAS.440.3532L, 2014ApJ...790...42S, 2014ApJ...797L..29S, 2018MNRAS.478..835Z}. Moreover, the $6^{\circ}$ obliquity of the Sun is of the requisite magnitude, and recent data suggests that spin-orbit misalignments in systems of small Kepler planets are more common than previously thought, particularly among hot stars \citep{2021AJ....161...68L}. The oblateness-driven inclination excitation theory is also consistent with the relatively large (up to $\sim10^{\circ}$) observed inclinations of ultra-short period planets \citep{2018ApJ...864L..38D, 2020ApJ...890L..31L}, provided their inward migration occurs early enough \citep{2020ApJ...905...71M}. The biggest unknown with the theory is that it requires rapid disk dispersal timescales, such that a primordial star-disk misalignment can naturally become a spin-orbit misalignment \citep{2020AJ....160..105S}.

In addition to the host star, another gravitational source that can drive differential precession is a distant ($\gtrsim1$~AU) giant planet (or ``cold Jupiter'') on an inclined orbit \citep{2017AJ....153...42L, 2017MNRAS.467.1531H, 2017MNRAS.468..549B, 2018MNRAS.478..197P}. The giant can similarly generate mutual inclinations among the inner planets up to roughly the magnitude of the giant's orbital tilt, depending on the coupling of the inner planets to one another relative to the perturbations from the giant. The inclination excitation can, however, become much larger than the giant’s inclination when there exists a secular resonance between a nodal precession frequency and a system eigenfrequency \citep{2018AJ....155..139G, 2018MNRAS.478..197P}. In addition to dynamically quiet secular interactions, an outer system containing \textit{multiple} giant planets can undergo a violent epoch of planet-planet scattering and collisions that reduce the multiplicity and/or dynamically heat the inner system \citep[e.g.][]{2017MNRAS.464.1709G, 2017MNRAS.468.3000M, 2020MNRAS.491.5595P}. Recently, \cite{2020AJ....159...38M} showed that cold Jupiters have $\sim10^{\circ}$ mutual inclinations relative to inner transiting systems, with lower inclinations relative to the multis than singles. This is consistent with a scenario in which inclined giants perturb inner systems. However, it is equally consistent with other mechanisms that generate inclinations among the inner planets and produce an inclination with respect to the cold Jupiter as a by-product.

Cumulatively, the data suggest that distant giants likely play a role in dynamically heating some subset of inner planetary systems, with prime examples including HAT-P-11 \citep{2018AJ....155..255Y}, $\pi$ Mensae \citep[e.g.][]{2020MNRAS.497.2096X, 2021MNRAS.502.2893K}, and WASP-107 \citep[e.g.][]{2021AJ....161...70P}, all of which have short-period Neptunes (or sub-Neptunes) with large stellar obliquities and eccentric cold Jupiters. However, as far as generating the Kepler dichotomy through dynamical excitation of inner systems, distant giant planets are unlikely to be the dominant solution. There are several reasons for this. First, roughly three quarters of Kepler planetary systems of super-Earths/sub-Neptunes have just a single transiting planet, while roughly one third of these systems contain distant giant planets \citep{2018AJ....156...92Z, 2019AJ....157...52B}. Provided the distant giant fraction is similar around transit singles and transit multis \citep{2020AJ....159...38M}, the statistical constraints make it difficult to generate a sufficient number of transit singles through giant planet perturbations. Moreover, it is unlikely that the cold Jupiter occurrence is significantly higher among the transit singles compared to the transit multis, given that the stellar metallicity distributions are indistinguishable \citep{2018AJ....155..134M}. 

A second consideration is that, even when one or more cold Jupiters are present, they do not always (in the absence of secular resonances) lead to enhanced mutual inclinations in inner systems, since the inner planet orbits are often tightly coupled \citep[e.g.][]{2017MNRAS.464.1709G, 2017MNRAS.468.3000M}. However, when mutual inclinations do result, they are generally larger than the $i \lesssim 5^{\circ}-10^{\circ}$ scale required by \citetalias{2020AJ....160..276H}, particularly when there are scattering interactions in the outer system \citep{2017MNRAS.468.3000M}. Finally, the gravitational perturbations from distant giants are often overpowered by stellar oblateness soon after disk dispersal \citep{2020AJ....160..105S}, such that a planetary system initially dominated by the star can evade giant-induced inclination excitation by adiabatically realigning to the giant's orbital plane during stellar spin-down.

\section{Conclusion}

The distribution of inclinations in multiple-planet systems encodes fundamental clues about planet formation. Unfortunately, it is inherently difficult to infer from radial velocity and transit observations. In particular, the observed transiting multiplicity distribution is strongly affected by the underlying mutual inclination distribution, but it also depends on the intrinsic multiplicity distribution, yielding a near degeneracy that makes inferences of inclinations difficult.
In this work, we used Transit Duration Variations (TDVs) of the Kepler planet population to break the near degeneracy and constrain the mutual inclination distribution of close-in, multi-planet systems. TDVs often arise when a transiting planet’s transit chord drifts due to orbital precession induced by torques from perturbing planets (Figure \ref{fig: geometric diagram}). The signal is sensitive to the mutual orbital inclinations and can yield detectable drifts on the order of $\dot{T}_{\mathrm{dur}} \sim 1-10$ min/yr. Several dozen Kepler planets exhibit TDV drift signals (Table \ref{tab: KOIs with TDVs}; Figure \ref{fig: Kepler-9 TDVs} for an example). Our work is the first to exploit these detections statistically to characterize the mutual inclination distribution. 

We compared the observed TDV detections of Kepler planets to expectations from simulated planet populations subject to different assumptions about the mutual inclination distribution. These simulated planetary systems were drawn from two population models built using  the ``SysSim'' empirically-calibrated forward modeling framework: (1) the ``two-Rayleigh model'' \citep{2019MNRAS.490.4575H}, which assumes a dichotomous mutual inclination distribution with low ($\sigma_{i, \mathrm{low}} \sim 1^{\circ} – 2^{\circ}$) and high ($\sigma_{i, \mathrm{high}} \sim 30^{\circ} – 65^{\circ}$) inclination components, and (2) the ``maximum AMD model'' \citep{2020AJ....160..276H}, in which the mutual inclination distribution is broad, continuous, and multiplicity-dependent with small inclinations on the scale of a few degrees. (See Figure \ref{fig: i vs e scatterplots} for a comparison of the two models.) To analyze the simulated planet population, we considered both analytic and $N$-body calculations of $\dot{T}_{\mathrm{dur}}$, along with a simulated TDV detection pipeline to identify which planets would have TDVs that are both measurable (requiring transits with sufficiently large signal-to-noise) and detectable (requiring a significant TDV slope, $|\dot{T}_{\mathrm{dur}}|> 3 \  \sigma_{\dot{T}_{\mathrm{dur}}}$).  

Our main result is shown in Figure \ref{fig: histograms with number of detected TDVs}, where the key diagnostic is the number of planets with detected TDVs. The maximum AMD model yields a quantity ($22^{+10}_{-6}$) that is in good agreement with the observed number from Kepler (16 after cuts have been applied). The two-Rayleigh model, by contrast, consistently overpredicts the number of planets with detected TDVs ($43^{+18}_{-13}$). This is because it has too many high inclination systems, which generally produce larger TDV signals (Figure \ref{fig: TDV slope vs inc}). These results are robust with respect to model assumptions, such as the upper limit of $R_p$. When restricting to planets with $R_p < 4 \ R_{\oplus}$ (rather than $R_p < 10 \ R_{\oplus}$), there are 6 observed Kepler planets with detected TDVs, compared to $8^{+4}_{-3}$ with the maximum AMD model and $16^{+7}_{-5}$ with the two-Rayleigh model (Figure \ref{fig: histograms with number of detected TDVs for 4 Rearth cut}).

Given these results, our key takeaway is that the TDV statistics support a continuous distribution of relatively low mutual inclinations ($i \lesssim 5^{\circ} – 10^{\circ}$) rather than a dichotomous distribution with many high inclination systems. These results are consistent with \cite{2018ApJ...860..101Z}, who considered TTV statistics rather than TDVs. Moreover, small inclinations are also supported by studies that combined Kepler and RV survey statistics \citep{2012AJ....143...94T, 2012A&A...541A.139F}. Cumulatively, this work is further evidence that the apparent excess of single-transiting systems relative to expectations from the multis, an observation known as the ``Kepler dichotomy'', does not actually provide evidence for a dichotomy in the underlying architectures \citep{2020AJ....160..276H}. Rather, the observations are naturally explained by and more consistent with a ``Kepler continuum'' of intrinsic multiplicities and low mutual inclinations. 

Even with predominantly small (few-degree) mutual inclinations, there must have been one or more physical processes that disrupted these close-in systems from perfect coplanarity. Primordial excitation during the planet assembly phase is perhaps the most favored mechanism for producing a non-dichotomous, multiplicity-dependent distribution of low mutual inclinations, although the optimal disk surface densities, gas damping timescales, and planetary embryo properties are still uncertain. Stellar oblateness can also drive small mutual inclinations if $\sim5^{\circ}-10^{\circ}$ spin-orbit misalignments (like the $6^{\circ}$ solar obliquity) are widespread and disk-dispersal timescales are rapid. Both of these are observable properties that will become better understood over time. Distant giant planets are almost certainly responsible for dynamical heating of a subset of inner planetary systems, but they are probably not the dominant origin of the transit-reducing mutual inclinations. Regardless, the results in this paper can be exploited in future work to constrain the prevalence of these various mechanisms.

Statistical characterization of TDV signals will improve in the future as observational time baselines lengthen. In particular, the PLATO Mission \citep{2014ExA....38..249R} will enable the detection of many more Kepler planet TDVs when it revisits the Kepler field. Moreover, some planets may be seen to transit into or out of view \citep{2018A&A...618A..41F}, as has already been observed in a handful of cases \citep{2019AJ....158..133H, 2020AJ....160..195J}. Modeling this expanded set of TDV observations will allow an even more detailed characterization of the mutual inclination distribution. 

In summary, TDVs provide a window into the three-dimensional properties of planetary systems, which are otherwise difficult to probe. This is another axis with which to study our Solar System in the context of exoplanetary systems, and, in this case, it appears that the near-coplanarity of the Solar System is indeed the rule.

\section{Acknowledgements}
We thank Tsevi Mazeh and Dimitri Veras for comments and questions that improved the paper. We also thank the anonymous referee for their insightful and helpful report. S.C.M. was supported by NASA through the NASA Hubble Fellowship grant \#HST-HF2-51465 awarded by the Space Telescope Science Institute, which is operated by the Association of Universities for Research in Astronomy, Inc., for NASA, under contract NAS5-26555. 
M.Y.H. acknowledges the support of the Natural Sciences and Engineering Research Council of Canada (NSERC), funding reference number PGSD3 - 516712 - 2018.
This work was supported by a grant from the Simons Foundation/SFARI (675601, E.B.F.).
E.B.F. acknowledges the support of the Ambrose Monell Foundation and the Institute for Advanced Study.
M.Y.H. and E.B.F. acknowledge support from the Penn State Eberly College of Science and Department of Astronomy \& Astrophysics, the Center for Exoplanets and Habitable Worlds, and the Center for Astrostatistics.  

\newpage
\appendix

\section{Fourth-order expansion of the secular disturbing function}
\label{sec: Appendix disturbing function expansion}

Here we provide the expression for the secular disturbing function expansion of two point-mass planets with masses $m$ and $m'$. The semi-major axes are $a$ and $a'$, with $a < a'$. The disturbing functions for the inner planet and outer planet, respectively, are given by
\begin{equation}
\label{eq: R and R'}
\begin{split}
\mathcal{R} &= \frac{\mu'}{a'}\mathcal{R}_{\mathrm{D}} + \frac{\mu'}{a'}\alpha\mathcal{R}_{\mathrm{E}} \\ 
\mathcal{R}' &= \frac{\mu}{a'}\mathcal{R}_{\mathrm{D}} + \frac{\mu}{a'}\frac{1}{\alpha^2}\mathcal{R}_{\mathrm{I}},
\end{split}
\end{equation}
where $\mu \equiv \mathcal{G}m$ and $\alpha \equiv a/a'$. $\mathcal{R}_{\mathrm{D}}$ is the direct part of the disturbing function, and $\mathcal{R}_{\mathrm{E}}$ and $\mathcal{R}_{\mathrm{I}}$ are the indirect parts. In our case, the indirect parts are zero because there are no secular terms. As for $\mathcal{R}_{\mathrm{D}}$, we retain secular terms up to fourth-order in $e$ and $s \equiv \sin(i/2)$. These terms can be found in the Appendix tables of \cite{1999ssd..book.....M}, yielding 
\begin{equation}
\begin{split}
\label{eq: Rd}
\mathcal{R}_D &= f_1 + f_2(e^2 + e'^2) + f_3(s^2 + s'^2) + f_4e^4 + f_5e^2e'^2 + f_6e'^4 + f_7(e^2s^2 + e'^2s^2 + e^2s'^2 + e'^2s'^2)  \\
&+ f_8(s^4 + s'^4) + f_9s^2s'^2 + [f_{10}e e' + f_{11}e^3e' + f_{12}e e'^3 + f_{13} e e'(s^2 + s'^2)]\cos(\varpi'-\varpi) \\
&+ [f_{14}s s' + f_{15}s s'(e^2 + e'^2) + f_{16} s s'(s^2 + s'^2)]\cos(\Omega' - \Omega) + f_{17}e^2 e'^2\cos(2\varpi' - 2\varpi) \\
&+ f_{18}e^2 s^2\cos(2\varpi - 2\Omega) + f_{19}e e's^2\cos(\varpi' + \varpi - 2\Omega) + f_{20}e'^2 s^2\cos(2\varpi'-2\Omega) \\
&+ f_{21}e^2 s s'\cos(2\varpi - \Omega' - \Omega) + f_{22}e e' s s'\cos(\varpi' - \varpi - \Omega' + \Omega) \\
&+ f_{23}e e' s s'\cos(\varpi' - \varpi + \Omega' - \Omega) + f_{24}e e' s s'\cos(\varpi' + \varpi - \Omega' - \Omega).
\end{split}
\end{equation}
The $f_i$ coefficients in this expression are functions of $\alpha$ that depend on the Laplace coefficients and their derivatives.

The disturbing function expansion in equation \ref{eq: Rd} is written explicitly for two planets. It is straightforward to generalize this treatment to $N$-planet systems. For each planet in the system, the disturbing functions associated with each pairwise planet-planet interaction are summed, using $\mathcal{R}$ when the perturbing planet is external and $\mathcal{R}'$ when the perturbing planet is internal.

\section{Comparison of analytic and $N$-body TDV calculation}
\label{sec: Appendix comparison of analytic TDV to N-body}

To assess the performance of the analytic calculation of $\dot{T}_{\mathrm{dur}}$, here we show how it compares to a computation using a direct $N$-body integration. The analytic approach was described in Section \ref{sec: analytic calculation}. The $N$-body calculation uses the 
\texttt{REBOUND} gravitational dynamics software \citep{2012A&A...537A.128R} and was described in Section \ref{sec: TDV calculation}. Rather than build a set of systems on which to test the calculations, we simply use the SysSim systems from both the maximum AMD model and the two-Rayleigh model.

\begin{figure}[h!]
\epsscale{0.6}
\plotone{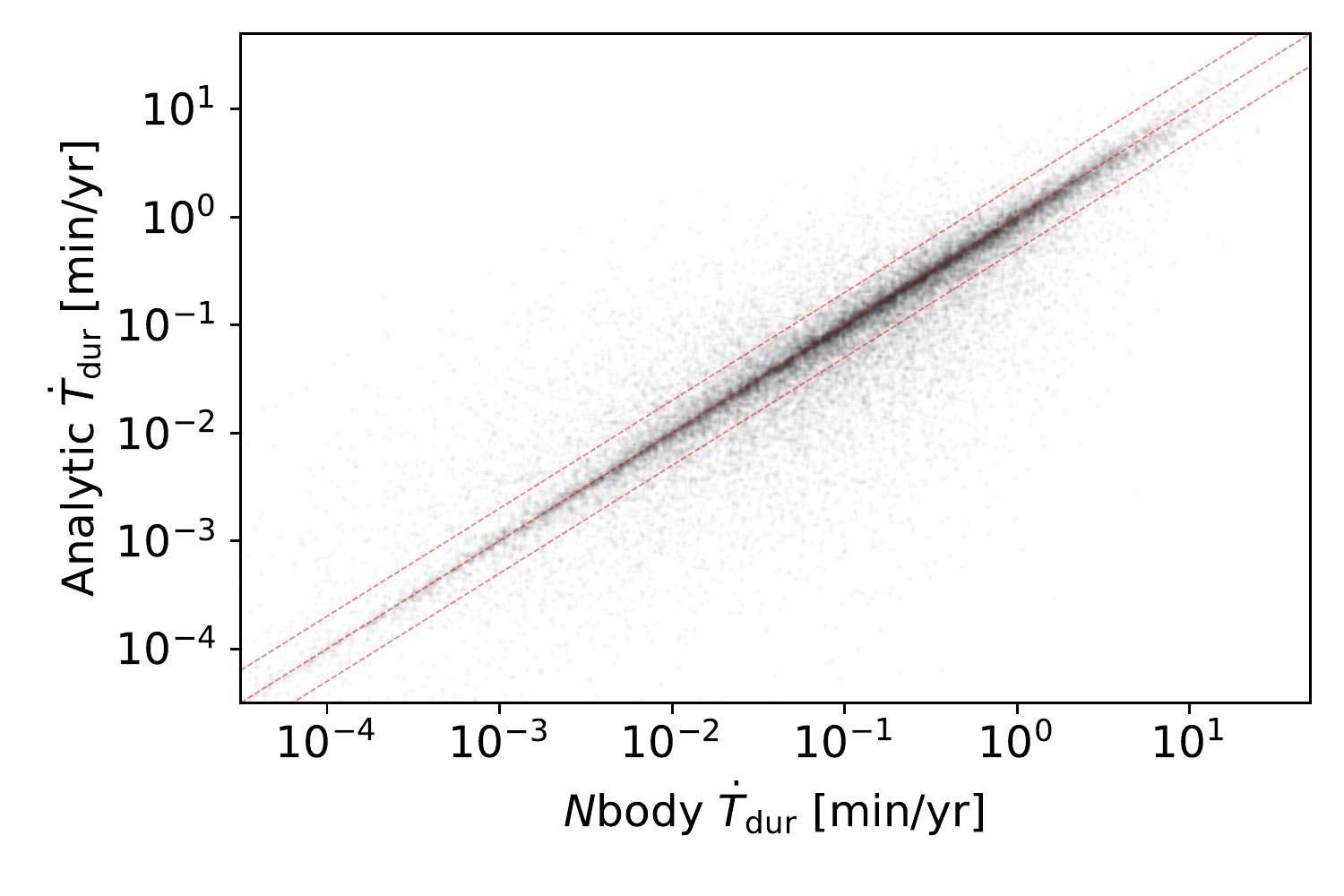}
\caption{Analytic vs. $N$-body calculation of $\dot{T}_{\mathrm{dur}}$. Each scatter point corresponds to a planet within a SysSim system. We consider 20 system catalogs from the maximum AMD model. The red dashed lines show the lines where analytic = $N$-body, analytic = $2\times N$-body, and analytic = $0.5\times N$-body.} 
\label{fig: AMD_model_TDV_analytic_vs_Nbody}
\end{figure}

\begin{figure}[h!]
\epsscale{0.6}
\plotone{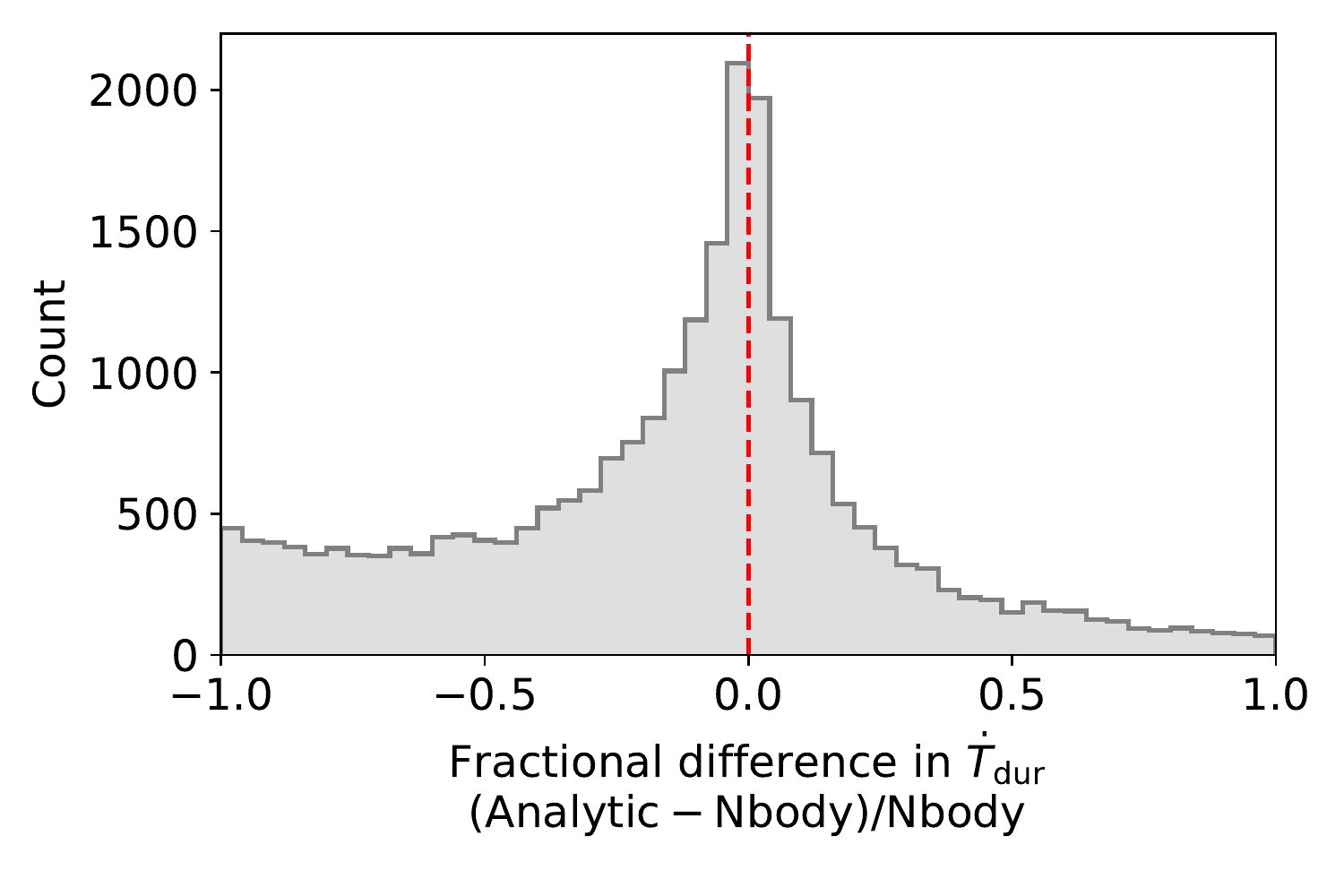}
\caption{Histogram of the fractional difference of the analytic vs. $N$-body calculation of $\dot{T}_{\mathrm{dur}}$. We consider planets in 20 system catalogs from the maximum AMD model (same as in Figure \ref{fig: AMD_model_TDV_analytic_vs_Nbody}).}
\label{fig: AMD_model_frac_diff_TDV_hist}
\end{figure}

\begin{figure}[h!]
\epsscale{0.6}
\plotone{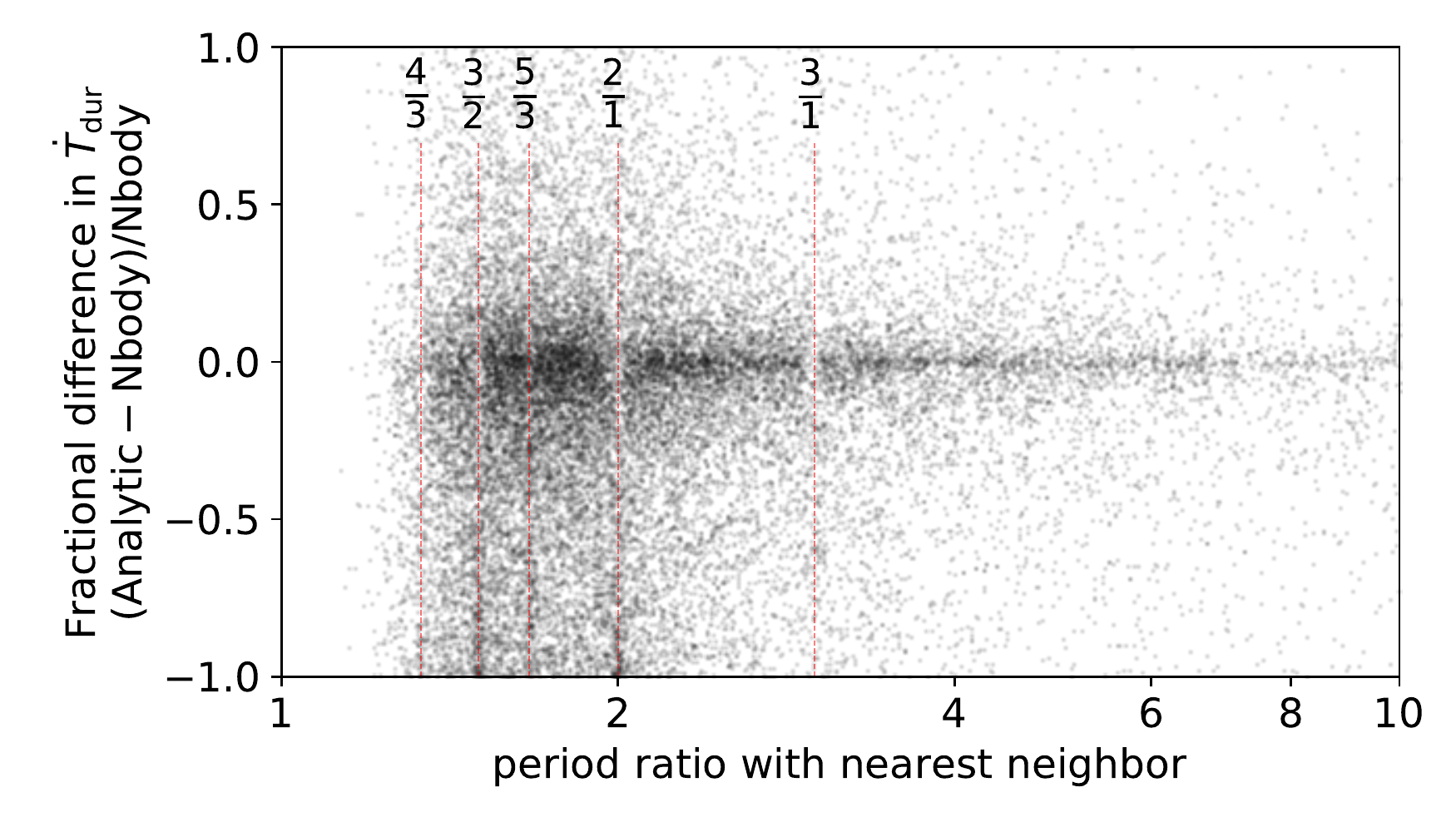}
\caption{Fractional difference of the analytic vs. $N$-body calculation of $\dot{T}_{\mathrm{dur}}$, plotted as a function of the period ratio between the planet and its nearest neighbor. We consider planets in 20 system catalogs from the maximum AMD model (same as in Figures \ref{fig: AMD_model_TDV_analytic_vs_Nbody} and \ref{fig: AMD_model_frac_diff_TDV_hist}). The locations of several low-order mean-motion resonances are shown with thin red lines.} 
\label{fig: AMD_model_frac_diff_TDV_vs_period_ratio}
\end{figure}

\begin{figure}[h!]
\epsscale{0.6}
\plotone{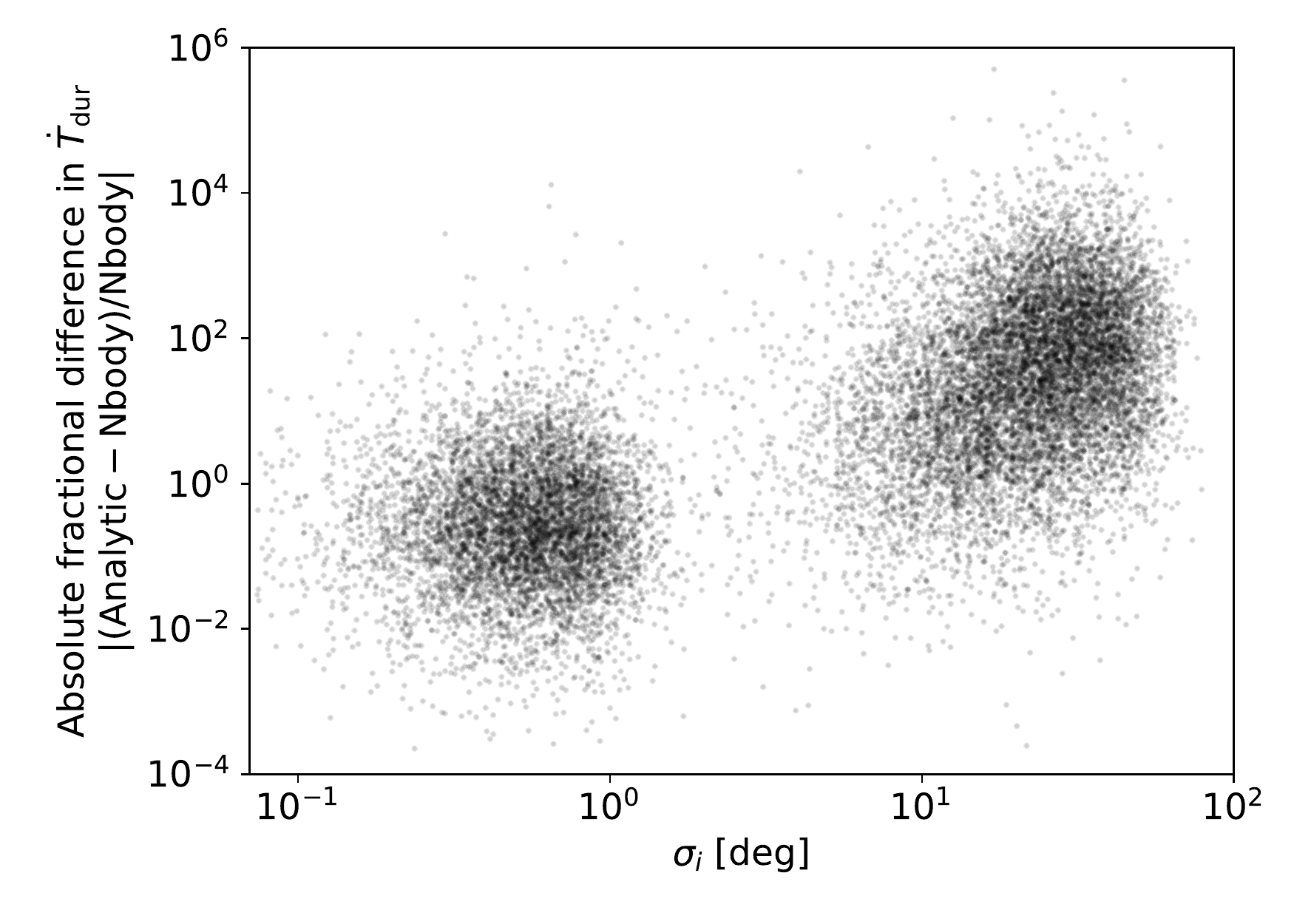}
\caption{Absolute fractional difference of the analytic vs. $N$-body calculation of $\dot{T}_{\mathrm{dur}}$, plotted as a function of the standard deviation of orbital inclinations within the system. We consider planets in 20 system catalogs from the two-Rayleigh model (different from Figures \ref{fig: AMD_model_TDV_analytic_vs_Nbody}, \ref{fig: AMD_model_frac_diff_TDV_hist}, and \ref{fig: AMD_model_frac_diff_TDV_vs_period_ratio}). } 
\label{fig: Two_Rayleigh_model_abs_frac_diff_TDV_vs_sigma_i}
\end{figure}

Figure \ref{fig: AMD_model_TDV_analytic_vs_Nbody} shows the analytic $\dot{T}_{\mathrm{dur}}$ versus the $N$-body $\dot{T}_{\mathrm{dur}}$ for a set of systems generated by the maximum AMD model, and Figure \ref{fig: AMD_model_frac_diff_TDV_hist} shows a histogram of the fractional difference. Overall, we observe a fairly strong agreement. The best half of the distribution has analytic $\dot{T}_{\mathrm{dur}}$ values within $25\%$ of the $N$-body values; the best three quarters of the distribution agree within $50\%$. Figure \ref{fig: AMD_model_frac_diff_TDV_vs_period_ratio} shows the fractional difference between the analytic and $N$-body values as a function of the period ratio between the planet for which $\dot{T}_{\mathrm{dur}}$ is calculated and its nearest neighbor. The analytic calculation is less accurate for smaller period ratios and for planets near mean-motion resonances. This is expected because the disturbing function expansion leaves out resonant terms. Figure \ref{fig: Two_Rayleigh_model_abs_frac_diff_TDV_vs_sigma_i} shows the comparisons for the two-Rayleigh model. There are orders-of-magnitude discrepancies between the analytic and $N$-body calculations at large inclinations, where the accuracy of the secular expansion breaks down.

\newpage

\bibliographystyle{aasjournal}
\bibliography{main}

\begin{thebibliography}{}
\expandafter\ifx\csname natexlab\endcsname\relax\def\natexlab#1{#1}\fi
\providecommand{\url}[1]{\href{#1}{#1}}
\providecommand{\dodoi}[1]{doi:~\href{http://doi.org/#1}{\nolinkurl{#1}}}
\providecommand{\doeprint}[1]{\href{http://ascl.net/#1}{\nolinkurl{http://ascl.net/#1}}}
\providecommand{\doarXiv}[1]{\href{https://arxiv.org/abs/#1}{\nolinkurl{https://arxiv.org/abs/#1}}}

\bibitem[{{Armitage}(2011)}]{2011ARA&A..49..195A}
{Armitage}, P.~J. 2011, \araa, 49, 195,
  \dodoi{10.1146/annurev-astro-081710-102521}

\bibitem[{{Ballard} \& {Johnson}(2016)}]{2016ApJ...816...66B}
{Ballard}, S., \& {Johnson}, J.~A. 2016, \apj, 816, 66,
  \dodoi{10.3847/0004-637X/816/2/66}

\bibitem[{{Batalha} {et~al.}(2013){Batalha}, {Rowe}, {Bryson}, {Barclay},
  {Burke}, {Caldwell}, {Christiansen}, {Mullally}, {Thompson}, {Brown},
  {Dupree}, {Fabrycky}, {Ford}, {Fortney}, {Gilliland}, {Isaacson}, {Latham},
  {Marcy}, {Quinn}, {Ragozzine}, {Shporer}, {Borucki}, {Ciardi}, {Gautier},
  {Haas}, {Jenkins}, {Koch}, {Lissauer}, {Rapin}, {Basri}, {Boss}, {Buchhave},
  {Carter}, {Charbonneau}, {Christensen-Dalsgaard}, {Clarke}, {Cochran},
  {Demory}, {Desert}, {Devore}, {Doyle}, {Esquerdo}, {Everett}, {Fressin},
  {Geary}, {Girouard}, {Gould}, {Hall}, {Holman}, {Howard}, {Howell},
  {Ibrahim}, {Kinemuchi}, {Kjeldsen}, {Klaus}, {Li}, {Lucas}, {Meibom},
  {Morris}, {Pr{\v{s}}a}, {Quintana}, {Sanderfer}, {Sasselov}, {Seader},
  {Smith}, {Steffen}, {Still}, {Stumpe}, {Tarter}, {Tenenbaum}, {Torres},
  {Twicken}, {Uddin}, {Van Cleve}, {Walkowicz}, \&
  {Welsh}}]{2013ApJS..204...24B}
{Batalha}, N.~M., {Rowe}, J.~F., {Bryson}, S.~T., {et~al.} 2013, \apjs, 204,
  24, \dodoi{10.1088/0067-0049/204/2/24}

\bibitem[{{Batygin}(2012)}]{2012Natur.491..418B}
{Batygin}, K. 2012, \nat, 491, 418, \dodoi{10.1038/nature11560}

\bibitem[{{Becker} \& {Adams}(2017)}]{2017MNRAS.468..549B}
{Becker}, J.~C., \& {Adams}, F.~C. 2017, \mnras, 468, 549,
  \dodoi{10.1093/mnras/stx461}

\bibitem[{{Berger} {et~al.}(2020){Berger}, {Huber}, {Gaidos}, {van Saders}, \&
  {Weiss}}]{2020AJ....160..108B}
{Berger}, T.~A., {Huber}, D., {Gaidos}, E., {van Saders}, J.~L., \& {Weiss},
  L.~M. 2020, \aj, 160, 108, \dodoi{10.3847/1538-3881/aba18a}

\bibitem[{{Bezanson} {et~al.}(2014){Bezanson}, {Edelman}, {Karpinski}, \&
  {Shah}}]{2014arXiv1411.1607B}
{Bezanson}, J., {Edelman}, A., {Karpinski}, S., \& {Shah}, V.~B. 2014, arXiv
  e-prints, arXiv:1411.1607.
\newblock \doarXiv{1411.1607}

\bibitem[{{Boley} {et~al.}(2020){Boley}, {Van Laerhoven}, \& {Granados
  Contreras}}]{2020AJ....159..207B}
{Boley}, A.~C., {Van Laerhoven}, C., \& {Granados Contreras}, A.~P. 2020, \aj,
  159, 207, \dodoi{10.3847/1538-3881/ab8067}

\bibitem[{{Borsato} {et~al.}(2019){Borsato}, {Malavolta}, {Piotto}, {Buchhave},
  {Mortier}, {Rice}, {Collier Cameron}, {Coffinet}, {Sozzetti}, {Charbonneau},
  {Cosentino}, {Dumusque}, {Figueira}, {Latham}, {Lopez-Morales}, {Mayor},
  {Micela}, {Molinari}, {Pepe}, {Phillips}, {Poretti}, {Udry}, \&
  {Watson}}]{2019MNRAS.484.3233B}
{Borsato}, L., {Malavolta}, L., {Piotto}, G., {et~al.} 2019, \mnras, 484, 3233,
  \dodoi{10.1093/mnras/stz181}

\bibitem[{{Borucki} {et~al.}(2010){Borucki}, {Koch}, {Basri}, {Batalha},
  {Brown}, {Caldwell}, {Caldwell}, {Christensen-Dalsgaard}, {Cochran},
  {DeVore}, {Dunham}, {Dupree}, {Gautier}, {Geary}, {Gilliland}, {Gould},
  {Howell}, {Jenkins}, {Kondo}, {Latham}, {Marcy}, {Meibom}, {Kjeldsen},
  {Lissauer}, {Monet}, {Morrison}, {Sasselov}, {Tarter}, {Boss}, {Brownlee},
  {Owen}, {Buzasi}, {Charbonneau}, {Doyle}, {Fortney}, {Ford}, {Holman},
  {Seager}, {Steffen}, {Welsh}, {Rowe}, {Anderson}, {Buchhave}, {Ciardi},
  {Walkowicz}, {Sherry}, {Horch}, {Isaacson}, {Everett}, {Fischer}, {Torres},
  {Johnson}, {Endl}, {MacQueen}, {Bryson}, {Dotson}, {Haas}, {Kolodziejczak},
  {Van Cleve}, {Chandrasekaran}, {Twicken}, {Quintana}, {Clarke}, {Allen},
  {Li}, {Wu}, {Tenenbaum}, {Verner}, {Bruhweiler}, {Barnes}, \&
  {Prsa}}]{2010Sci...327..977B}
{Borucki}, W.~J., {Koch}, D., {Basri}, G., {et~al.} 2010, Science, 327, 977,
  \dodoi{10.1126/science.1185402}

\bibitem[{{Bryan} {et~al.}(2019){Bryan}, {Knutson}, {Lee}, {Fulton}, {Batygin},
  {Ngo}, \& {Meshkat}}]{2019AJ....157...52B}
{Bryan}, M.~L., {Knutson}, H.~A., {Lee}, E.~J., {et~al.} 2019, \aj, 157, 52,
  \dodoi{10.3847/1538-3881/aaf57f}

\bibitem[{{Carrera} {et~al.}(2019){Carrera}, {Ford}, \&
  {Izidoro}}]{2019MNRAS.486.3874C}
{Carrera}, D., {Ford}, E.~B., \& {Izidoro}, A. 2019, \mnras, 486, 3874,
  \dodoi{10.1093/mnras/stz974}

\bibitem[{{Carrera} {et~al.}(2018){Carrera}, {Ford}, {Izidoro},
  {Jontof-Hutter}, {Raymond}, \& {Wolfgang}}]{2018ApJ...866..104C}
{Carrera}, D., {Ford}, E.~B., {Izidoro}, A., {et~al.} 2018, \apj, 866, 104,
  \dodoi{10.3847/1538-4357/aadf8a}

\bibitem[{{Coleman} \& {Nelson}(2014)}]{2014MNRAS.445..479C}
{Coleman}, G. A.~L., \& {Nelson}, R.~P. 2014, \mnras, 445, 479,
  \dodoi{10.1093/mnras/stu1715}

\bibitem[{{Cossou} {et~al.}(2014){Cossou}, {Raymond}, {Hersant}, \&
  {Pierens}}]{2014A&A...569A..56C}
{Cossou}, C., {Raymond}, S.~N., {Hersant}, F., \& {Pierens}, A. 2014, \aap,
  569, A56, \dodoi{10.1051/0004-6361/201424157}

\bibitem[{{Cranmer} {et~al.}(2021){Cranmer}, {Tamayo}, {Rein}, {Battaglia},
  {Hadden}, {Armitage}, {Ho}, \& {Spergel}}]{2021arXiv210104117C}
{Cranmer}, M., {Tamayo}, D., {Rein}, H., {et~al.} 2021, arXiv e-prints,
  arXiv:2101.04117.
\newblock \doarXiv{2101.04117}

\bibitem[{{Dai} {et~al.}(2018){Dai}, {Masuda}, \& {Winn}}]{2018ApJ...864L..38D}
{Dai}, F., {Masuda}, K., \& {Winn}, J.~N. 2018, \apjl, 864, L38,
  \dodoi{10.3847/2041-8213/aadd4f}

\bibitem[{{Dawson}(2020)}]{2020AJ....159..223D}
{Dawson}, R.~I. 2020, \aj, 159, 223, \dodoi{10.3847/1538-3881/ab7fa5}

\bibitem[{{Dawson} {et~al.}(2016){Dawson}, {Lee}, \&
  {Chiang}}]{2016ApJ...822...54D}
{Dawson}, R.~I., {Lee}, E.~J., \& {Chiang}, E. 2016, \apj, 822, 54,
  \dodoi{10.3847/0004-637X/822/1/54}

\bibitem[{{Fabrycky} {et~al.}(2014){Fabrycky}, {Lissauer}, {Ragozzine}, {Rowe},
  {Steffen}, {Agol}, {Barclay}, {Batalha}, {Borucki}, {Ciardi}, {Ford},
  {Gautier}, {Geary}, {Holman}, {Jenkins}, {Li}, {Morehead}, {Morris},
  {Shporer}, {Smith}, {Still}, \& {Van Cleve}}]{2014ApJ...790..146F}
{Fabrycky}, D.~C., {Lissauer}, J.~J., {Ragozzine}, D., {et~al.} 2014, \apj,
  790, 146, \dodoi{10.1088/0004-637X/790/2/146}

\bibitem[{{Fang} \& {Margot}(2012)}]{2012ApJ...761...92F}
{Fang}, J., \& {Margot}, J.-L. 2012, \apj, 761, 92,
  \dodoi{10.1088/0004-637X/761/2/92}

\bibitem[{{Figueira} {et~al.}(2012){Figueira}, {Marmier}, {Bou{\'e}}, {Lovis},
  {Santos}, {Montalto}, {Udry}, {Pepe}, \& {Mayor}}]{2012A&A...541A.139F}
{Figueira}, P., {Marmier}, M., {Bou{\'e}}, G., {et~al.} 2012, \aap, 541, A139,
  \dodoi{10.1051/0004-6361/201219017}

\bibitem[{{Ford} {et~al.}(2011){Ford}, {Rowe}, {Fabrycky}, {Carter}, {Holman},
  {Lissauer}, {Ragozzine}, {Steffen}, {Batalha}, {Borucki}, {Bryson},
  {Caldwell}, {Dunham}, {Gautier}, {Jenkins}, {Koch}, {Li}, {Lucas}, {Marcy},
  {McCauliff}, {Mullally}, {Quintana}, {Still}, {Tenenbaum}, {Thompson}, \&
  {Twicken}}]{2011ApJS..197....2F}
{Ford}, E.~B., {Rowe}, J.~F., {Fabrycky}, D.~C., {et~al.} 2011, \apjs, 197, 2,
  \dodoi{10.1088/0067-0049/197/1/2}

\bibitem[{{Freudenthal} {et~al.}(2018){Freudenthal}, {von Essen}, {Dreizler},
  {Wedemeyer}, {Agol}, {Morris}, {Becker}, {Mallonn}, {Hoyer}, {Ofir},
  {Tal-Or}, {Deeg}, {Herrero}, {Ribas}, {Khalafinejad}, {Hern{\'a}ndez}, \&
  {Rodr{\'\i}guez S.}}]{2018A&A...618A..41F}
{Freudenthal}, J., {von Essen}, C., {Dreizler}, S., {et~al.} 2018, \aap, 618,
  A41, \dodoi{10.1051/0004-6361/201833436}

\bibitem[{{Granados Contreras} \& {Boley}(2018)}]{2018AJ....155..139G}
{Granados Contreras}, A.~P., \& {Boley}, A.~C. 2018, \aj, 155, 139,
  \dodoi{10.3847/1538-3881/aaac82}

\bibitem[{{Gratia} \& {Fabrycky}(2017)}]{2017MNRAS.464.1709G}
{Gratia}, P., \& {Fabrycky}, D. 2017, \mnras, 464, 1709,
  \dodoi{10.1093/mnras/stw2180}

\bibitem[{{Hamann} {et~al.}(2019){Hamann}, {Montet}, {Fabrycky}, {Agol}, \&
  {Kruse}}]{2019AJ....158..133H}
{Hamann}, A., {Montet}, B.~T., {Fabrycky}, D.~C., {Agol}, E., \& {Kruse}, E.
  2019, \aj, 158, 133, \dodoi{10.3847/1538-3881/ab32e3}

\bibitem[{{Hansen}(2017)}]{2017MNRAS.467.1531H}
{Hansen}, B. M.~S. 2017, \mnras, 467, 1531, \dodoi{10.1093/mnras/stx182}

\bibitem[{{Hansen} \& {Murray}(2013)}]{2013ApJ...775...53H}
{Hansen}, B. M.~S., \& {Murray}, N. 2013, \apj, 775, 53,
  \dodoi{10.1088/0004-637X/775/1/53}

\bibitem[{{He} {et~al.}(2019){He}, {Ford}, \&
  {Ragozzine}}]{2019MNRAS.490.4575H}
{He}, M.~Y., {Ford}, E.~B., \& {Ragozzine}, D. 2019, \mnras, 490, 4575,
  \dodoi{10.1093/mnras/stz2869}

\bibitem[{{He} {et~al.}(2021){He}, {Ford}, \&
  {Ragozzine}}]{2021AJ....161...16H}
---. 2021, \aj, 161, 16, \dodoi{10.3847/1538-3881/abc68b}

\bibitem[{{He} {et~al.}(2020){He}, {Ford}, {Ragozzine}, \&
  {Carrera}}]{2020AJ....160..276H}
{He}, M.~Y., {Ford}, E.~B., {Ragozzine}, D., \& {Carrera}, D. 2020, \aj, 160,
  276, \dodoi{10.3847/1538-3881/abba18}

\bibitem[{{Holczer} {et~al.}(2016){Holczer}, {Mazeh}, {Nachmani},
  {Jontof-Hutter}, {Ford}, {Fabrycky}, {Ragozzine}, {Kane}, \&
  {Steffen}}]{2016ApJS..225....9H}
{Holczer}, T., {Mazeh}, T., {Nachmani}, G., {et~al.} 2016, \apjs, 225, 9,
  \dodoi{10.3847/0067-0049/225/1/9}

\bibitem[{{Hsu} {et~al.}(2019){Hsu}, {Ford}, {Ragozzine}, \&
  {Ashby}}]{2019AJ....158..109H}
{Hsu}, D.~C., {Ford}, E.~B., {Ragozzine}, D., \& {Ashby}, K. 2019, \aj, 158,
  109, \dodoi{10.3847/1538-3881/ab31ab}

\bibitem[{{Hsu} {et~al.}(2018){Hsu}, {Ford}, {Ragozzine}, \&
  {Morehead}}]{2018AJ....155..205H}
{Hsu}, D.~C., {Ford}, E.~B., {Ragozzine}, D., \& {Morehead}, R.~C. 2018, \aj,
  155, 205, \dodoi{10.3847/1538-3881/aab9a8}

\bibitem[{{Izidoro} {et~al.}(2019){Izidoro}, {Bitsch}, {Raymond}, {Johansen},
  {Morbidelli}, {Lambrechts}, \& {Jacobson}}]{2019arXiv190208772I}
{Izidoro}, A., {Bitsch}, B., {Raymond}, S.~N., {et~al.} 2019, arXiv e-prints,
  arXiv:1902.08772.
\newblock \doarXiv{1902.08772}

\bibitem[{{Izidoro} {et~al.}(2017){Izidoro}, {Ogihara}, {Raymond},
  {Morbidelli}, {Pierens}, {Bitsch}, {Cossou}, \&
  {Hersant}}]{2017MNRAS.470.1750I}
{Izidoro}, A., {Ogihara}, M., {Raymond}, S.~N., {et~al.} 2017, \mnras, 470,
  1750, \dodoi{10.1093/mnras/stx1232}

\bibitem[{{Johansen} {et~al.}(2012){Johansen}, {Davies}, {Church}, \&
  {Holmelin}}]{2012ApJ...758...39J}
{Johansen}, A., {Davies}, M.~B., {Church}, R.~P., \& {Holmelin}, V. 2012, \apj,
  758, 39, \dodoi{10.1088/0004-637X/758/1/39}

\bibitem[{{Judkovsky} {et~al.}(2020){Judkovsky}, {Ofir}, \&
  {Aharonson}}]{2020AJ....160..195J}
{Judkovsky}, Y., {Ofir}, A., \& {Aharonson}, O. 2020, \aj, 160, 195,
  \dodoi{10.3847/1538-3881/abb406}

\bibitem[{{Kane} {et~al.}(2019){Kane}, {Ragozzine}, {Flowers}, {Holczer},
  {Mazeh}, \& {Relles}}]{2019AJ....157..171K}
{Kane}, M., {Ragozzine}, D., {Flowers}, X., {et~al.} 2019, \aj, 157, 171,
  \dodoi{10.3847/1538-3881/ab0d91}

\bibitem[{Kant(1755)}]{Kant1755}
Kant, I. 1755, General History of Nature and Theory of the Heavens
  ((K\"onigsberg: Petersen))

\bibitem[{{Kunovac Hod{\v{z}}i{\'c}} {et~al.}(2021){Kunovac Hod{\v{z}}i{\'c}},
  {Triaud}, {Cegla}, {Chaplin}, \& {Davies}}]{2021MNRAS.502.2893K}
{Kunovac Hod{\v{z}}i{\'c}}, V., {Triaud}, A. H.~M.~J., {Cegla}, H.~M.,
  {Chaplin}, W.~J., \& {Davies}, G.~R. 2021, \mnras, 502, 2893,
  \dodoi{10.1093/mnras/stab237}

\bibitem[{{Lai}(2014)}]{2014MNRAS.440.3532L}
{Lai}, D. 2014, \mnras, 440, 3532, \dodoi{10.1093/mnras/stu485}

\bibitem[{{Lai} \& {Pu}(2017)}]{2017AJ....153...42L}
{Lai}, D., \& {Pu}, B. 2017, \aj, 153, 42, \dodoi{10.3847/1538-3881/153/1/42}

\bibitem[{Laplace(1796)}]{Laplace1796}
Laplace, P.~S. 1796, Exposition dy Syst\`eme du Monde (Paris: Cerie-Social)
  ((Paris: Cerie-Social))

\bibitem[{{Laskar} \& {Petit}(2017)}]{2017A&A...605A..72L}
{Laskar}, J., \& {Petit}, A.~C. 2017, \aap, 605, A72,
  \dodoi{10.1051/0004-6361/201630022}

\bibitem[{{Laughlin} {et~al.}(2002){Laughlin}, {Chambers}, \&
  {Fischer}}]{2002ApJ...579..455L}
{Laughlin}, G., {Chambers}, J., \& {Fischer}, D. 2002, \apj, 579, 455,
  \dodoi{10.1086/342746}

\bibitem[{{Li} {et~al.}(2020){Li}, {Dai}, \& {Becker}}]{2020ApJ...890L..31L}
{Li}, G., {Dai}, F., \& {Becker}, J. 2020, \apjl, 890, L31,
  \dodoi{10.3847/2041-8213/ab72f4}

\bibitem[{{Lissauer} {et~al.}(2011){Lissauer}, {Ragozzine}, {Fabrycky},
  {Steffen}, {Ford}, {Jenkins}, {Shporer}, {Holman}, {Rowe}, {Quintana},
  {Batalha}, {Borucki}, {Bryson}, {Caldwell}, {Carter}, {Ciardi}, {Dunham},
  {Fortney}, {Gautier}, {Howell}, {Koch}, {Latham}, {Marcy}, {Morehead}, \&
  {Sasselov}}]{2011ApJS..197....8L}
{Lissauer}, J.~J., {Ragozzine}, D., {Fabrycky}, D.~C., {et~al.} 2011, \apjs,
  197, 8, \dodoi{10.1088/0067-0049/197/1/8}

\bibitem[{{Louden} {et~al.}(2021){Louden}, {Winn}, {Petigura}, {Isaacson},
  {Howard}, {Masuda}, {Albrecht}, \& {Kosiarek}}]{2021AJ....161...68L}
{Louden}, E.~M., {Winn}, J.~N., {Petigura}, E.~A., {et~al.} 2021, \aj, 161, 68,
  \dodoi{10.3847/1538-3881/abcebd}

\bibitem[{{Masuda}(2017)}]{2017AJ....154...64M}
{Masuda}, K. 2017, \aj, 154, 64, \dodoi{10.3847/1538-3881/aa7aeb}

\bibitem[{{Masuda} {et~al.}(2020){Masuda}, {Winn}, \&
  {Kawahara}}]{2020AJ....159...38M}
{Masuda}, K., {Winn}, J.~N., \& {Kawahara}, H. 2020, \aj, 159, 38,
  \dodoi{10.3847/1538-3881/ab5c1d}

\bibitem[{{Matsumoto} \& {Kokubo}(2017)}]{2017AJ....154...27M}
{Matsumoto}, Y., \& {Kokubo}, E. 2017, \aj, 154, 27,
  \dodoi{10.3847/1538-3881/aa74c7}

\bibitem[{{Mazeh} {et~al.}(2015){Mazeh}, {Perets}, {McQuillan}, \&
  {Goldstein}}]{2015ApJ...801....3M}
{Mazeh}, T., {Perets}, H.~B., {McQuillan}, A., \& {Goldstein}, E.~S. 2015,
  \apj, 801, 3, \dodoi{10.1088/0004-637X/801/1/3}

\bibitem[{{Millholland} \& {Laughlin}(2019)}]{2019NatAs...3..424M}
{Millholland}, S., \& {Laughlin}, G. 2019, Nature Astronomy, 3, 424,
  \dodoi{10.1038/s41550-019-0701-7}

\bibitem[{{Millholland} {et~al.}(2018){Millholland}, {Laughlin}, {Teske},
  {Butler}, {Burt}, {Holden}, {Vogt}, {Crane}, {Shectman}, \&
  {Thompson}}]{2018AJ....155..106M}
{Millholland}, S., {Laughlin}, G., {Teske}, J., {et~al.} 2018, \aj, 155, 106,
  \dodoi{10.3847/1538-3881/aaa894}

\bibitem[{{Millholland} \& {Spalding}(2020)}]{2020ApJ...905...71M}
{Millholland}, S.~C., \& {Spalding}, C. 2020, \apj, 905, 71,
  \dodoi{10.3847/1538-4357/abc4e5}

\bibitem[{{Mills} \& {Fabrycky}(2017)}]{2017AJ....153...45M}
{Mills}, S.~M., \& {Fabrycky}, D.~C. 2017, \aj, 153, 45,
  \dodoi{10.3847/1538-3881/153/1/45}

\bibitem[{{Miralda-Escud{\'e}}(2002)}]{2002ApJ...564.1019M}
{Miralda-Escud{\'e}}, J. 2002, \apj, 564, 1019, \dodoi{10.1086/324279}

\bibitem[{{Moriarty} \& {Ballard}(2016)}]{2016ApJ...832...34M}
{Moriarty}, J., \& {Ballard}, S. 2016, \apj, 832, 34,
  \dodoi{10.3847/0004-637X/832/1/34}

\bibitem[{{Mulders} {et~al.}(2018){Mulders}, {Pascucci}, {Apai}, \&
  {Ciesla}}]{2018AJ....156...24M}
{Mulders}, G.~D., {Pascucci}, I., {Apai}, D., \& {Ciesla}, F.~J. 2018, \aj,
  156, 24, \dodoi{10.3847/1538-3881/aac5ea}

\bibitem[{{Munoz Romero} \& {Kempton}(2018)}]{2018AJ....155..134M}
{Munoz Romero}, C.~E., \& {Kempton}, E. M.~R. 2018, \aj, 155, 134,
  \dodoi{10.3847/1538-3881/aaab5e}

\bibitem[{{Murray} \& {Dermott}(1999)}]{1999ssd..book.....M}
{Murray}, C.~D., \& {Dermott}, S.~F. 1999, {Solar system dynamics}

\bibitem[{{Mustill} {et~al.}(2017){Mustill}, {Davies}, \&
  {Johansen}}]{2017MNRAS.468.3000M}
{Mustill}, A.~J., {Davies}, M.~B., \& {Johansen}, A. 2017, \mnras, 468, 3000,
  \dodoi{10.1093/mnras/stx693}

\bibitem[{{Nelson} {et~al.}(2014){Nelson}, {Ford}, {Wright}, {Fischer}, {von
  Braun}, {Howard}, {Payne}, \& {Dindar}}]{2014MNRAS.441..442N}
{Nelson}, B.~E., {Ford}, E.~B., {Wright}, J.~T., {et~al.} 2014, \mnras, 441,
  442, \dodoi{10.1093/mnras/stu450}

\bibitem[{{Nelson} {et~al.}(2016){Nelson}, {Robertson}, {Payne}, {Pritchard},
  {Deck}, {Ford}, {Wright}, \& {Isaacson}}]{2016MNRAS.455.2484N}
{Nelson}, B.~E., {Robertson}, P.~M., {Payne}, M.~J., {et~al.} 2016, \mnras,
  455, 2484, \dodoi{10.1093/mnras/stv2367}

\bibitem[{{Petit} {et~al.}(2017){Petit}, {Laskar}, \&
  {Bou{\'e}}}]{2017A&A...607A..35P}
{Petit}, A.~C., {Laskar}, J., \& {Bou{\'e}}, G. 2017, \aap, 607, A35,
  \dodoi{10.1051/0004-6361/201731196}

\bibitem[{{Piaulet} {et~al.}(2021){Piaulet}, {Benneke}, {Rubenzahl}, {Howard},
  {Lee}, {Thorngren}, {Angus}, {Peterson}, {Schlieder}, {Werner}, {Kreidberg},
  {Jaouni}, {Crossfield}, {Ciardi}, {Petigura}, {Livingston}, {Dressing},
  {Fulton}, {Beichman}, {Christiansen}, {Gorjian}, {Hardegree-Ullman}, {Krick},
  \& {Sinukoff}}]{2021AJ....161...70P}
{Piaulet}, C., {Benneke}, B., {Rubenzahl}, R.~A., {et~al.} 2021, \aj, 161, 70,
  \dodoi{10.3847/1538-3881/abcd3c}

\bibitem[{{Poon} {et~al.}(2020){Poon}, {Nelson}, {Jacobson}, \&
  {Morbidelli}}]{2020MNRAS.491.5595P}
{Poon}, S. T.~S., {Nelson}, R.~P., {Jacobson}, S.~A., \& {Morbidelli}, A. 2020,
  \mnras, 491, 5595, \dodoi{10.1093/mnras/stz3296}

\bibitem[{{Pu} \& {Lai}(2018)}]{2018MNRAS.478..197P}
{Pu}, B., \& {Lai}, D. 2018, \mnras, 478, 197, \dodoi{10.1093/mnras/sty1098}

\bibitem[{{Rauer} {et~al.}(2014){Rauer}, {Catala}, {Aerts}, {Appourchaux},
  {Benz}, {Brandeker}, {Christensen-Dalsgaard}, {Deleuil}, {Gizon}, {Goupil},
  {G{\"u}del}, {Janot-Pacheco}, {Mas-Hesse}, {Pagano}, {Piotto}, {Pollacco},
  {Santos}, {Smith}, {Su{\'a}rez}, {Szab{\'o}}, {Udry}, {Adibekyan}, {Alibert},
  {Almenara}, {Amaro-Seoane}, {Eiff}, {Asplund}, {Antonello}, {Barnes},
  {Baudin}, {Belkacem}, {Bergemann}, {Bihain}, {Birch}, {Bonfils}, {Boisse},
  {Bonomo}, {Borsa}, {Brand{\~a}o}, {Brocato}, {Brun}, {Burleigh}, {Burston},
  {Cabrera}, {Cassisi}, {Chaplin}, {Charpinet}, {Chiappini}, {Church},
  {Csizmadia}, {Cunha}, {Damasso}, {Davies}, {Deeg}, {D{\'\i}az}, {Dreizler},
  {Dreyer}, {Eggenberger}, {Ehrenreich}, {Eigm{\"u}ller}, {Erikson}, {Farmer},
  {Feltzing}, {de Oliveira Fialho}, {Figueira}, {Forveille}, {Fridlund},
  {Garc{\'\i}a}, {Giommi}, {Giuffrida}, {Godolt}, {Gomes da Silva}, {Granzer},
  {Grenfell}, {Grotsch-Noels}, {G{\"u}nther}, {Haswell}, {Hatzes},
  {H{\'e}brard}, {Hekker}, {Helled}, {Heng}, {Jenkins}, {Johansen},
  {Khodachenko}, {Kislyakova}, {Kley}, {Kolb}, {Krivova}, {Kupka}, {Lammer},
  {Lanza}, {Lebreton}, {Magrin}, {Marcos-Arenal}, {Marrese}, {Marques},
  {Martins}, {Mathis}, {Mathur}, {Messina}, {Miglio}, {Montalban}, {Montalto},
  {Monteiro}, {Moradi}, {Moravveji}, {Mordasini}, {Morel}, {Mortier},
  {Nascimbeni}, {Nelson}, {Nielsen}, {Noack}, {Norton}, {Ofir}, {Oshagh},
  {Ouazzani}, {P{\'a}pics}, {Parro}, {Petit}, {Plez}, {Poretti}, {Quirrenbach},
  {Ragazzoni}, {Raimondo}, {Rainer}, {Reese}, {Redmer}, {Reffert},
  {Rojas-Ayala}, {Roxburgh}, {Salmon}, {Santerne}, {Schneider}, {Schou},
  {Schuh}, {Schunker}, {Silva-Valio}, {Silvotti}, {Skillen}, {Snellen}, {Sohl},
  {Sousa}, {Sozzetti}, {Stello}, {Strassmeier}, {{\v{S}}vanda}, {Szab{\'o}},
  {Tkachenko}, {Valencia}, {Van Grootel}, {Vauclair}, {Ventura}, {Wagner},
  {Walton}, {Weingrill}, {Werner}, {Wheatley}, \&
  {Zwintz}}]{2014ExA....38..249R}
{Rauer}, H., {Catala}, C., {Aerts}, C., {et~al.} 2014, Experimental Astronomy,
  38, 249, \dodoi{10.1007/s10686-014-9383-4}

\bibitem[{{Rein} \& {Liu}(2012)}]{2012A&A...537A.128R}
{Rein}, H., \& {Liu}, S.~F. 2012, \aap, 537, A128,
  \dodoi{10.1051/0004-6361/201118085}

\bibitem[{{Rein} \& {Tamayo}(2015)}]{2015MNRAS.452..376R}
{Rein}, H., \& {Tamayo}, D. 2015, \mnras, 452, 376,
  \dodoi{10.1093/mnras/stv1257}

\bibitem[{{Sanchis-Ojeda} {et~al.}(2012){Sanchis-Ojeda}, {Fabrycky}, {Winn},
  {Barclay}, {Clarke}, {Ford}, {Fortney}, {Geary}, {Holman}, {Howard},
  {Jenkins}, {Koch}, {Lissauer}, {Marcy}, {Mullally}, {Ragozzine}, {Seader},
  {Still}, \& {Thompson}}]{2012Natur.487..449S}
{Sanchis-Ojeda}, R., {Fabrycky}, D.~C., {Winn}, J.~N., {et~al.} 2012, \nat,
  487, 449, \dodoi{10.1038/nature11301}

\bibitem[{{Sandford} {et~al.}(2019){Sandford}, {Kipping}, \&
  {Collins}}]{2019MNRAS.489.3162S}
{Sandford}, E., {Kipping}, D., \& {Collins}, M. 2019, \mnras, 489, 3162,
  \dodoi{10.1093/mnras/stz2350}

\bibitem[{{Shahaf} {et~al.}(2021){Shahaf}, {Mazeh}, {Zucker}, \&
  {Fabrycky}}]{2021MNRAS.tmp.1312S}
{Shahaf}, S., {Mazeh}, T., {Zucker}, S., \& {Fabrycky}, D. 2021, \mnras,
  \dodoi{10.1093/mnras/stab1359}

\bibitem[{{Spalding} \& {Batygin}(2014)}]{2014ApJ...790...42S}
{Spalding}, C., \& {Batygin}, K. 2014, \apj, 790, 42,
  \dodoi{10.1088/0004-637X/790/1/42}

\bibitem[{{Spalding} \& {Batygin}(2016)}]{2016ApJ...830....5S}
---. 2016, \apj, 830, 5, \dodoi{10.3847/0004-637X/830/1/5}

\bibitem[{{Spalding} {et~al.}(2014){Spalding}, {Batygin}, \&
  {Adams}}]{2014ApJ...797L..29S}
{Spalding}, C., {Batygin}, K., \& {Adams}, F.~C. 2014, \apjl, 797, L29,
  \dodoi{10.1088/2041-8205/797/2/L29}

\bibitem[{{Spalding} \& {Millholland}(2020)}]{2020AJ....160..105S}
{Spalding}, C., \& {Millholland}, S.~C. 2020, \aj, 160, 105,
  \dodoi{10.3847/1538-3881/aba629}

\bibitem[{{Steffen} {et~al.}(2010){Steffen}, {Batalha}, {Borucki}, {Buchhave},
  {Caldwell}, {Cochran}, {Endl}, {Fabrycky}, {Fressin}, {Ford}, {Fortney},
  {Haas}, {Holman}, {Howell}, {Isaacson}, {Jenkins}, {Koch}, {Latham},
  {Lissauer}, {Moorhead}, {Morehead}, {Marcy}, {MacQueen}, {Quinn},
  {Ragozzine}, {Rowe}, {Sasselov}, {Seager}, {Torres}, \&
  {Welsh}}]{2010ApJ...725.1226S}
{Steffen}, J.~H., {Batalha}, N.~M., {Borucki}, W.~J., {et~al.} 2010, \apj, 725,
  1226, \dodoi{10.1088/0004-637X/725/1/1226}

\bibitem[{{Terquem} \& {Papaloizou}(2007)}]{2007ApJ...654.1110T}
{Terquem}, C., \& {Papaloizou}, J. C.~B. 2007, \apj, 654, 1110,
  \dodoi{10.1086/509497}

\bibitem[{{Thompson} {et~al.}(2018){Thompson}, {Coughlin}, {Hoffman},
  {Mullally}, {Christiansen}, {Burke}, {Bryson}, {Batalha}, {Haas},
  {Catanzarite}, {Rowe}, {Barentsen}, {Caldwell}, {Clarke}, {Jenkins}, {Li},
  {Latham}, {Lissauer}, {Mathur}, {Morris}, {Seader}, {Smith}, {Klaus},
  {Twicken}, {Van Cleve}, {Wohler}, {Akeson}, {Ciardi}, {Cochran}, {Henze},
  {Howell}, {Huber}, {Pr{\v{s}}a}, {Ram{\'\i}rez}, {Morton}, {Barclay},
  {Campbell}, {Chaplin}, {Charbonneau}, {Christensen-Dalsgaard}, {Dotson},
  {Doyle}, {Dunham}, {Dupree}, {Ford}, {Geary}, {Girouard}, {Isaacson},
  {Kjeldsen}, {Quintana}, {Ragozzine}, {Shabram}, {Shporer}, {Silva Aguirre},
  {Steffen}, {Still}, {Tenenbaum}, {Welsh}, {Wolfgang}, {Zamudio}, {Koch}, \&
  {Borucki}}]{2018ApJS..235...38T}
{Thompson}, S.~E., {Coughlin}, J.~L., {Hoffman}, K., {et~al.} 2018, \apjs, 235,
  38, \dodoi{10.3847/1538-4365/aab4f9}

\bibitem[{{Tremaine} \& {Dong}(2012)}]{2012AJ....143...94T}
{Tremaine}, S., \& {Dong}, S. 2012, \aj, 143, 94,
  \dodoi{10.1088/0004-6256/143/4/94}

\bibitem[{{Winn} \& {Fabrycky}(2015)}]{2015ARA&A..53..409W}
{Winn}, J.~N., \& {Fabrycky}, D.~C. 2015, \araa, 53, 409,
  \dodoi{10.1146/annurev-astro-082214-122246}

\bibitem[{{Winn} {et~al.}(2017){Winn}, {Petigura}, {Morton}, {Weiss}, {Dai},
  {Schlaufman}, {Howard}, {Isaacson}, {Marcy}, {Justesen}, \&
  {Albrecht}}]{2017AJ....154..270W}
{Winn}, J.~N., {Petigura}, E.~A., {Morton}, T.~D., {et~al.} 2017, \aj, 154,
  270, \dodoi{10.3847/1538-3881/aa93e3}

\bibitem[{{Xuan} \& {Wyatt}(2020)}]{2020MNRAS.497.2096X}
{Xuan}, J.~W., \& {Wyatt}, M.~C. 2020, \mnras, 497, 2096,
  \dodoi{10.1093/mnras/staa2033}

\bibitem[{{Yang} {et~al.}(2020){Yang}, {Xie}, \& {Zhou}}]{2020AJ....159..164Y}
{Yang}, J.-Y., {Xie}, J.-W., \& {Zhou}, J.-L. 2020, \aj, 159, 164,
  \dodoi{10.3847/1538-3881/ab7373}

\bibitem[{{Yee} {et~al.}(2018){Yee}, {Petigura}, {Fulton}, {Knutson},
  {Batygin}, {Bakos}, {Hartman}, {Hirsch}, {Howard}, {Isaacson}, {Kosiarek},
  {Sinukoff}, \& {Weiss}}]{2018AJ....155..255Y}
{Yee}, S.~W., {Petigura}, E.~A., {Fulton}, B.~J., {et~al.} 2018, \aj, 155, 255,
  \dodoi{10.3847/1538-3881/aabfec}

\bibitem[{{Zanazzi} \& {Lai}(2018)}]{2018MNRAS.478..835Z}
{Zanazzi}, J.~J., \& {Lai}, D. 2018, \mnras, 478, 835,
  \dodoi{10.1093/mnras/sty1075}

\bibitem[{{Zhu} \& {Dong}(2021)}]{2021arXiv210302127Z}
{Zhu}, W., \& {Dong}, S. 2021, arXiv e-prints, arXiv:2103.02127.
\newblock \doarXiv{2103.02127}

\bibitem[{{Zhu} {et~al.}(2018){Zhu}, {Petrovich}, {Wu}, {Dong}, \&
  {Xie}}]{2018ApJ...860..101Z}
{Zhu}, W., {Petrovich}, C., {Wu}, Y., {Dong}, S., \& {Xie}, J. 2018, \apj, 860,
  101, \dodoi{10.3847/1538-4357/aac6d5}

\bibitem[{{Zhu} \& {Wu}(2018)}]{2018AJ....156...92Z}
{Zhu}, W., \& {Wu}, Y. 2018, \aj, 156, 92, \dodoi{10.3847/1538-3881/aad22a}

\bibitem[{{Zink} {et~al.}(2019){Zink}, {Christiansen}, \&
  {Hansen}}]{2019MNRAS.483.4479Z}
{Zink}, J.~K., {Christiansen}, J.~L., \& {Hansen}, B. M.~S. 2019, \mnras, 483,
  4479, \dodoi{10.1093/mnras/sty3463}

\end{thebibliography}

\end{document}